\documentclass[letter,12pt]{article}
\pdfoutput=1 
\usepackage{jheppub} 
\usepackage{subfigure}

\title{Searches of exotic Higgs bosons in general mass spectra of the Georgi-Machacek model at the LHC}

\author[a,b,c]{Cheng-Wei~Chiang,}
\author[a]{An-Li~Kuo,}
\author[a,d]{and Toshifumi~Yamada}

\affiliation[a]{Department of Physics and Center for Mathematics and Theoretical Physics, \\
National Central University, Chungli, Taiwan 32001, Republic of China}
\affiliation[b]{Institute of Physics, Academia Sinica, Taipei, Taiwan 11529, Republic of China}
\affiliation[c]{Physics Division, National Center for Theoretical Sciences, Hsinchu, Taiwan 30013, R.O.C.}
\affiliation[d]{School of Physics, Korea Institute for Advanced Study, Seoul 130-722, Republic of Korea}

\emailAdd{chengwei@ncu.edu.tw}
\emailAdd{101222028@cc.ncu.edu.tw}
\emailAdd{toshifumi@kias.re.kr}

\abstract{
We derive the most general sets of viable mass spectra of the exotic Higgs bosons in the Georgi-Machacek model that are consistent with the theoretical constraints of vacuum stability and perturbative unitarity and the experimental constraints of electroweak precision observables, $Zb \bar b$ coupling and Higgs boson signal strengths.  Branching ratios of various cascade decay channels of the doubly-charged Higgs boson in the ${\bf 5}$ representation, the singly-charged Higgs boson in ${\bf 3}$, and the singlet Higgs boson are further computed.  
As one of the most promising channels for discovering the model, we study the prospects for detecting the doubly-charged Higgs boson that is produced via the vector boson fusion process and decays into final states containing a pair of same-sign leptons at the 14-TeV LHC and a 100-TeV future $pp$ collider.  
For this purpose, we evaluate acceptance times efficiency for signals of the doubly-charged Higgs boson with general viable mass spectra and compare it with the standard model background estimates.
}

\preprint{\\ \rightline{}}
\keywords{}

\begin{document} 
\maketitle
\flushbottom

\section{Introduction}

Proposed in the mid 80s, the Georgi-Machacek (GM) model~\cite{Georgi:1985nv,Chanowitz:1985ug} augments the SM Higgs sector by adding a complex triplet of hypercharge $Y = 1$ and a real triplet of $Y = 0$ under the SM $SU(2)_L \times U(1)_Y$ gauge symmetry.  It has many intriguing properties.  First, the triplet fields can develop vacuum expectation values (VEV's), as automatically induced by SM electroweak symmetry breaking (EWSB) through a trilinear interaction term with the SM Higgs doublet field.  With the triplet VEV's, it is possible to give Majorana mass to the left-handed neutrinos through the so-called type-II seesaw mechanism.

Secondly, the model predicts the existence of several Higgs multiplets under the custodial symmetry: two singlets, one triplet, and one quintet~\cite{Gunion:1989ci}.  In particular, the quintet contains a doubly-charged Higgs boson that can mediate lepton number-violating or even lepton flavor-violating processes.  Recently, there have been many phenomenological studies about searching for the exotic Higgs bosons at colliders~\cite{Haber:1999zh,Godfrey:2010qb,Chiang:2012dk,Chiang:2012cn,Englert:2013zpa,Englert:2013wga,Chiang:2013rua,Hartling:2014zca,Chiang:2014bia,Hartling:2014aga,Chiang:2015rva} and their effects in enhancing the strength of phase transition in electroweak baryogenesis~\cite{Chiang:2014hia}.

Thirdly, with the assumption of vacuum alignment between the complex and real triplet VEV's in the tree potential, the model preserves the electroweak $\rho$ parameter at unity even with a VEV as large as up to a few tens of GeV.~\footnote{It is noted that divergences for the $\rho$ parameter and certain mixings among the Higgs bosons have been studied at loop levels and found to have a similar naturalness issue as the SM Higgs mass~\cite{Gunion:1990dt}.}  The possibility of a large triplet VEV leads to enhanced couplings between the exotic Higgs bosons and the weak gauge bosons and thus a plethora of interesting collider phenomena.  For example, without a significant mass hierarchy among different Higgs multiplets, the doubly-charged Higgs boson decays dominantly into a pair of like-sign $W$ bosons rather than like-sign leptons.

Yet another feature impossible for models extended with only $SU(2)_L$ singlet and/or doublet fields is that the coupling between the SM-like Higgs boson and the weak gauge bosons can be stronger than in the SM as a result of mixing between the SM doublet and the triplet fields~\cite{Logan:2010en,Falkowski:2012vh,Chang:2012gn,Chiang:2013rua}.  This, for example, can be tested through a precise determination of the SM-like Higgs signal strengths at the LHC.
Finally, the model predicts the existence of a singly charged Higgs boson coupling with the $W$ and $Z$ bosons at tree level through mixing, while such a vertex is induced only at loop levels in singlet- and/or doublet-extended models~\cite{Grifols}, such as the two-Higgs doublet model~\cite{hwz1,hwz2}.

The structure of this paper is as follows.  Section~\ref{sec:GM} briefly reviews the GM model, paying particular attention to the mass spectrum, some tree-level theoretical constraints, and indirect experimental constraints.  In section~\ref{sec:parameter_search}, we perform a comprehensive scan of the mass spectrum allowed by the above-mentioned constraints.  We here incorporate the most general case in which there can be a mass hierarchy among the different Higgs multiplets.  The Higgs masses, signal strengths of the SM-like Higgs boson decays into $\gamma\gamma$ and $\gamma Z$, and branching ratios of (cascade) decays of the exotic Higgs bosons are plotted. 
In section~\ref{sec:numerical_results}, we concentrate on one of the signals of the GM model, namely, the vector boson fusion production of the doubly-charged Higgs boson that decays into final states with a same-sign lepton pair at the LHC, for which we evaluate the production cross section and the acceptance times efficiency with a certain set of selection criteria. Combining them with the branching ratios of the exotic Higgs boson decays evaluated in the previous section, and comparing them with SM background estimates, we obtain the prospect for the discovery of the GM model through this channel for most general mass spectra.
We also comment on the phenomenology at a 100-TeV hadron collider.  Finally, section~\ref{sec:summary} summarizes our findings in this work.

\section{Review on the Georgi-Machacek Model \label{sec:GM}}

In this section, we review the basics of the Higgs sector in the GM model, theoretical constraints of the vacuum stability and perturbative unitarity, both at tree level, and indirect experimental constraints, such as the oblique corrections, the $Zb\bar b$ vertex, and 125-GeV Higgs signal strengths.

\subsection{Higgs Sector and Mass Spectrum}

The EWSB sector of the GM model~\cite{Georgi:1985nv,Chanowitz:1985ug} comprises one isospin doublet scalar field with hypercharge $Y=1/2$, one isospin triplet scalar field with $Y=1$, and one isospin triplet scalar field with $Y=0$~\footnote{Here the normalization for the hypercharge quantum number $Y$ is such that the electric charge $Q = I_3 + Y$, where $I_3$ denotes the third component of the weak isospin number.}.
These fields are denoted respectively by~\footnote{Here we use the convention that $\chi^{--}=(\chi^{++})^*$, $\chi^{-}=(\chi^{+})^*$, $\xi^-= (\xi^+)^*$ and $\phi^-= (\phi^+)^*$.}
\begin{align}
\begin{split}
&
\phi=
\begin{pmatrix}
\phi^+ \\ \phi^0
\end{pmatrix}
~,~~
\chi=
\begin{pmatrix}
\chi^{++} \\ \chi^+ \\ \chi^0
\end{pmatrix}
~,~~
\xi=
\begin{pmatrix}
\xi^+ \\ \xi^0 \\ -(\xi^+)^*
\end{pmatrix}
~,
\\
&
\mbox{with }
\phi^0 = \frac{1}{\sqrt{2}}(h_{\phi}+ia_{\phi}) ~,~
\chi^0 = \frac{1}{\sqrt{2}}(h_{\chi}+ia_{\chi}) ~,~
\xi^0 = h_{\xi} ~,
\end{split}
\label{eq:component_fields}
\end{align}
where the neutral components have been further decomposed into CP-even ones ($h_{\phi}, \, h_{\chi}, \, h_{\xi}$) and CP-odd ones ($a_{\phi}, \, a_{\chi}$).
The global SU(2)$_L \times$SU(2)$_R$ symmetry is imposed on the Higgs potential at tree level, which is explicitly broken by the Yukawa and the hypercharge gauge interactions.
To make this symmetry manifest, it is convenient to introduce the SU(2)$_L \times$SU(2)$_R$-covariant forms of the fields:
\begin{align}
\begin{split}
\Phi \ &\equiv \ \left( \epsilon_2 \phi^*, \, \phi \right)
\ = \ \left(
\begin{array}{cc}
(\phi^0)^* & \phi^+ \\
-(\phi^+)^* & \phi^0
\end{array}
\right) ~,~~ \mbox{with }
\epsilon_2 = \left( \begin{array}{cc}
0 & 1 \\
-1 & 0
\end{array}
\right) ~,
\\
\Delta \ &\equiv \ \left( \epsilon_3 \chi^*, \, \xi, \, \chi \right)
\ = \ \left(
\begin{array}{ccc}
(\chi^0)^*    & \xi^+ & \chi^{++} \\
-(\chi^+)^*   & \xi^0 & \chi^+ \\
(\chi^{++})^* & -(\xi^+)^* & \chi^0
\end{array}
\right) ~,~~
\mbox{with }
\epsilon_3 = \left( \begin{array}{ccc}
0 & 0 & 1 \\
0 & -1 & 0 \\
1 & 0 & 0
\end{array}
\right) ~.
\end{split}
\end{align}
Under an SU(2)$_L$~$\times$~SU(2)$_R$ transformation, $\Phi \rightarrow U_{2L} \Phi U^{\dagger}_{2R}$ and $\Delta \rightarrow U_{3L} \Delta U^{\dagger}_{3R}$, where $U_{2L}$ $(U_{2R})$ is the two-dimensional representation of the SU(2)$_L$ (SU(2)$_R$) group component and $U_{3L}$ $(U_{3R})$ is the corresponding three-dimensional one.

Using $\Phi$ and $\Delta$, the Lagrangian of the EWSB sector is succinctly given by
\begin{align}
{\cal L} \ =& \ 
\frac{1}{2} {\rm tr}[ (D^{\mu}\Phi)^{\dagger} D_{\mu}\Phi ]
\ + \ 
\frac{1}{2} {\rm tr}[ (D^{\mu}\Delta)^{\dagger} D_{\mu}\Delta ] 
\ - \ V(\Phi, \, \Delta) ~,
\end{align}
where $D_{\mu}$ denotes the covariant derivative for $\Phi$ or $\Delta$.
The potential term, $V(\Phi, \, \Delta)$, is given by
\begin{align}
\begin{split}
V(\Phi, \, \Delta) \ =& \ \frac{1}{2} m_1^2 \, {\rm tr}[ \Phi^{\dagger} \Phi ] + 
\frac{1}{2} m_2^2 \, {\rm tr}[ \Delta^{\dagger} \Delta ]
 +  \lambda_1 \left( {\rm tr}[ \Phi^{\dagger} \Phi ] \right)^2 
 +  \lambda_2 \left( {\rm tr}[ \Delta^{\dagger} \Delta ] \right)^2 
\\
&
 +  \lambda_3 {\rm tr}\left[ \left( \Delta^{\dagger} \Delta \right)^2 \right] 
 +  \lambda_4 {\rm tr}[ \Phi^{\dagger} \Phi ] {\rm tr}[ \Delta^{\dagger} \Delta ]
 +  \lambda_5 {\rm tr}\left[ \Phi^{\dagger} \frac{\sigma^a}{2} \Phi \frac{\sigma^b}{2} \right] 
                  {\rm tr}[ \Delta^{\dagger} T^a \Delta T^b]
\\
 &+ \mu_1 {\rm tr}\left[ \Phi^{\dagger} \frac{\sigma^a}{2} \Phi \frac{\sigma^b}{2} \right]
                               (P^{\dagger} \Delta P)_{ab}
 + \mu_2 {\rm tr}[ \Delta^{\dagger} T^a \Delta T^b]
                               (P^{\dagger} \Delta P)_{ab} ~,
\end{split}
\label{potential}
\end{align}
where summations over $a,b = 1,2,3$ are understood, $\sigma$'s and $T's$ are the $2\times2$ (Pauli matrices) and $3\times3$ matrix representations of the SU(2) generators, respectively, and
\begin{align}
P &= \frac{1}{\sqrt{2}} \left( \begin{array}{ccc}
-1 & i & 0 \\
0 & 0 & \sqrt{2} \\
1 & i & 0
\end{array}
\right) \nonumber
\end{align}
diagonalizes the adjoint representation of the SU(2) generator.  It is noted that all parameters in the Higgs potential are real and do not allow CP violation.

The EWSB vacuum is derived from the tadpole conditions:
\begin{align}
\frac{\partial V(\Phi,\Delta)}{\partial h_{\phi}} \ &= \ \frac{\partial V(\Phi,\Delta)}{\partial h_{\chi}}
\ = \ \frac{\partial V(\Phi,\Delta)}{\partial h_{\xi}} \ = \ 0 ~,
\label{vacuum}
\end{align}
where the fields other than $h_{\phi},h_{\chi}$, and $h_{\xi}$ take zero VEV's.
In Eq.~(\ref{vacuum}), we select the solution satisfying the relation $\langle h_{\chi} \rangle=\sqrt{2} \langle h_{\xi} \rangle$, by which the EWSB vacuum maintains the diagonal SU(2)$_{L+R}$ or SU(2)$_V$ symmetry.
We denote the VEV's of $h_{\phi}, h_{\chi}, h_{\xi}$ by
 $\langle h_{\phi} \rangle = v_{\Phi}, \ \langle h_{\chi} \rangle = \sqrt{2}v_{\Delta}, \ \langle h_{\xi} \rangle = v_{\Delta}$, respectively, which are related to the SM Higgs boson VEV, $v \simeq 246$~GeV, by
 $\vert\langle h_{\phi} \rangle\vert^2 + 2\vert\langle h_{\chi} \rangle\vert^2 + 4\vert\langle h_{\xi} \rangle\vert^2 = v_{\Phi}^2+8v_{\Delta}^2 = v^2$.
In a fashion similar to the two-Higgs doublet model, we define $\tan \beta$ as the VEV ratio, $\tan \beta \equiv v_{\Phi}/ \left( 2\sqrt{2}v_{\Delta} \right)$.
Assuming $v_{\Phi},v_{\Delta} \neq 0$~\footnote{As alluded to earlier, the triplet VEV can be automatically induced by the $\mu_1$ term once the doublet gets a VEV to break the electroweak symmetry.}, we can rewrite $m_1^2, m_2^2$ in terms of $v_{\Phi},v_{\Delta}$ as
\begin{align}
m_1^2 \ &= \ -4\lambda_1 v_{\Phi}^2 - 6\lambda_4 v_{\Delta}^2 - 3\lambda_5 v_{\Delta}^2
- \frac{3}{2} \mu_1 v_{\Delta} ~, \nonumber \\
m_2^2 \ &= \ -12\lambda_2 v_{\Delta}^2 - 4\lambda_3 v_{\Delta}^2 - 2\lambda_4 v_{\Phi}^2
- \lambda_5 v_{\Phi}^2 - \mu_1 \frac{v_{\Phi}^2}{4v_{\Delta}} - 6\mu_2 v_{\Delta} ~.
\label{m1m2}
\end{align}

For later convenience, we define
\begin{align}
M_1^2 \ &\equiv \ -\frac{v}{\sqrt{2} \cos\beta} \mu_1 ~, 
\ \ \ M_2^2 \ \equiv \ -3\sqrt{2} \cos\beta \, v \mu_2 ~.
\end{align}
It is noted that $|\mu_1|$ or $M_1^2 \to \infty$ corresponds to the decoupling limit of the model~\cite{Chiang:2012cn,Hartling:2014zca}.  On the other hand, no decoupling limit exists once one imposes the $Z_2$ symmetry $\Delta \to - \Delta$.  Also, this symmetry does not allow the desired interaction between left-handed neutrinos and the triplet Higgs field for neutrino mass generation.

Because of the SU(2)$_V$ symmetry of the (tree-level) EWSB vacuum, the physical mass eigenstates form one 5-plet, one 3-plet and two singlets, where the components in each of the multiplets are degenerate in mass at the tree level.  Mass splitting within each multiplet due to custodial symmetry breaking is expected to be at the ${\cal O}(100)$~MeV level.
We denote the 5-plet, 3-plet and two singlets by $H_5 = (H_5^{++},H_5^{+},H_5^{0},H_5^{-},H_5^{--})^T$, $H_3=(H_3^{+},H_3^{0},H_3^{-})^T$, $H_1$ and $h$, respectively, with $h$ identified as the 125-GeV SM-like Higgs boson observed at the LHC.  In terms of the fields $\phi$, $\xi$ and $\chi$ introduced in Eq.~(\ref{eq:component_fields}), the physical states are expressed as follows:
\begin{align}
\begin{split}
&
H_5^{++} = \chi^{++} ~, \ \ \ H_5^+ = \frac{1}{\sqrt{2}} \left( \chi^+ - \xi^+ \right) ~, 
\ \ \ H_5^0 = \sqrt{\frac{1}{3}} h_{\chi} - \sqrt{\frac{2}{3}} h_{\xi} ~, 
\\
&
H_3^+ = -\cos \beta \, \phi^+ + \sin \beta \, \frac{1}{\sqrt{2}} \left( \chi^+ + \xi^+ \right) ~, 
\ \ \ H_3^0 = -\cos \beta \, a_{\phi} + \sin \beta \, a_{\chi} ~, 
\\
&
h = \cos \alpha \, h_{\phi} - \frac{\sin \alpha}{\sqrt3} \, \left( \sqrt{2} h_{\chi} + h_{\xi} \right) ~,
\ \ \ H_1 = \sin \alpha \, h_{\phi} + \frac{\cos \alpha}{\sqrt3} \, \left( \sqrt{2} h_{\chi} + h_{\xi} \right) ~,
\end{split}
\end{align}
where the mixing angle $\alpha$ between the singlets takes a value in the range $-\pi/2 \le \alpha \le \pi/2$, and is given through
\begin{align}
\tan 2\alpha &= \frac{2 (M^2)_{12}}{(M^2)_{22} - (M^2)_{11}} ~, 
\end{align}
 with
\begin{align}
\begin{split}
(M^2)_{11} &= 8 \lambda_1 v^2 \sin^2 \beta ~,
\\
(M^2)_{22} &= (3\lambda_2+\lambda_3) v^2 \cos^2 \beta + M_1^2 \sin^2 \beta - \frac{1}{2} M_2^2 ~,
\\
(M^2)_{12} &= \sqrt{\frac{3}{2}} \sin \beta \cos \beta \, \left[ (2\lambda_4+\lambda_5) v^2 - M_1^2 \right] ~.
\end{split}
\end{align}
 The mass eigenvalues are given by
\begin{align}
\begin{split}
m_{H_5}^2 &\equiv m_{H_5^{++}}^2=m_{H_5^+}^2=m_{H_5^0}^2
= (M_1^2-\frac{3}{2}\lambda_5v^2)\sin^2\beta+\lambda_3v^2\cos^2\beta+M_2^2 ~,
\\
m_{H_3}^2 & \equiv m_{H_3^+}^2 = m_{H_3^0}^2=M_1^2-\frac{1}{2}\lambda_5v^2 ~,
\\
m_{H_1^0}^2 &= M_{11}^2\sin^2 \alpha+M_{22}^2\cos^2 \alpha+2M_{12}^2\sin \alpha\cos\alpha ~,
\\
m_h^2 &=M_{11}^2\cos^2\alpha+M_{22}^2\sin^2\alpha-2M_{12}^2\sin \alpha\cos\alpha ~.
\end{split}
\label{massformulas}
\end{align}
It is noted that these masses are generally different, and the mass differences are of ${\cal O}(100)$~GeV if one na{\"i}vely takes $\mu_{1,2} \sim {\cal O}(100)$~GeV and the quartic couplings $\lambda$'s $\sim {\cal O}(1)$.  In our numerical analysis, we will assume a general mass hierarchy among these mass eigenvalues (but neglecting the smaller mass splitting within each multiplet), subject to the constraints to be discussed below, and analyze the prospects of detecting the exotic Higgs bosons at the 14-TeV LHC and future 100-TeV hadron collider.

\subsection{Theoretical Constraints}

We will take into account two theoretical constraints on the parameters of the GM Higgs potential.  One comes from the stability of the electroweak vacuum, and the other from the unitarity of the perturbation theory.  We satisfy ourselves with these constraints at the tree level for the consistency with the masses given above.

When requiring the electroweak vacuum to be stable ({\it i.e.}, bounded from below), one obtains the following constraints for the quartic couplings~\cite{arhrib,Chiang:2012cn}:
\begin{align}
\begin{split}
& \lambda_1 > 0 ~, \ \lambda_2+\lambda_3 > 0 ~, \ \lambda_2 + \frac{1}{2}\lambda_3 > 0 ~,
\ -\vert \lambda_4 \vert + 2\sqrt{\lambda_1(\lambda_2+\lambda_3)} > 0 ~,
\\
& \lambda_4 - \frac{1}{4}\vert \lambda_5 \vert + \sqrt{2\lambda_1(2\lambda_2+\lambda_3)} > 0 ~.
\end{split}
\label{stability}
\end{align}
From the perturbative unitarity, we have another set of constraints~\cite{Aoki:2007ah,Hartling:2014zca}:~
\footnote{See also Ref.~\cite{Grinstein:2013fia} for constraints on additional Higgs bosons based on Higgs data and unitarity.}
\begin{align}
\begin{split}
&
\left\vert \, 6 \lambda_1 + 7 \lambda_3 + 11\lambda_2\, \right\vert +\sqrt{(6\lambda_1-7\lambda_3-11\lambda_2)^2+36\lambda_4^2}< 4\pi ~,
\\
&
\left\vert \,  \lambda_4 - \lambda_5 \, \right\vert < 2\pi ~,
~~~ \left\vert \, 2 \lambda_3 + \lambda_2  \, \right\vert < \pi ~,
\\
&
\left\vert \, 2 \lambda_1 -  \lambda_3 + 2\lambda_2\, \right\vert 
+ \sqrt{(2\lambda_1+\lambda_3-2\lambda_2)^2+\lambda_5^2} < 4\pi ~.
\end{split}
\label{unitarity}
\end{align}
Since one can trade the quartic couplings and $M_{1,2}^2$ with the four physical masses, $\alpha$, $\beta$ and $v$, the above two sets of constraints can be turned into constraints on the unknown masses and mixing angles.

\subsection{Experimental Constraints}

We now turn to the discussion of constraints on the GM model derived from measurements of SM quantities in collider experiments.  These include the electroweak precision tests, the determination of the $Z b \bar{b}$ coupling, and the measurement of the Higgs boson signal strengths.

The GM model is subject to constraints from the $S$ and $U$ parameters~\cite{pt} of electroweak precision tests.
The $\rho$ parameter can take any value in the GM model if we add a term that explicitly breaks the SU(2)$_V$ symmetry, and hence the $T$ parameter does not impose any restriction on the model.
Since the absolute value of the $U$ parameter is found to be below $0.01$ in all the mass spectra generated in the next section, we will only consider the constraint from the $S$ parameter by fixing the $U = 0$ and taking the $T$ parameter to be free.
The latest experimental data~\cite{pdg} report the following $1\sigma$ range for the $S$ parameter:
\begin{align}
S \vert_{U=0, \, T \, {\rm free}} \ &= \ 0.00 \pm 0.08 ~.
\label{Sparameter}
\end{align}

In the GM model, the 3-plet Higgs bosons couple with the SM quarks through mixing with the Higgs doublet, as explicitly given, for example, in Ref.~\cite{Chiang:2012cn}.  Therefore, $H_3^+$ can give rise to significant radiative corrections to the $Z b \bar{b}$ coupling through a triangular one-loop diagram involving the top quark and $H_3^+$, as the $\bar{t} b H_3^+$ coupling has an overall factor proportional to $v_\Delta^2$ and a term proportional to the top quark Yukawa coupling.
The data on the $Z b \bar{b}$ coupling therefore impose a constraint on the triplet VEV $v_{\Delta}$ and the SU(2)$_V$ triplet mass $m_{H_3}$, which has been evaluated in Ref.~\cite{Chiang:2012cn}.
It was found that mass spectra with $v_{\Delta} \lesssim 50$~GeV and $m_{H_3}$ above 100~GeV were consistent at $2\sigma$ level with the current data~\cite{pdg}.

The signal strengths of the SM-like Higgs boson production and decay in various channels have been measured in the LHC 7-TeV and 8-TeV runs by the ATLAS \cite{ATLASsignalstrength} and CMS \cite{CMSsignalstrength} Collaborations, and provide significant constraints on the couplings of $h$ to SM particles in the GM model.
Here we consider the following six channels of the Higgs boson production and decay: the gluon-gluon fusion (GGF) production of $h$ decaying into $ZZ$, $WW$ and $\tau^+ \tau^-$, the vector boson fusion (VBF) production of $h$ decaying into $WW$ and $\tau^+ \tau^-$, and the vector boson associated (VBA) production of $h$ decaying into $b \bar{b}$.
The modification of the signal strengths in these channels depends only on the triplet VEV, $v_{\Delta}$ (or $\beta$), and the mixing angle of the SU(2)$_V$ singlets, $\alpha$.  Hence we can directly constrain $v_{\Delta}$ and $\alpha$ from the data, without specifying other parameters including the mass spectrum.  Note that we avoid using the diphoton signal strength because it is a loop-mediated process that has additional dependences on the masses of heavy charged Higgs bosons and their couplings with $h$.  Although such uncertainties in the diphoton channel will also enter the signal strengths of the above-mentioned six tree-level decay channels through modifications in the branching ratios, the effects are expected to be negligible because of the relatively small $h \to \gamma\gamma$ decay rate.  Throughout this paper, we employ the narrow width approximation when calculating the signal strengths.

\begin{figure}[ht]
\centering
\includegraphics[scale=0.5]{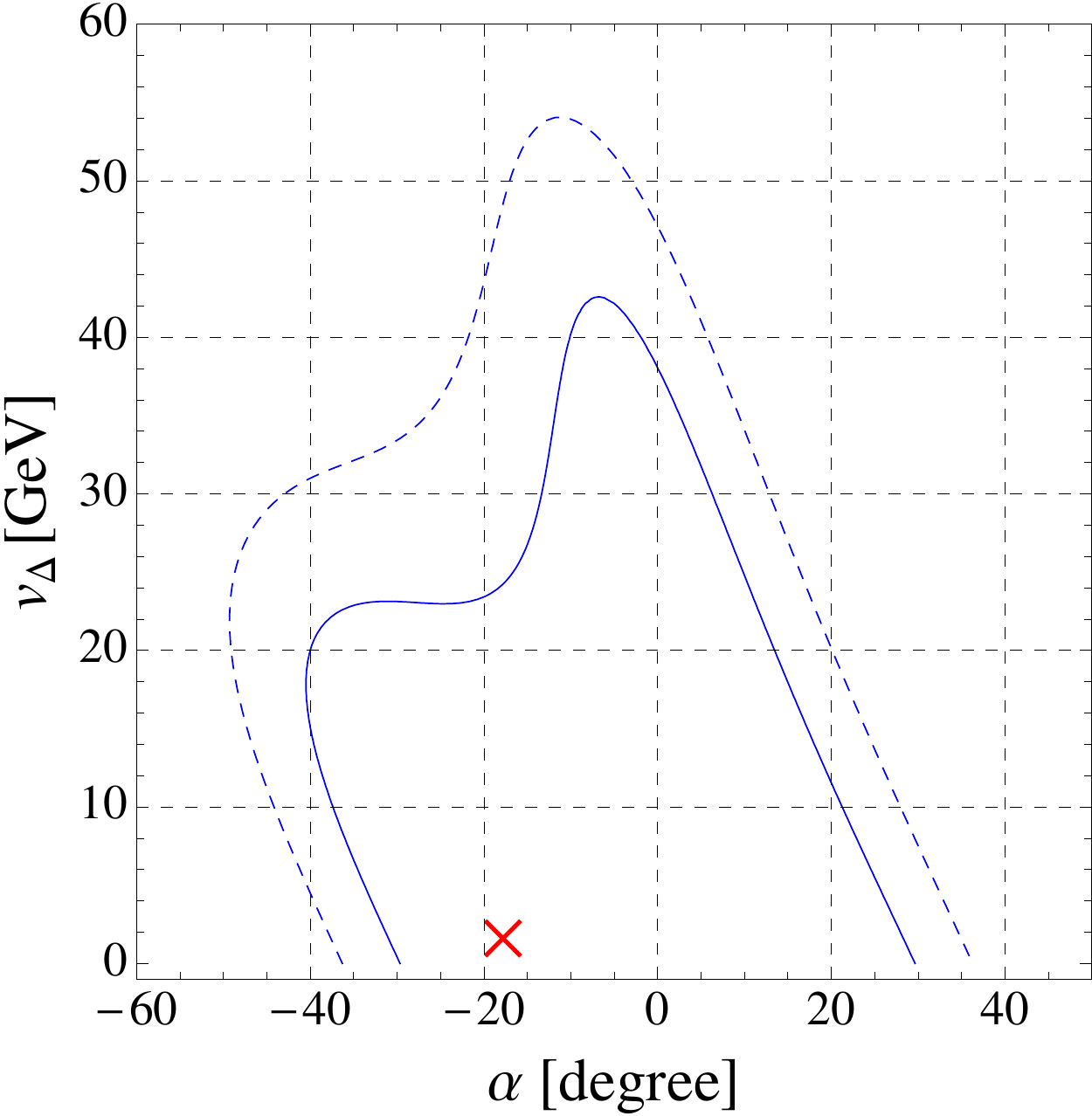}
\caption{$1\sigma$ (solid) and $2\sigma$ (dashed) contours on the $v_\Delta$-$\alpha$ plane through a $\chi^2$ fit to the current data of six Higgs signal strengths detailed in the main text.  The red cross marks the point with the $\chi^2$ minimum.}
\label{chisq}
\end{figure}

We perform a $\chi^2$ fit on $v_{\Delta}$-$\alpha$ plane by using the signal strength data of the above-mentioned six channels obtained in the LHC 7-TeV and 8-TeV runs~\cite{ATLASsignalstrength,CMSsignalstrength}.  The $1\sigma$ and $2\sigma$ contours along with the best-fit point are displayed in Fig.~\ref{chisq}.  From the figure, we select the following twelve sets of parameters that are consistent with the data at the $2\sigma$ level: $(v_\Delta,\alpha) = (10, -30^\circ)$, $(10, -10^\circ)$, $(10, +10^\circ)$, $(20, -30^\circ)$, $(20, -10^\circ)$, $(20, +10^\circ)$, $(30, -30^\circ)$, $(30, -10^\circ)$, $(30, +10^\circ)$, $(40, -10^\circ)$, $(50, -10^\circ)$, and $(1, 0^\circ)$ (close to the decoupling limit), where the values of $v_\Delta$ are given in units of GeV.  These parameter choices will be used in the next section for numerical studies.

\begin{figure}[ht]
\centering
\includegraphics[scale=0.5]{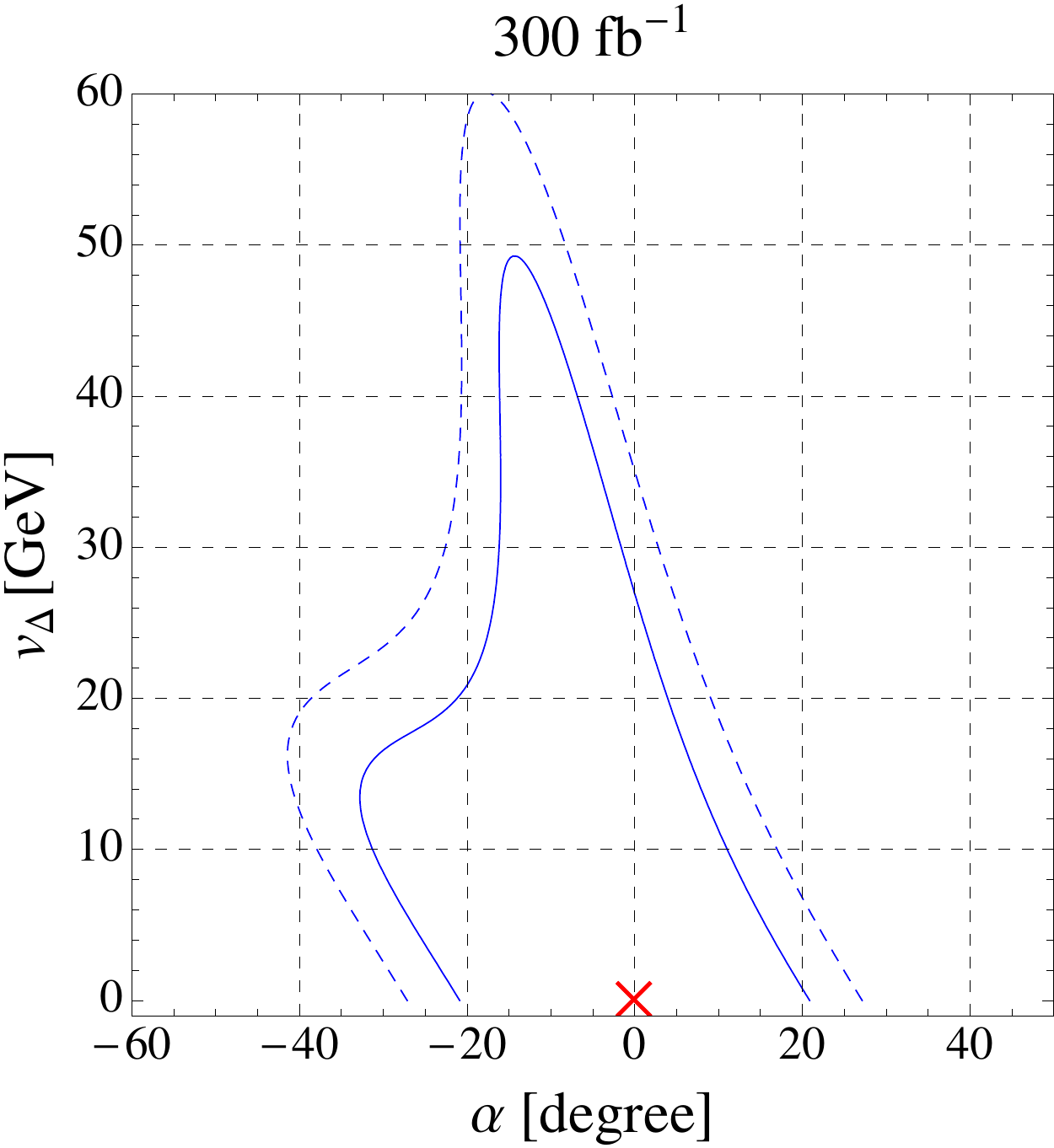}
\hspace{10mm}
\includegraphics[scale=0.5]{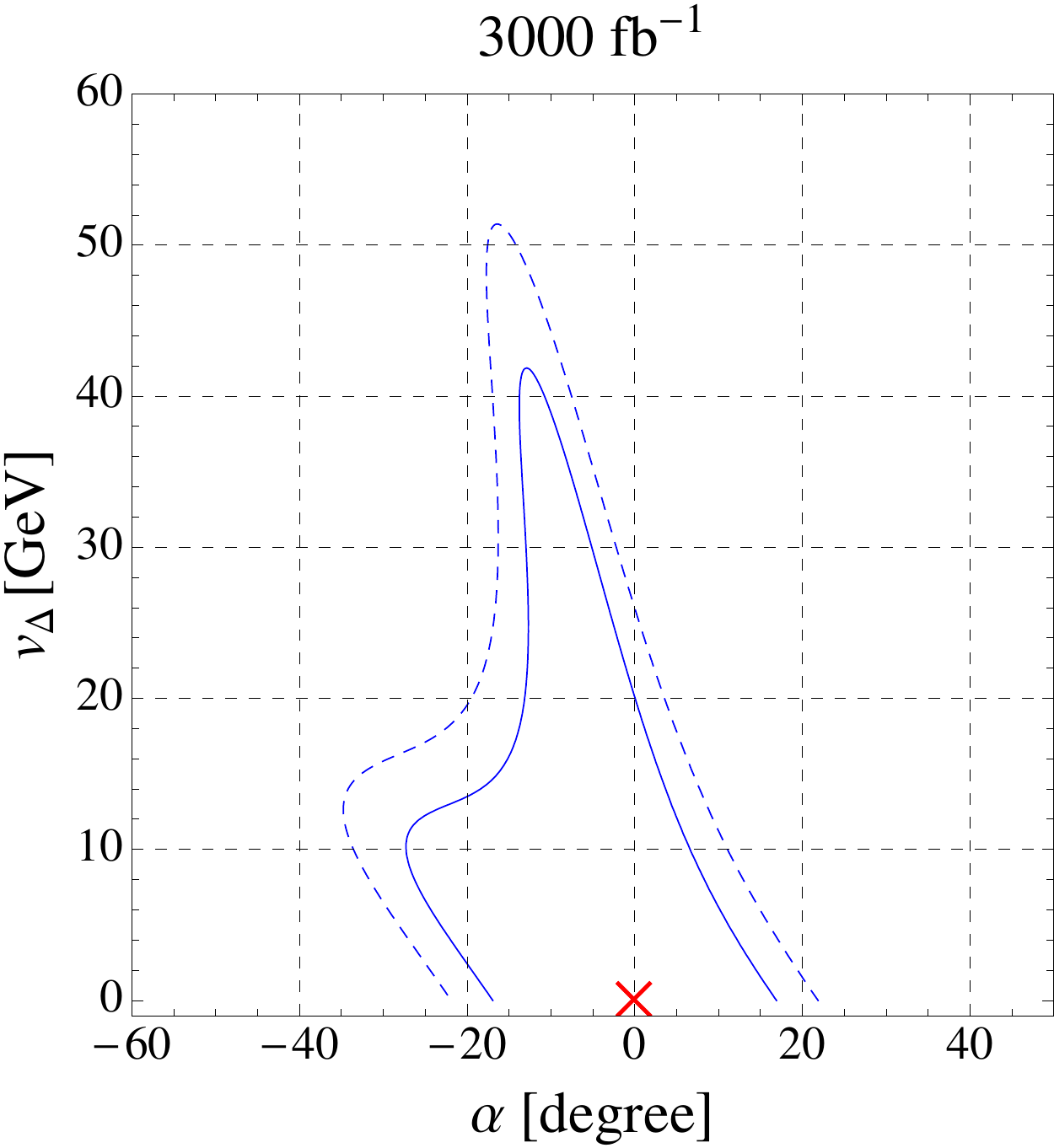}
\caption{$1\sigma$ (solid) and $2\sigma$ (dashed) contours on the $v_\Delta$-$\alpha$ plane through a $\chi^2$ fit to the SM signal strengths of the six channels with precisions expected to reach at 14-TeV LHC with 300~fb$^{-1}$ (left) and 3000~fb$^{-1}$ (right) of data.}
\label{chisqhigherfb}
\end{figure}

As a reference, we also present in Fig.~\ref{chisqhigherfb} the future prospects for confining the $v_{\Delta}$-$\alpha$ parameter space as derived from the same six signal strength measurements at the 14~TeV LHC with the integrated luminosities of 300~fb$^{-1}$ (left plot) and 3000~fb$^{-1}$ (right plot), assuming that the central values of the signal strength data are all unity.
We here use the uncertainty estimates given in Ref.~\cite{futurelhc}.  Although the constraint in the left plot of Fig.~\ref{chisqhigherfb} looks comparable to that in Fig.~\ref{chisq}, such a comparison is meaningless because the former assumes the SM signal strengths.  The constraint on $v_{\Delta}$, $\alpha$ does not improve significantly with 3000~fb$^{-1}$ of data compared to the case with 300~fb$^{-1}$ of data.  This is because, for the $h\to WW$, $ZZ$ and $\tau\tau$ channels, theoretical systematic uncertainties and experimental systematic uncertainties give major contributions to the overall uncertainty for the 300~fb$^{-1}$ data.  Hence, larger statistics does not lead to a significant reduction in uncertainties of Higgs boson signal strengths.

The signal strength of GGF production of the Higgs boson decaying into $\gamma \gamma$ has also been measured at the 7~TeV and 8~TeV LHC.  However, as alluded to before, the branching ratio of $h \to \gamma \gamma$ can be altered by the loop diagrams involving the charged Higgs bosons $(H_5^{++}, \, H_5^+, \, H_3^+)$ and hence depends significantly on details of the mass spectrum and triple scalar couplings.  Therefore, we will discuss the constraint from this channel after we perform a parameter scan for the Higgs mass spectrum in the next section.

We now comment on constraints from searches for an extra neutral Higgs boson through the $H_1\to\gamma\gamma$ process, as this mode yields the strongest bound.  The ATLAS Collaboration has already given a bound on this process for the mass range of 65 GeV to 600 GeV~\cite{ATLASdiphoton}.  However, we will not use this constraint in our analysis in the next section, because $BR(H_1\to\gamma\gamma)$ varies sensitively with the values of $M_1^2$ and $M_2^2$ while these parameters are taken to be free in our parameter scan.

\section{Search of Viable Exotic Higgs Boson Mass Spectra and Decay Branching Ratios of Exotic Higgs Bosons \label{sec:parameter_search}}

\begin{figure}[ht]
\centering
\includegraphics[scale=0.16]{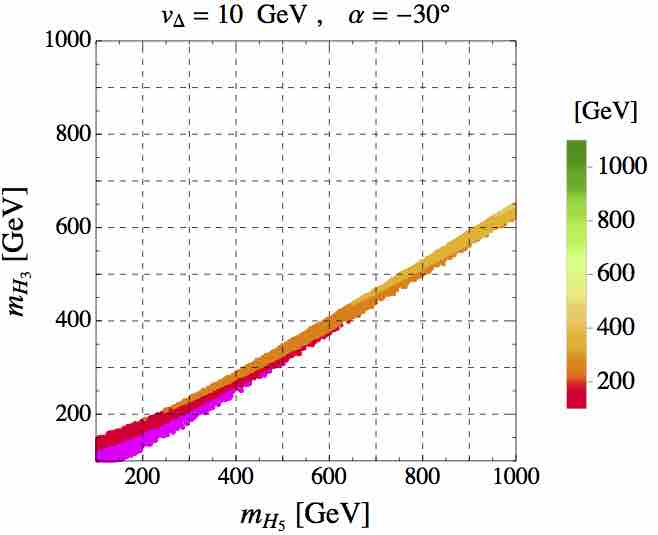}
\includegraphics[scale=0.16]{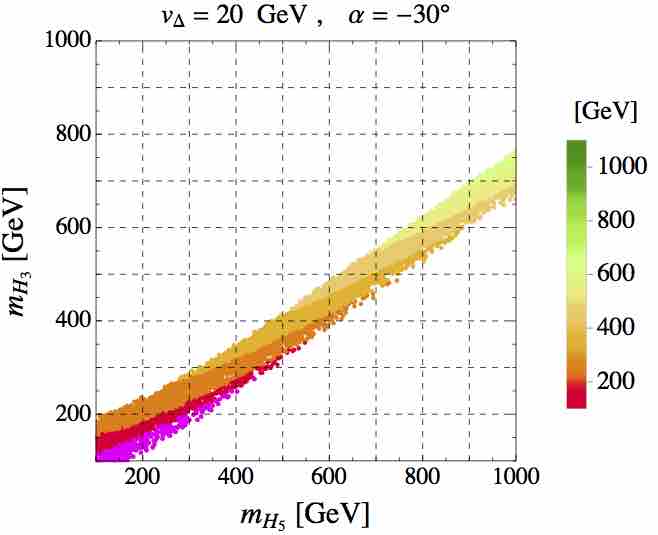}
\includegraphics[scale=0.16]{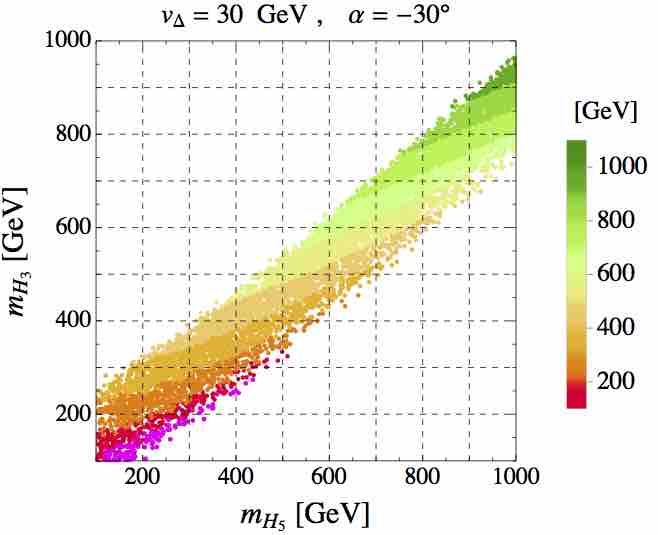}
\includegraphics[scale=0.16]{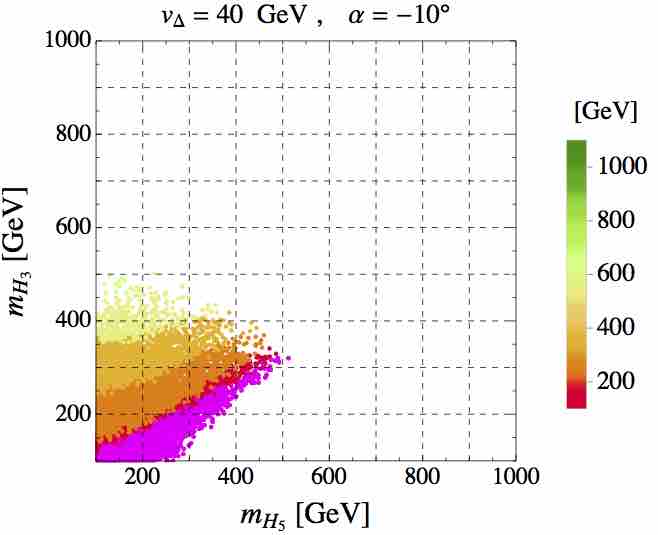}
\vspace{2mm}
\\
\includegraphics[scale=0.16]{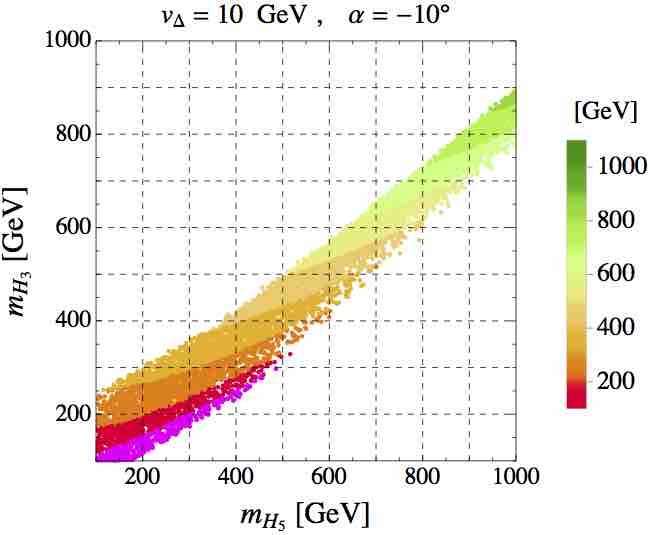}
\includegraphics[scale=0.16]{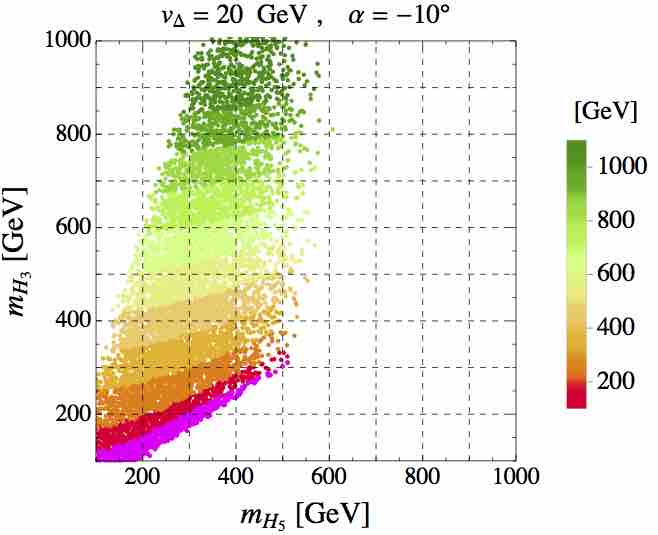}
\includegraphics[scale=0.16]{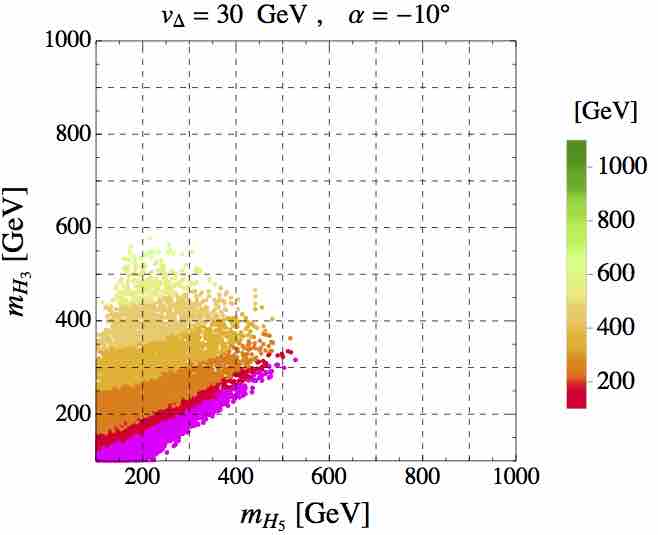}
\includegraphics[scale=0.16]{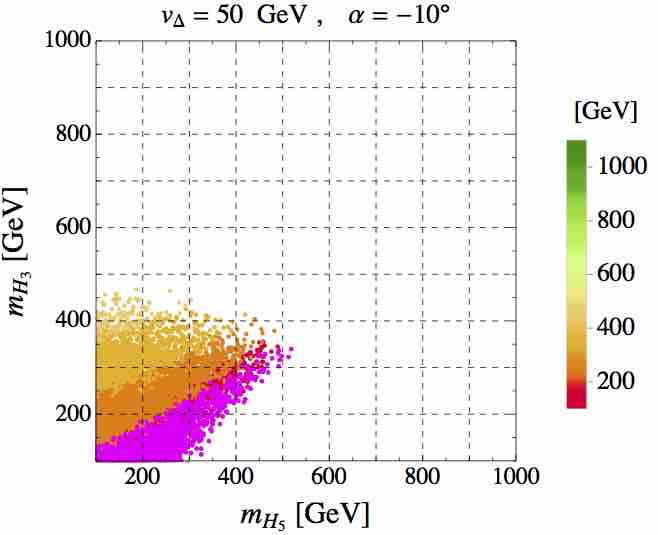}
\vspace{2mm}
\\
\includegraphics[scale=0.16]{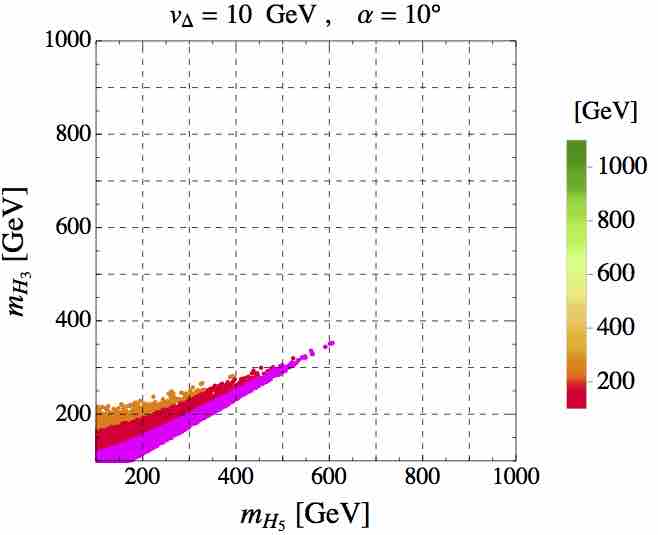}
\includegraphics[scale=0.16]{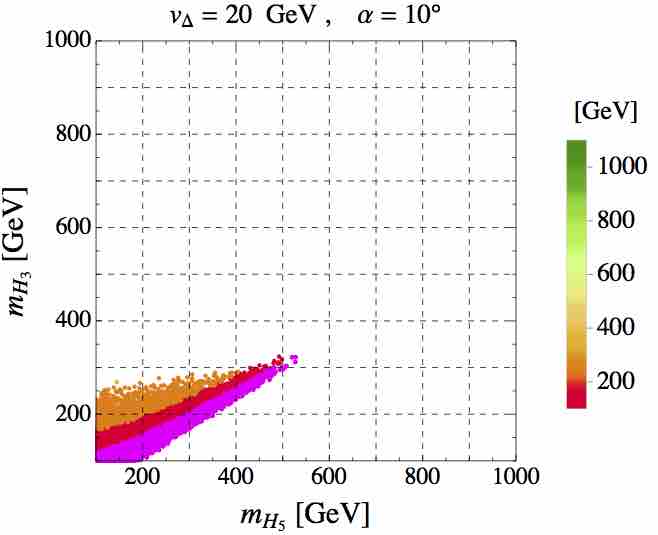}
\includegraphics[scale=0.16]{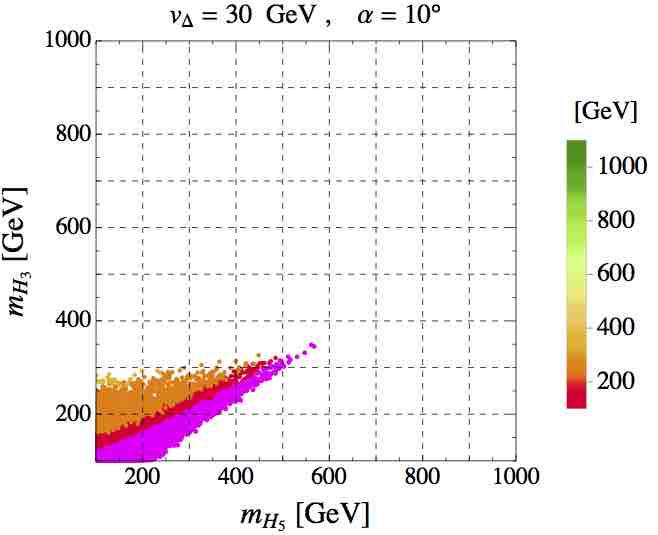}
\includegraphics[scale=0.16]{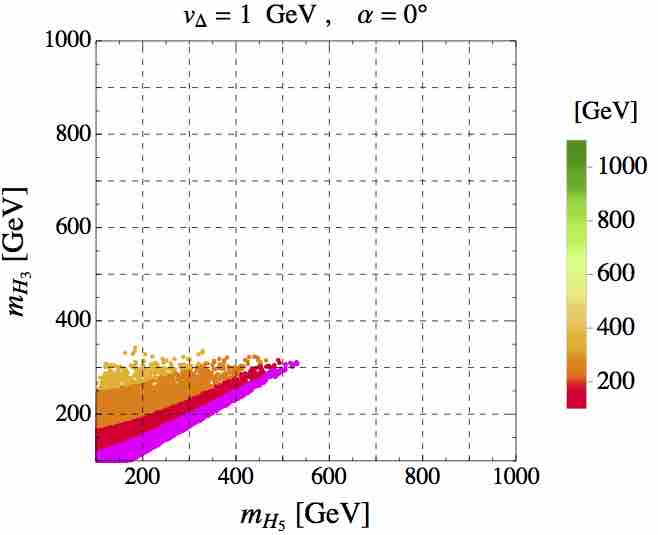}
\caption{$m_{H_1}$ for various values of ($v_\Delta,\alpha$). The values of $m_{H_1}$ for those magenta points are either equal or smaller than $m_h$.}
\label{h1mass}
\end{figure}

\begin{figure}[ht]
\centering
\includegraphics[scale=0.16]{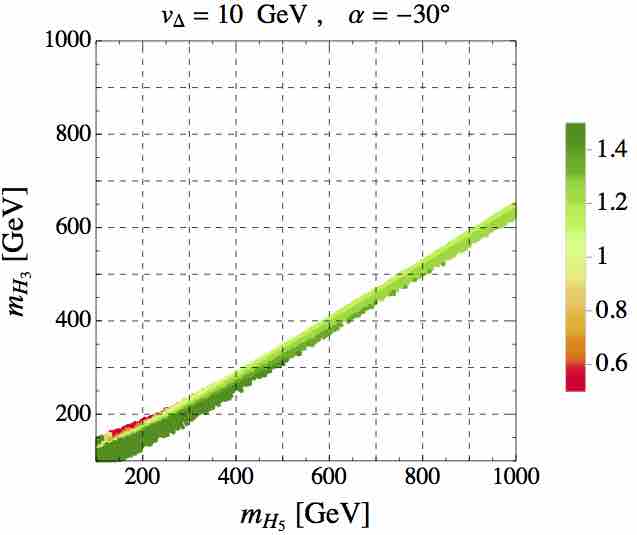}
\includegraphics[scale=0.16]{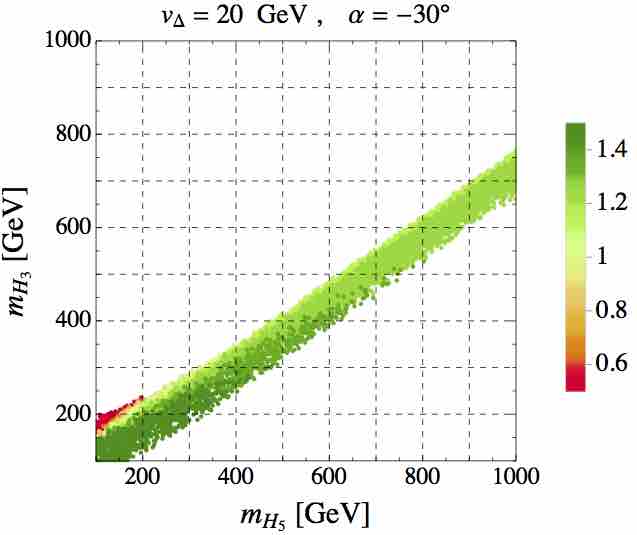}
\includegraphics[scale=0.16]{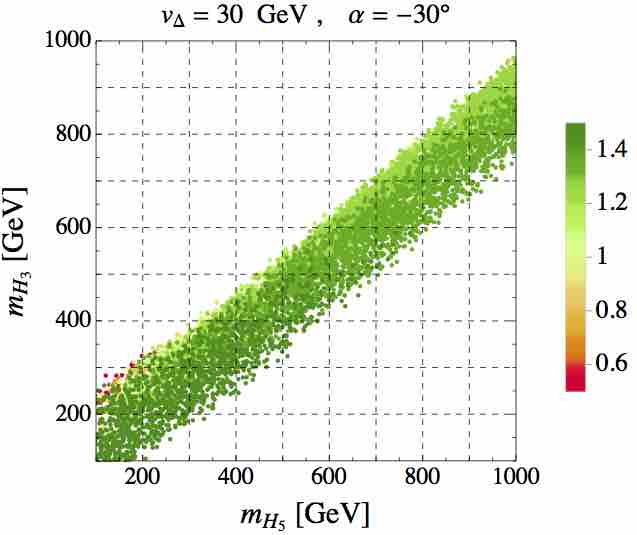}
\includegraphics[scale=0.16]{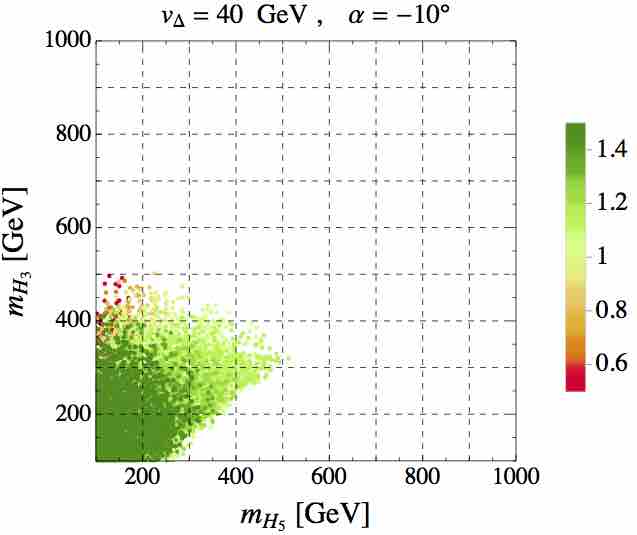}
\vspace{2mm}
\\
\includegraphics[scale=0.16]{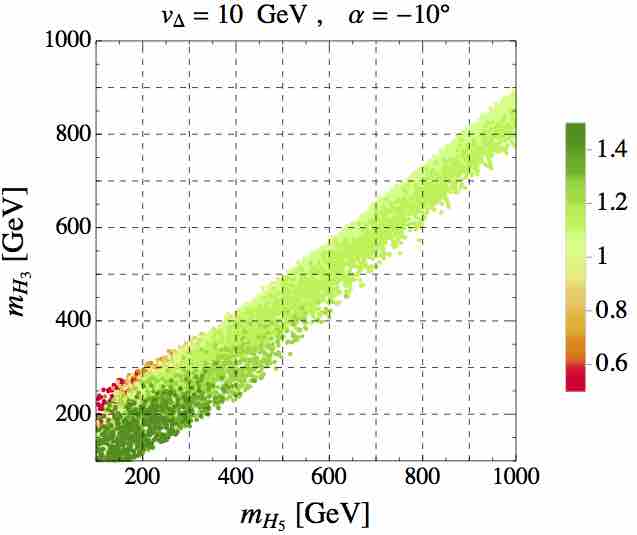}
\includegraphics[scale=0.16]{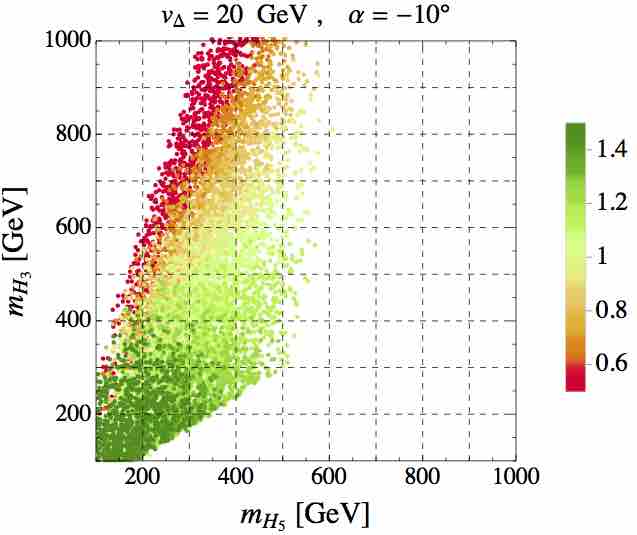}
\includegraphics[scale=0.16]{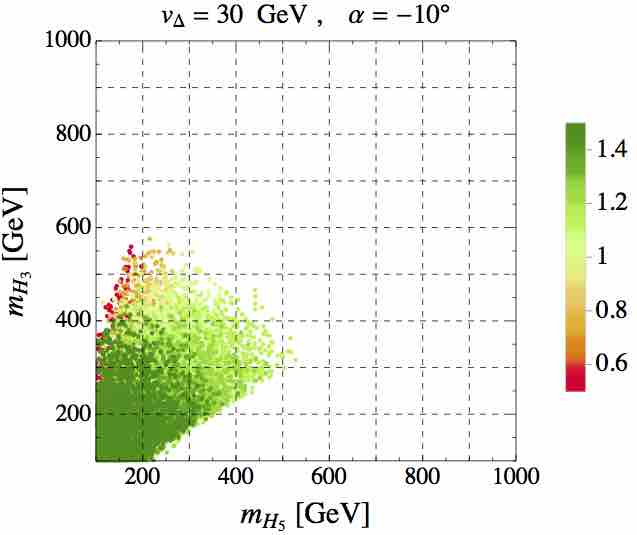}
\includegraphics[scale=0.16]{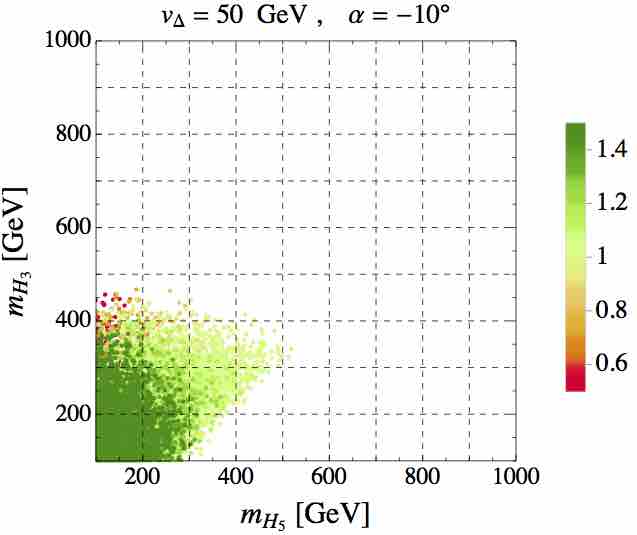}
\vspace{2mm}
\\
\includegraphics[scale=0.16]{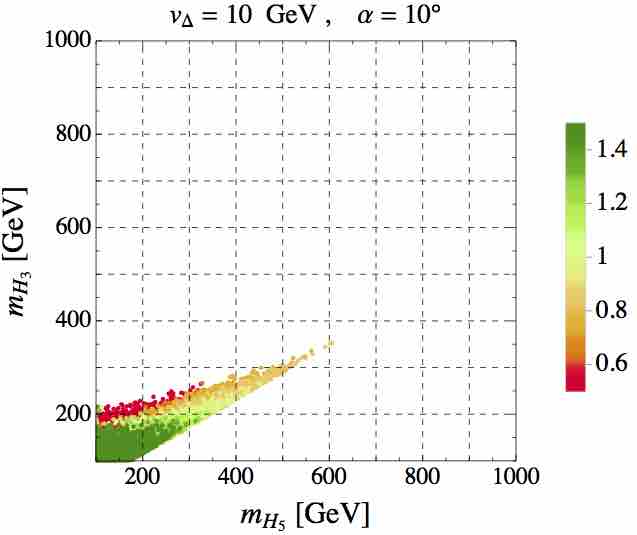}
\includegraphics[scale=0.16]{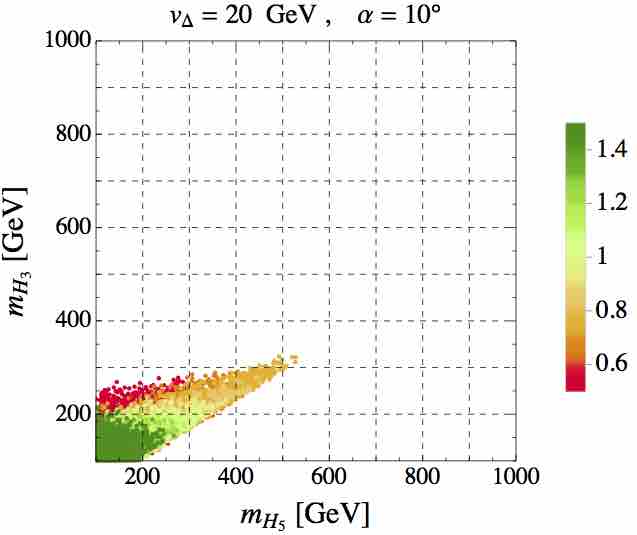}
\includegraphics[scale=0.16]{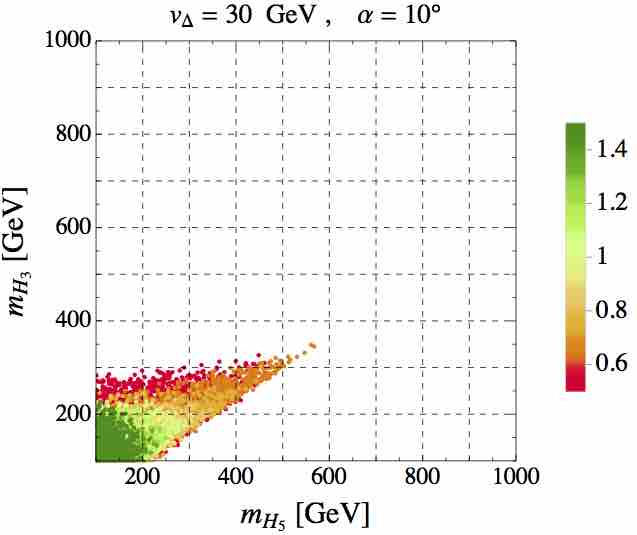}
\includegraphics[scale=0.16]{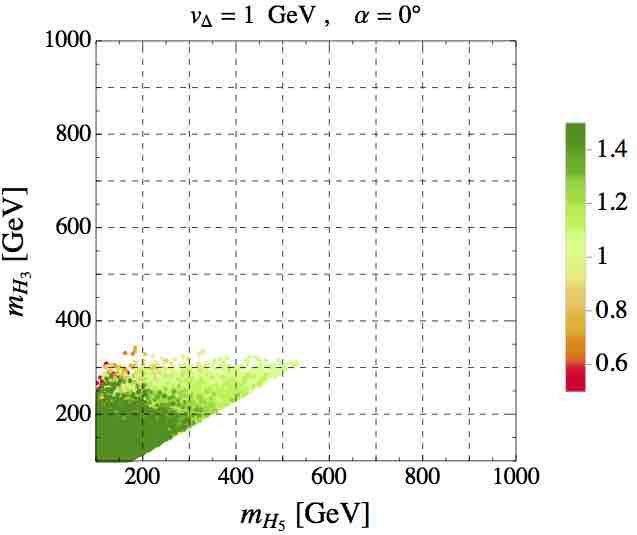}
\caption{$\mu^{\text{GGF}}_{h\gamma\gamma}$ for various values of ($v_\Delta,\alpha$).}
\label{muhdiphoton}
\end{figure}

\begin{figure}[ht]
\centering
\includegraphics[scale=0.16]{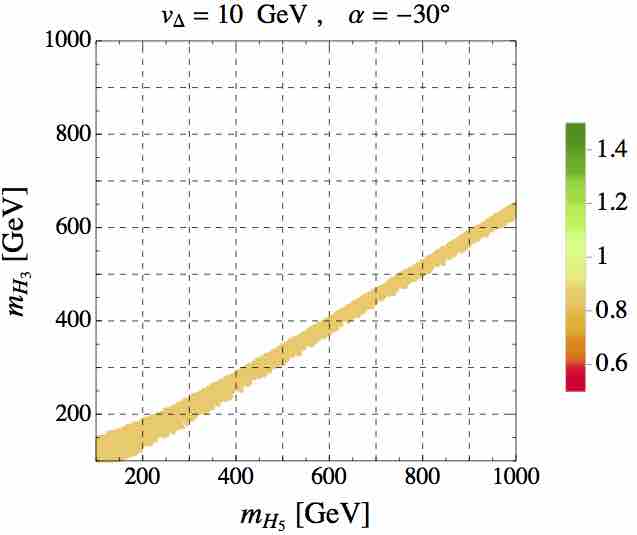}
\includegraphics[scale=0.16]{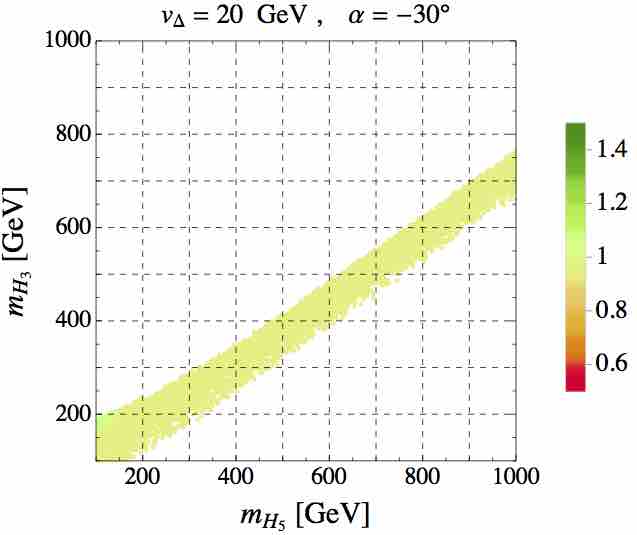}
\includegraphics[scale=0.16]{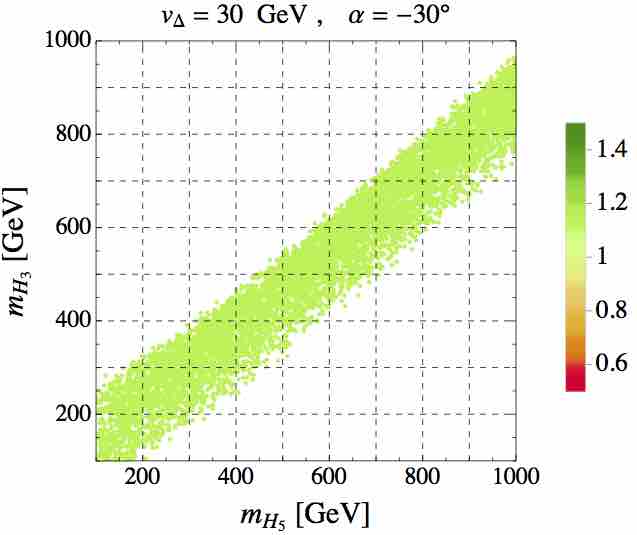}
\includegraphics[scale=0.16]{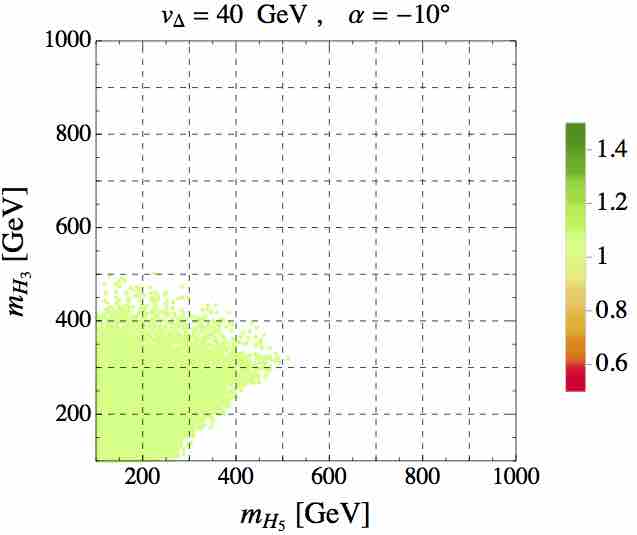}
\vspace{2mm}
\\
\includegraphics[scale=0.16]{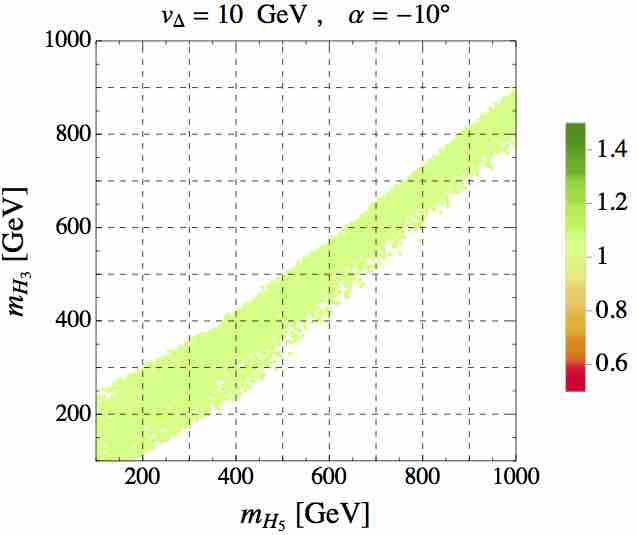}
\includegraphics[scale=0.16]{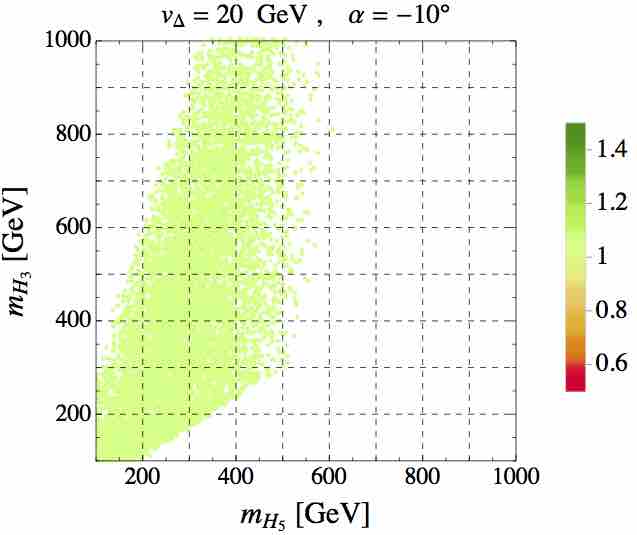}
\includegraphics[scale=0.16]{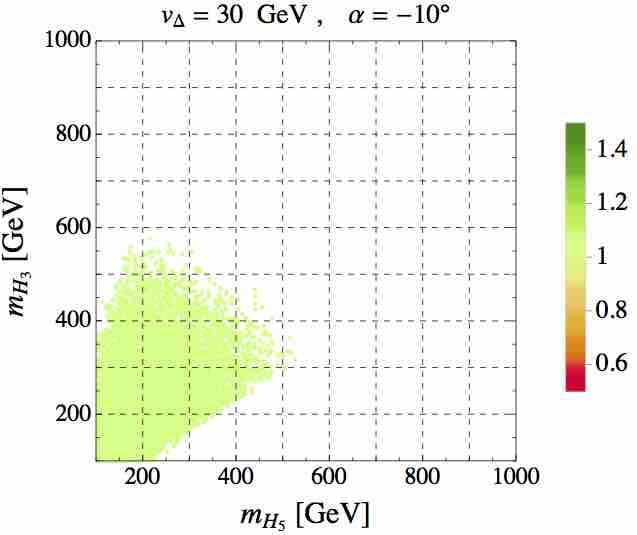}
\includegraphics[scale=0.16]{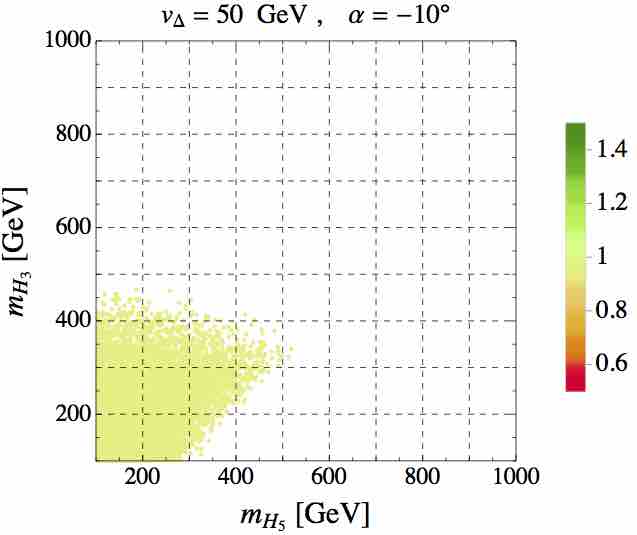}
\vspace{2mm}
\\
\includegraphics[scale=0.16]{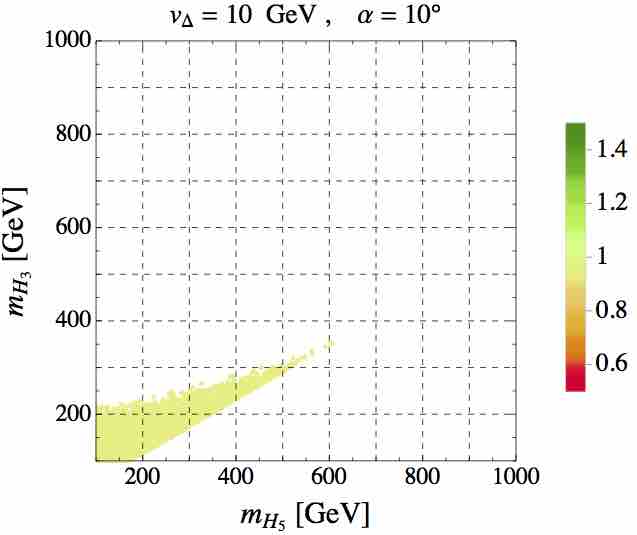}
\includegraphics[scale=0.16]{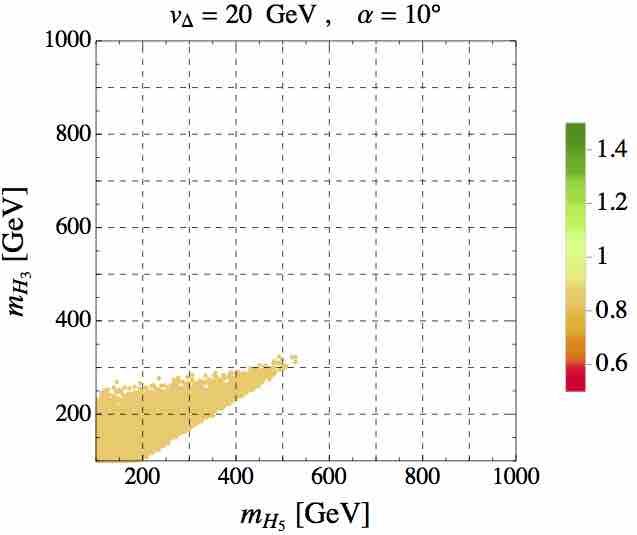}
\includegraphics[scale=0.16]{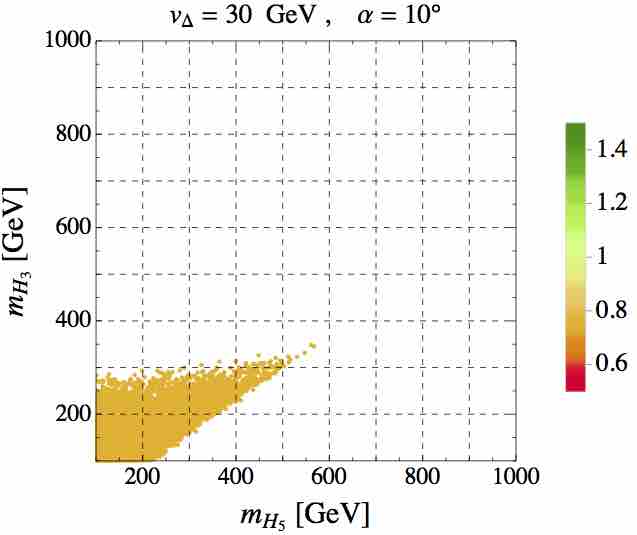}
\includegraphics[scale=0.16]{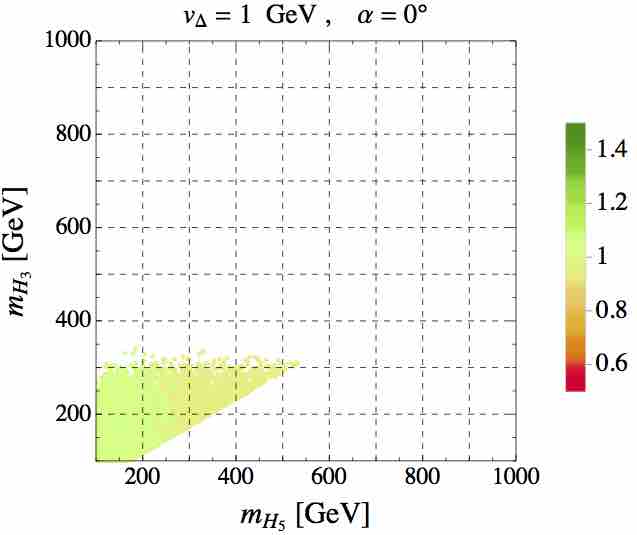}
\caption{$\mu^{\text{GGF}}_{h\gamma Z}$ for various values of ($v_\Delta,\alpha$).}
\label{muhzphoton}
\end{figure}

\begin{figure}[ht]
\centering
\includegraphics[scale=0.16]{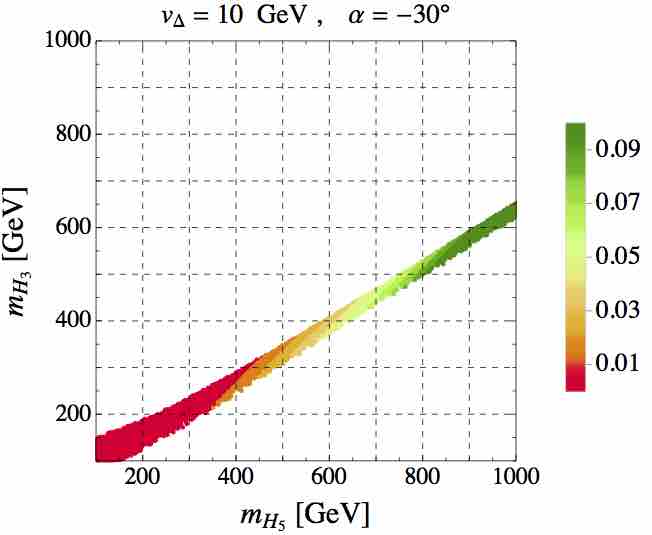}
\includegraphics[scale=0.16]{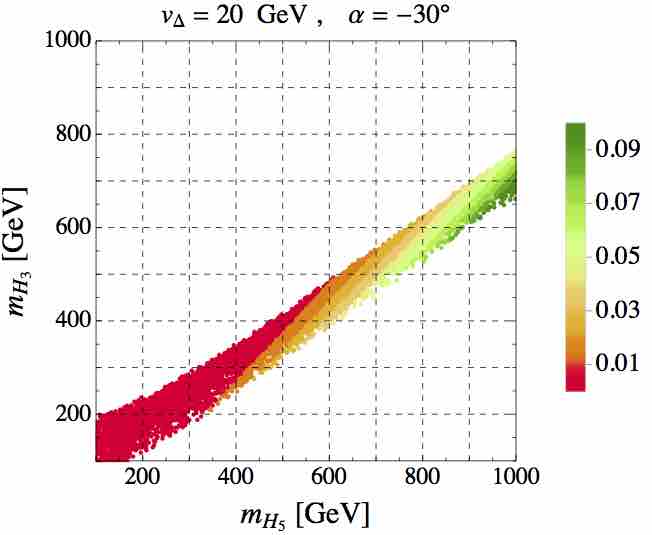}
\includegraphics[scale=0.16]{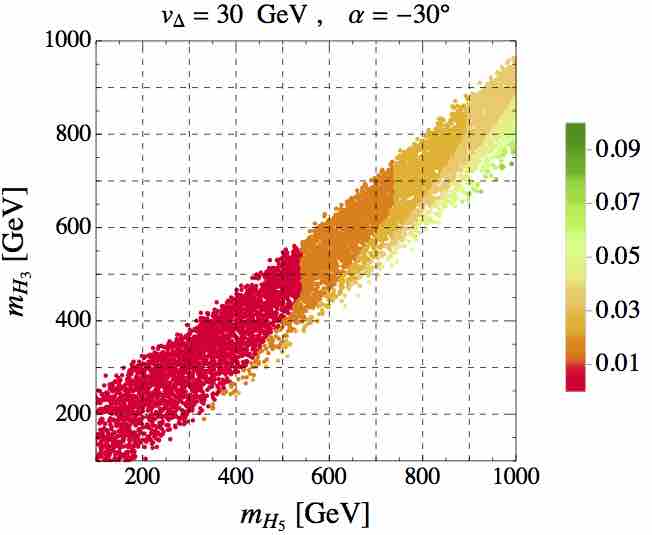}
\includegraphics[scale=0.16]{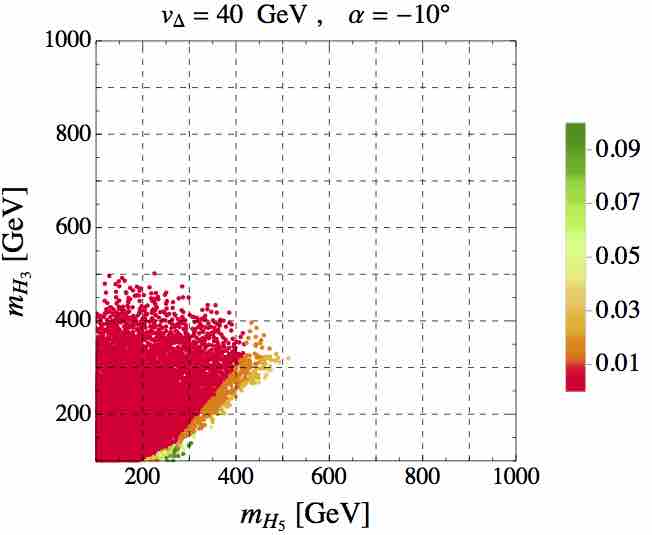}
\vspace{2mm}
\\
\includegraphics[scale=0.16]{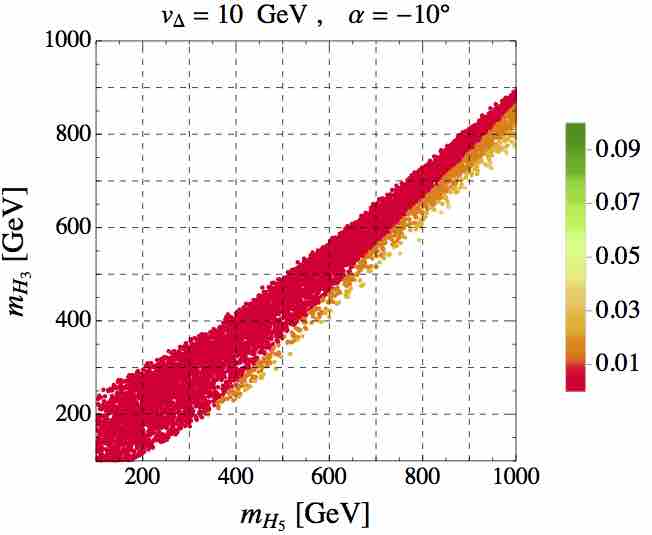}
\includegraphics[scale=0.16]{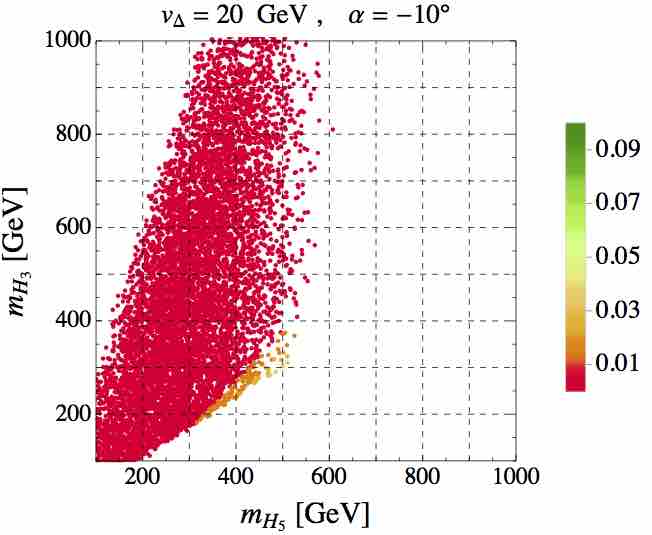}
\includegraphics[scale=0.16]{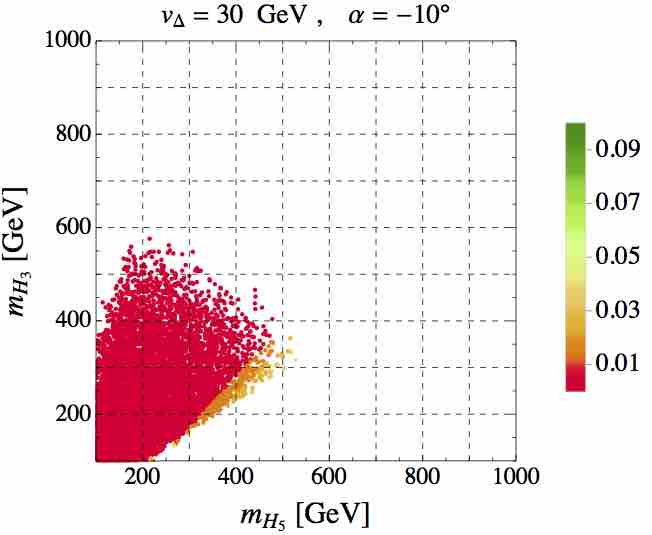}
\includegraphics[scale=0.16]{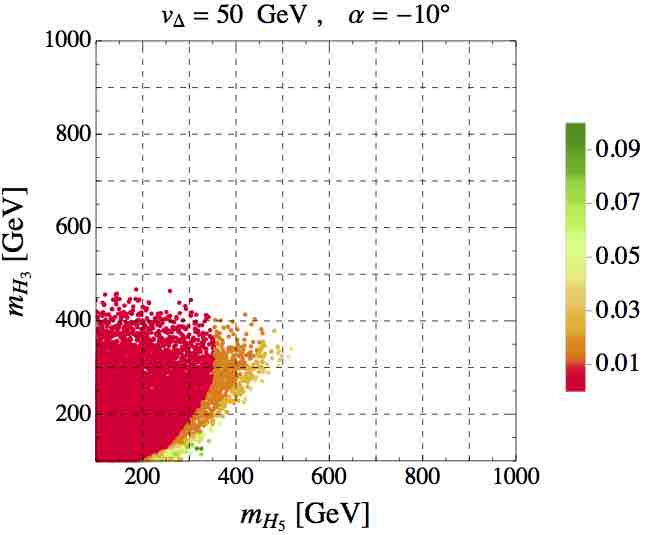}
\vspace{2mm}
\\
\includegraphics[scale=0.16]{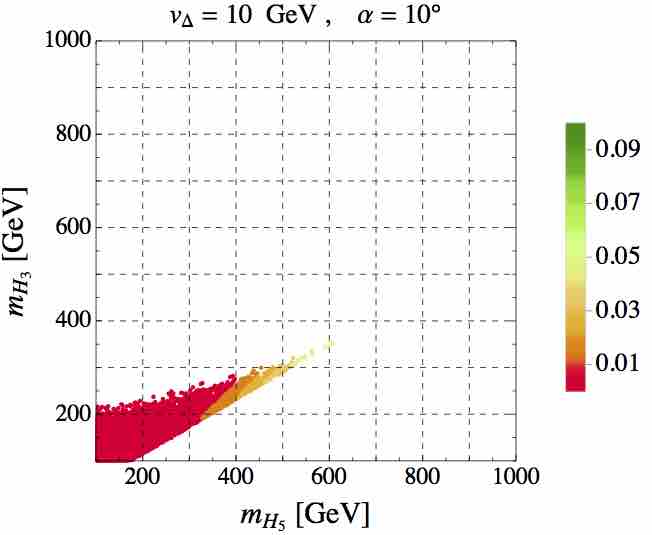}
\includegraphics[scale=0.16]{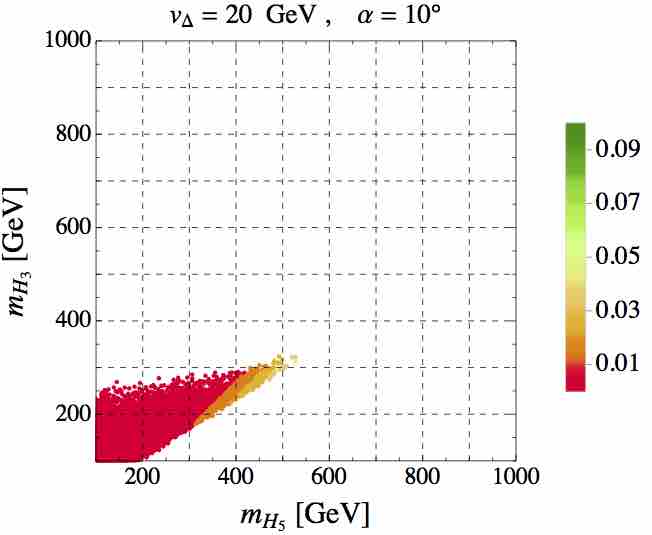}
\includegraphics[scale=0.16]{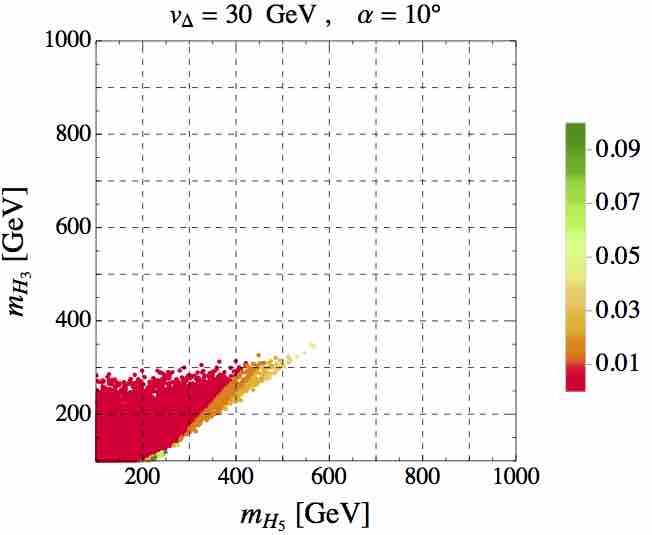}
\includegraphics[scale=0.16]{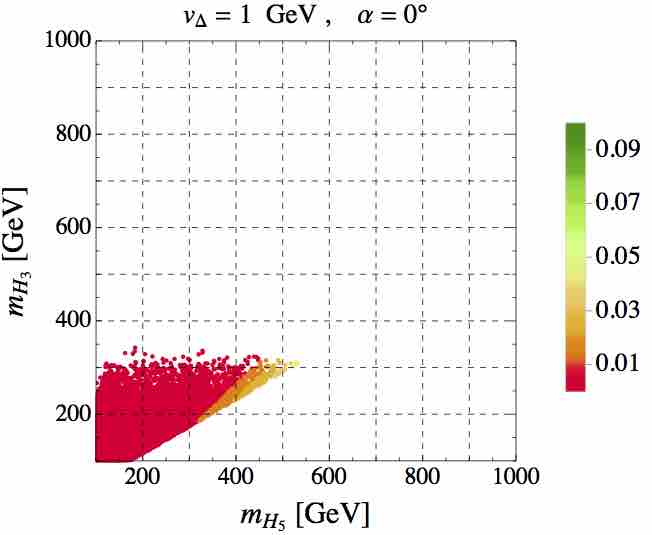}
\caption{The total decay width of $H_5^{++}$ divided by $m_{H_5}$ for various values of ($v_\Delta,\alpha$).}
\label{H5ppwidth}
\end{figure}

\begin{figure}[ht]
\centering
\includegraphics[scale=0.16]{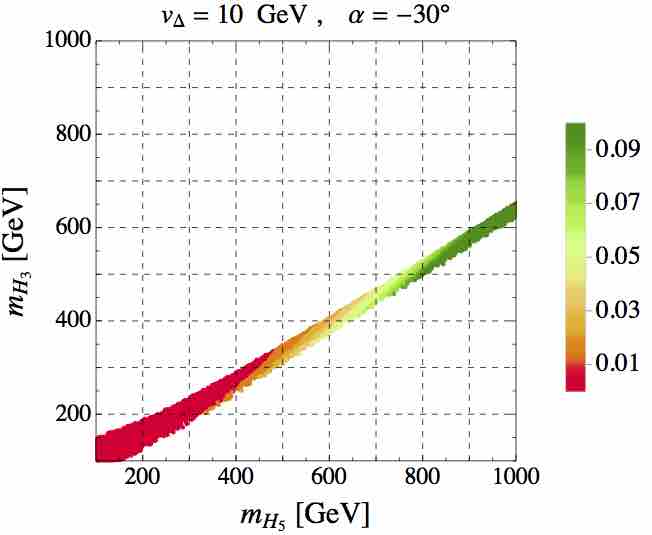}
\includegraphics[scale=0.16]{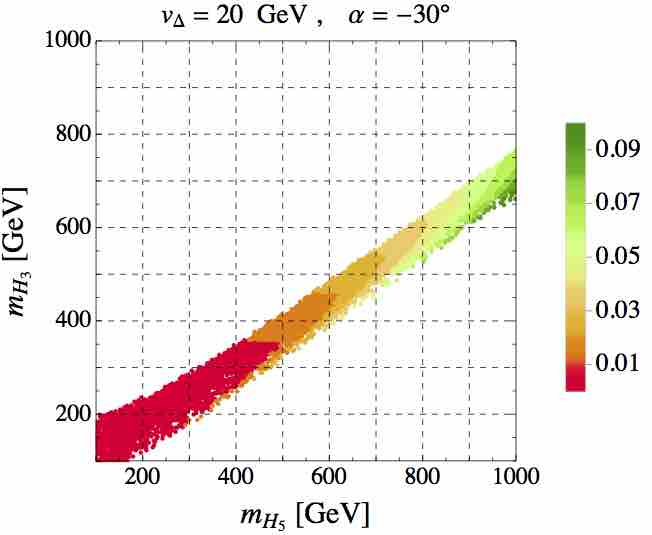}
\includegraphics[scale=0.16]{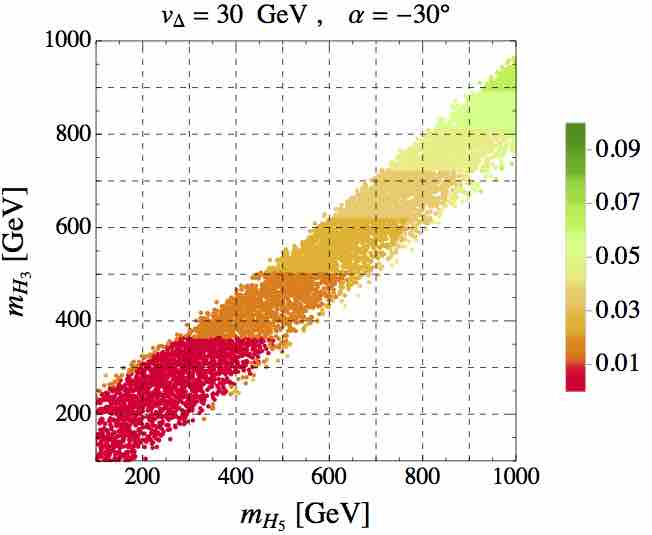}
\includegraphics[scale=0.16]{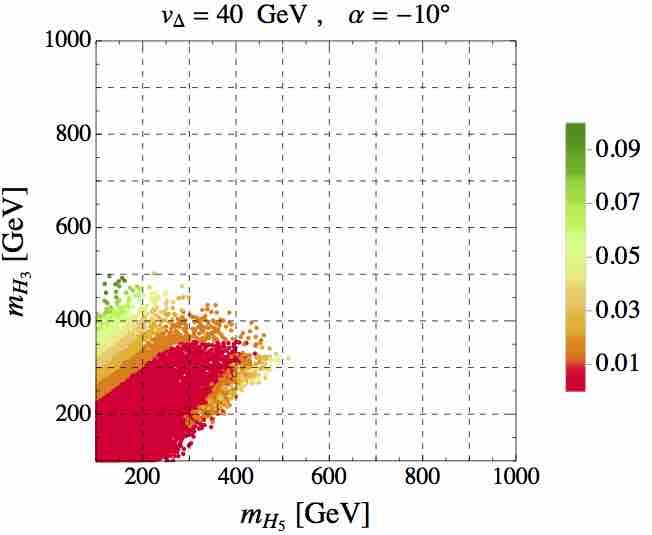}
\vspace{2mm}
\\
\includegraphics[scale=0.16]{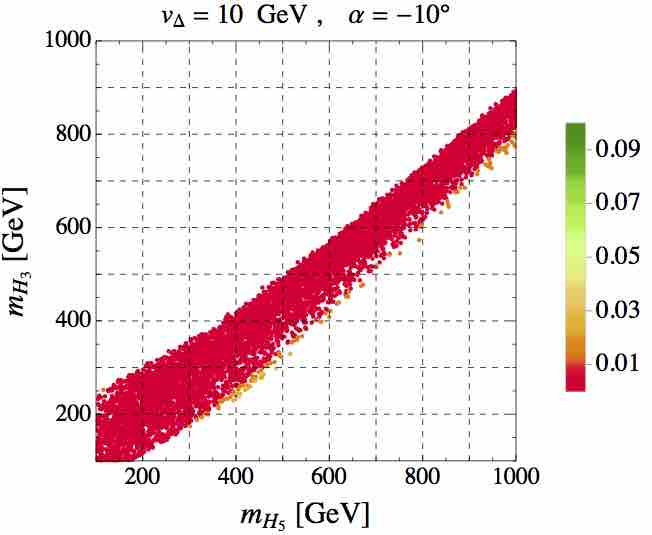}
\includegraphics[scale=0.16]{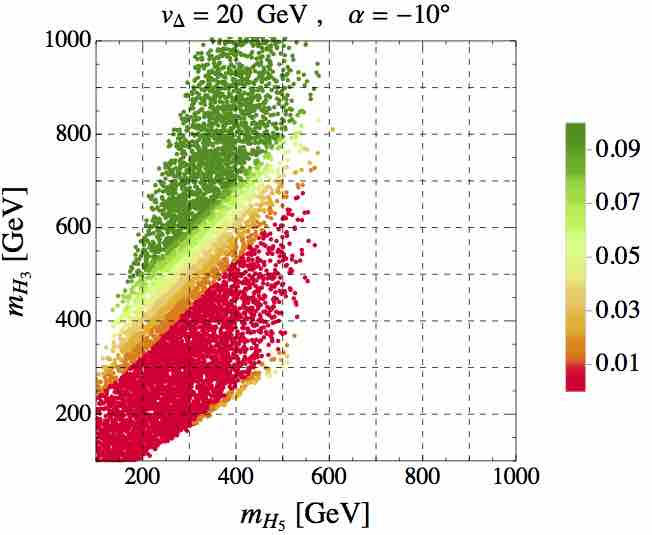}
\includegraphics[scale=0.16]{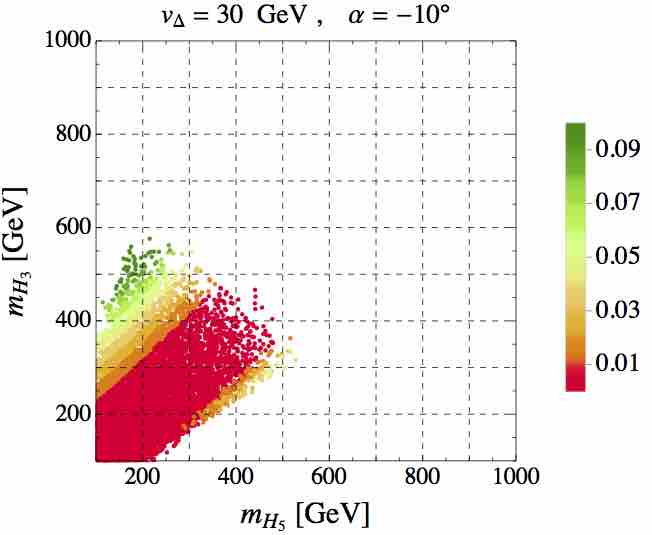}
\includegraphics[scale=0.16]{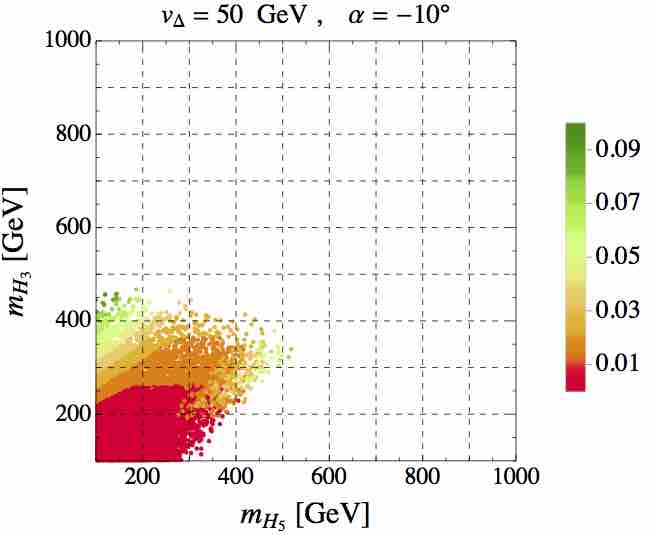}
\vspace{2mm}
\\
\includegraphics[scale=0.16]{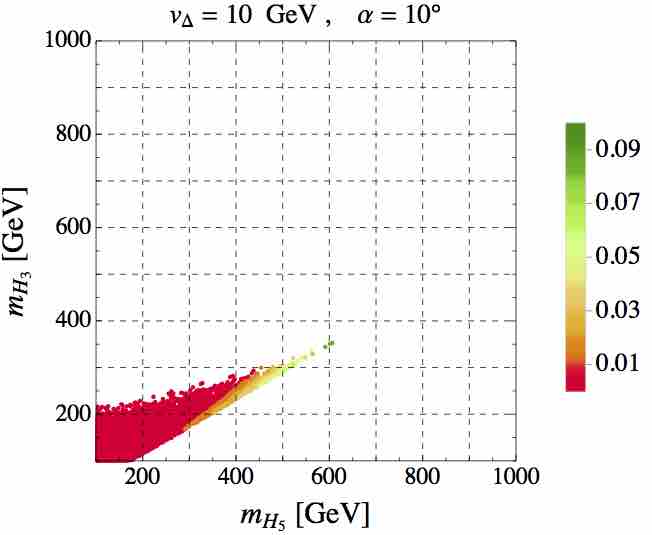}
\includegraphics[scale=0.16]{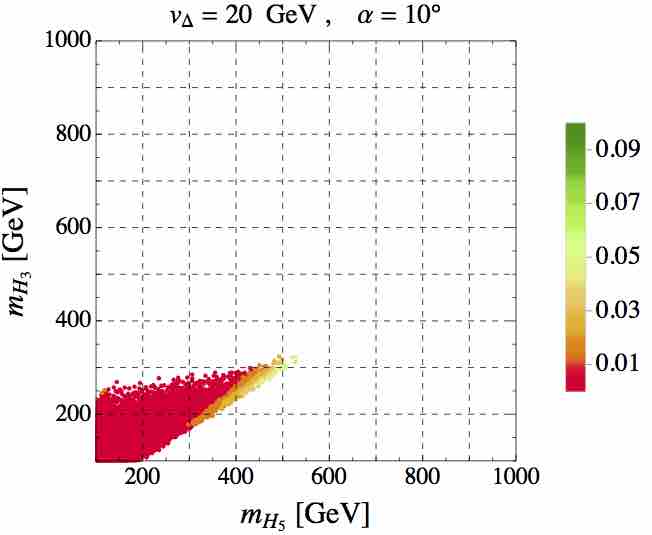}
\includegraphics[scale=0.16]{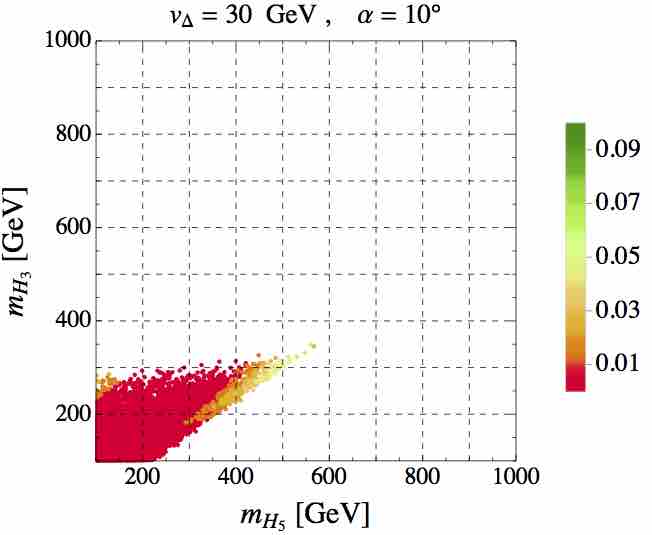}
\includegraphics[scale=0.16]{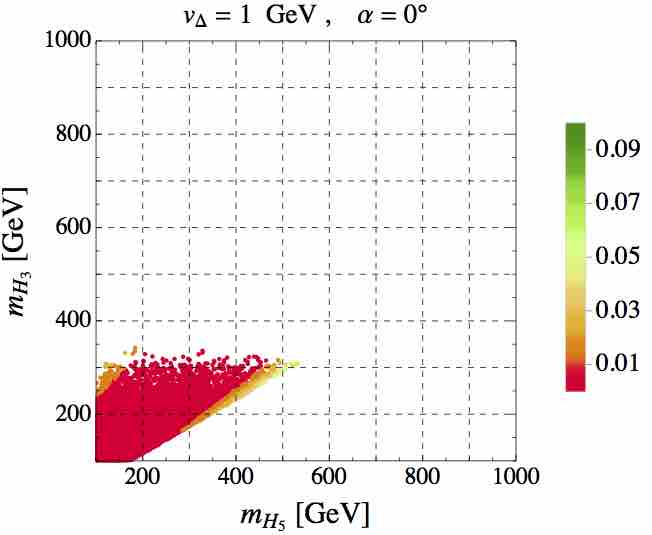}
\caption{The total decay width of $H_3^+$ divided by $m_{H_3}$ for various values of ($v_\Delta,\alpha$).}
\label{H3pwidth}
\end{figure}

\begin{figure}[ht]
\centering
\includegraphics[scale=0.16]{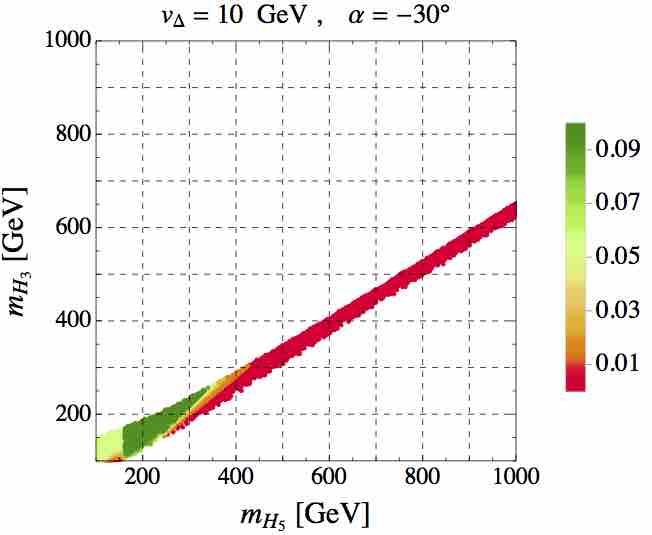}
\includegraphics[scale=0.16]{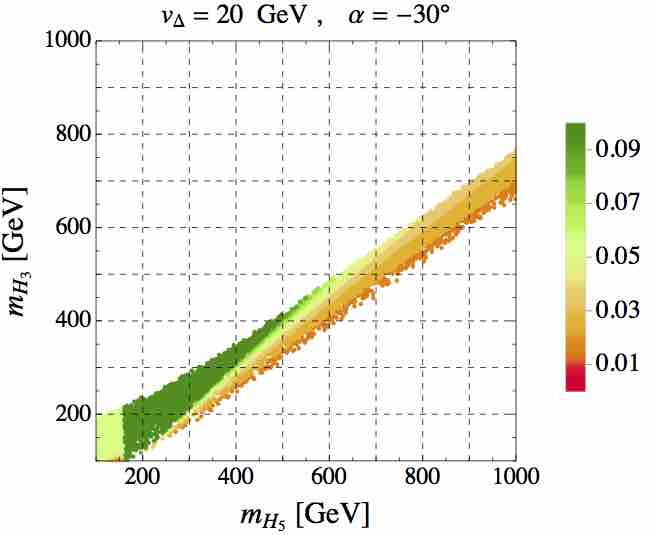}
\includegraphics[scale=0.16]{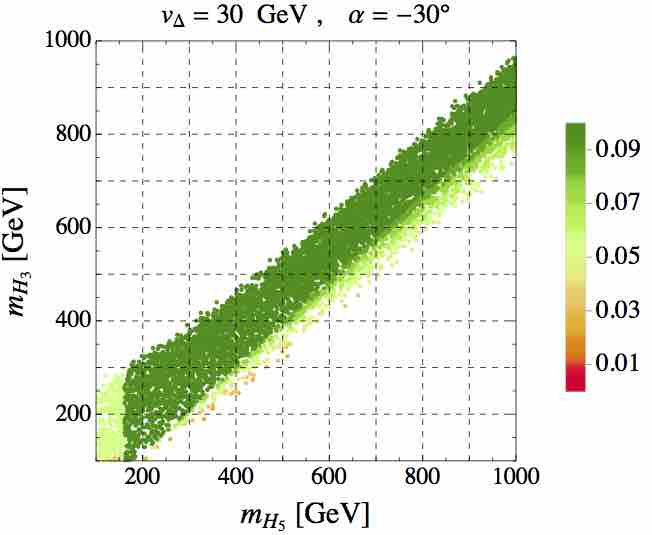}
\includegraphics[scale=0.16]{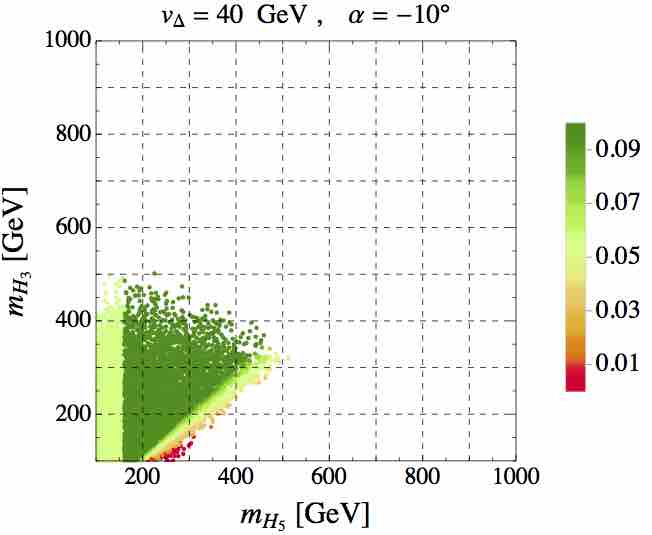}
\vspace{2mm}
\\
\includegraphics[scale=0.16]{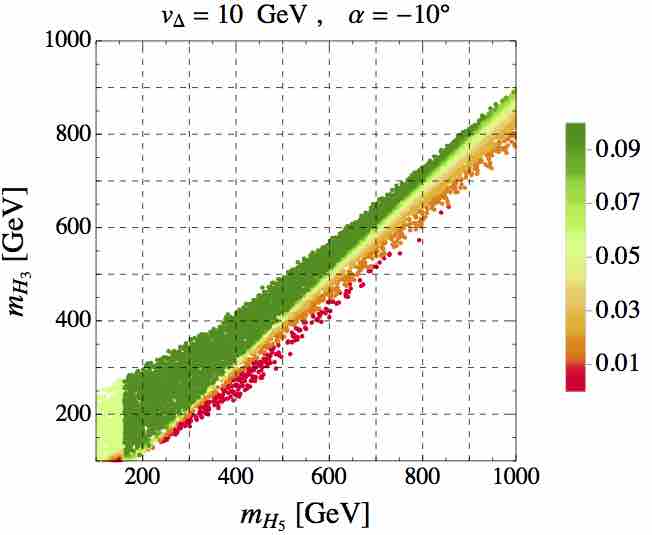}
\includegraphics[scale=0.16]{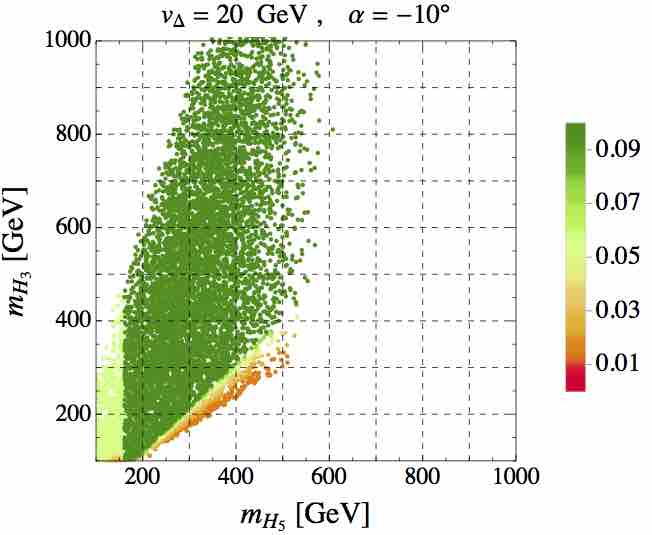}
\includegraphics[scale=0.16]{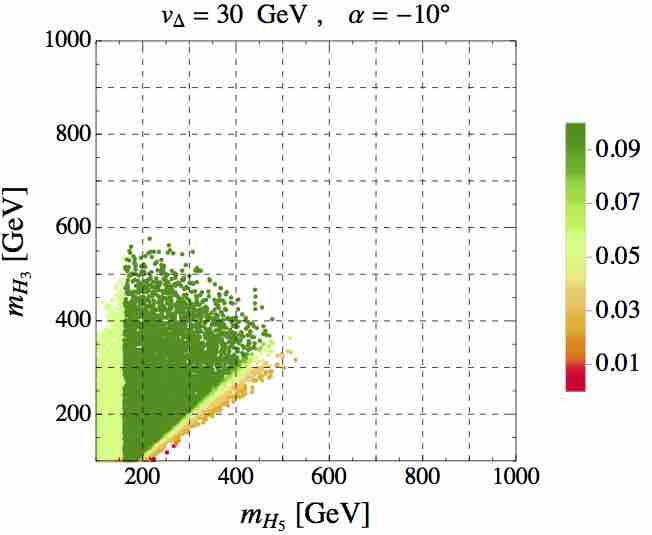}
\includegraphics[scale=0.16]{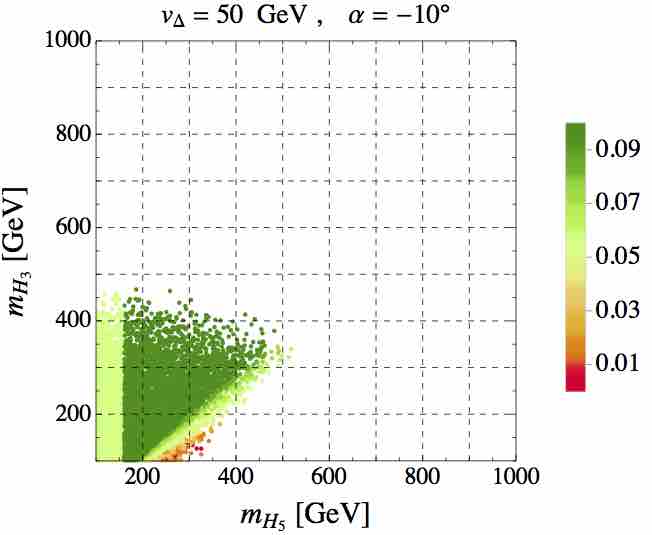}
\vspace{2mm}
\\
\includegraphics[scale=0.16]{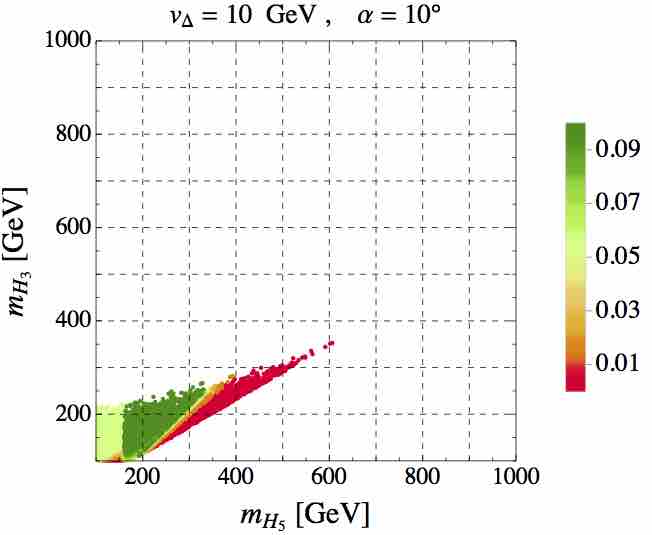}
\includegraphics[scale=0.16]{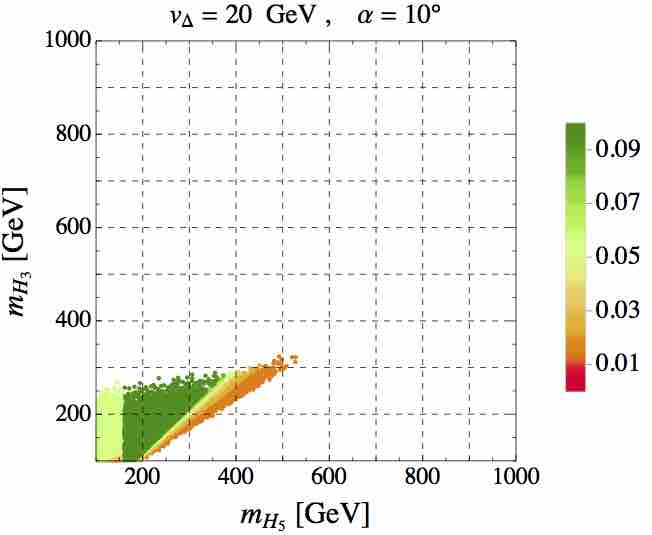}
\includegraphics[scale=0.16]{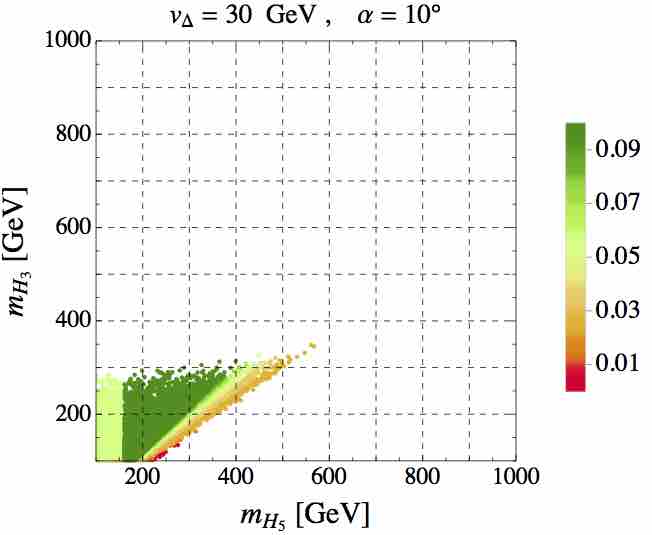}
\includegraphics[scale=0.16]{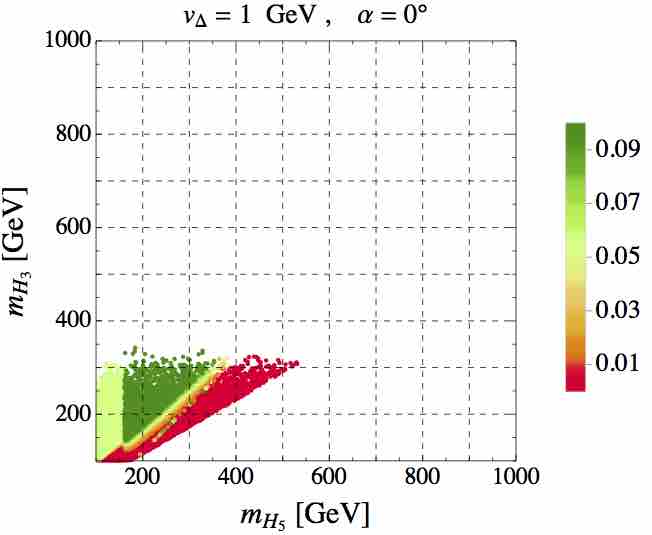}
\caption{$BR(H_5^{++} \rightarrow W^+(\rightarrow \ell^+ \nu_{\ell}) \, W^+(\rightarrow \ell^{\prime +} \nu_{\ell^{\prime}}) )$ for various values of ($v_\Delta,\alpha$).}
\label{h5ww}
\end{figure}

\begin{figure}[ht]
\centering
\includegraphics[scale=0.16]{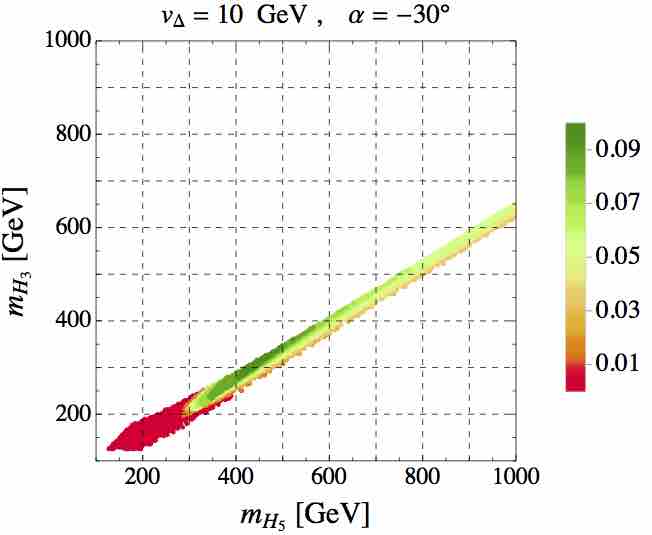}
\includegraphics[scale=0.16]{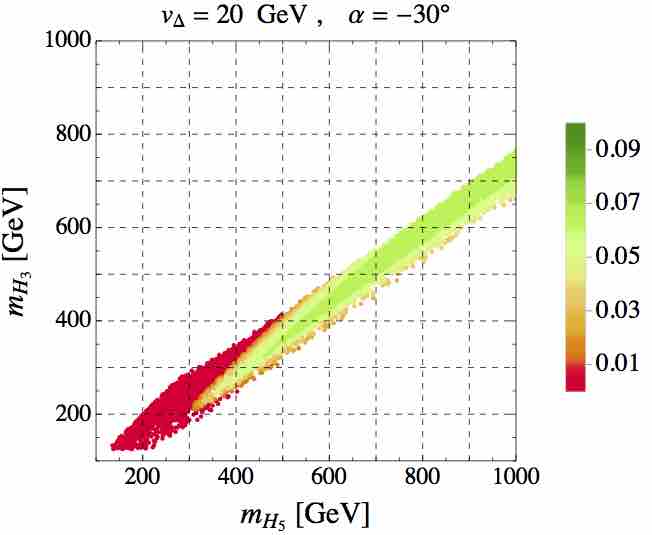}
\includegraphics[scale=0.16]{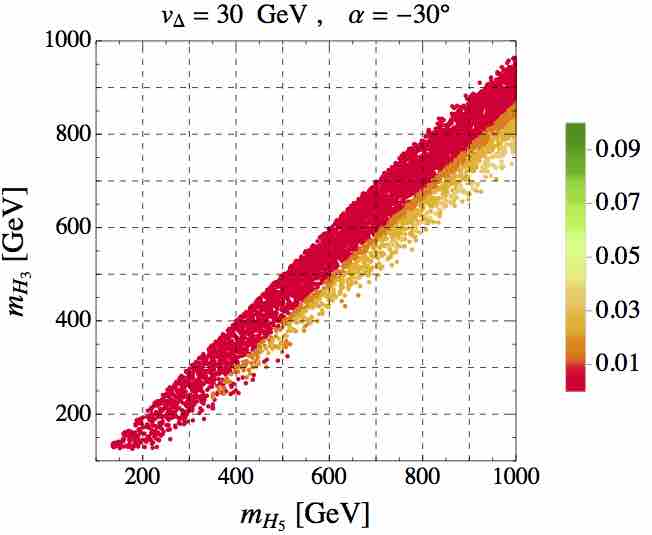}
\includegraphics[scale=0.16]{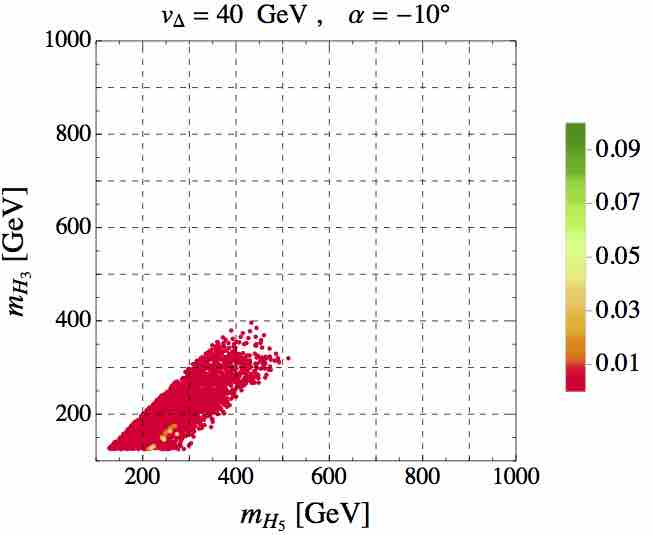}
\vspace{2mm}
\\
\includegraphics[scale=0.16]{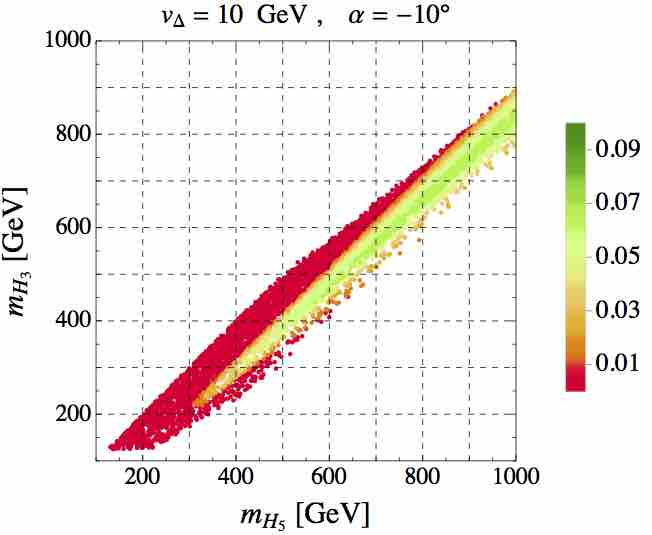}
\includegraphics[scale=0.16]{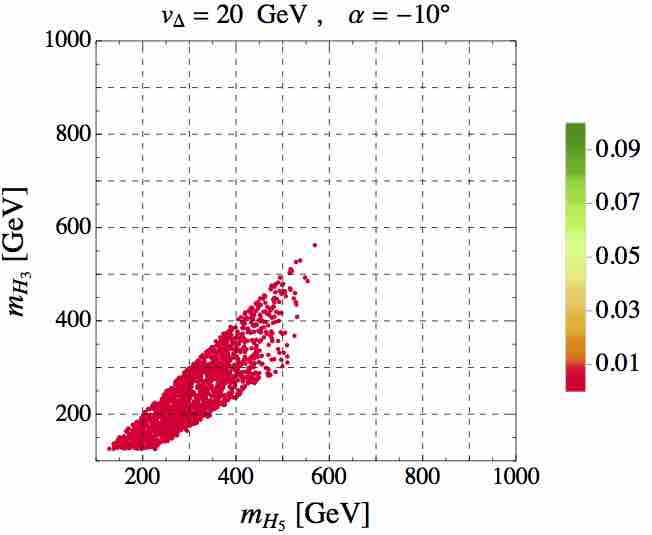}
\includegraphics[scale=0.16]{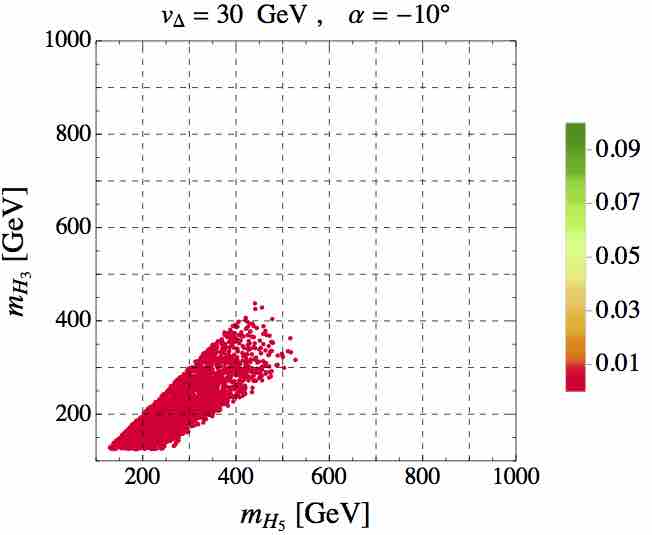}
\includegraphics[scale=0.16]{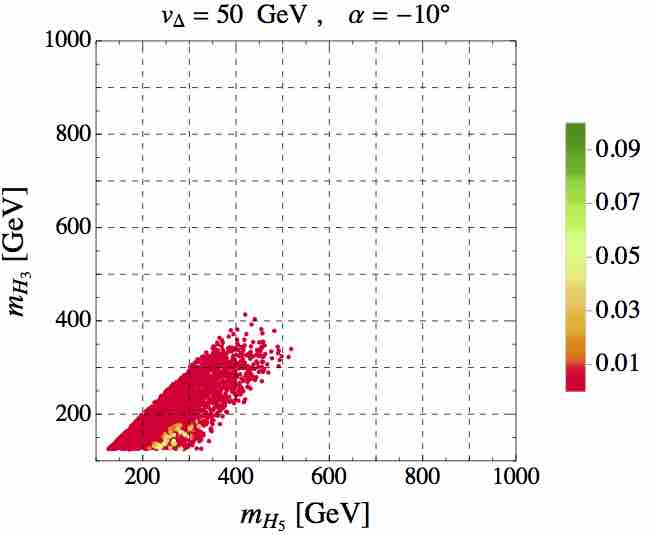}
\vspace{2mm}
\\
\includegraphics[scale=0.16]{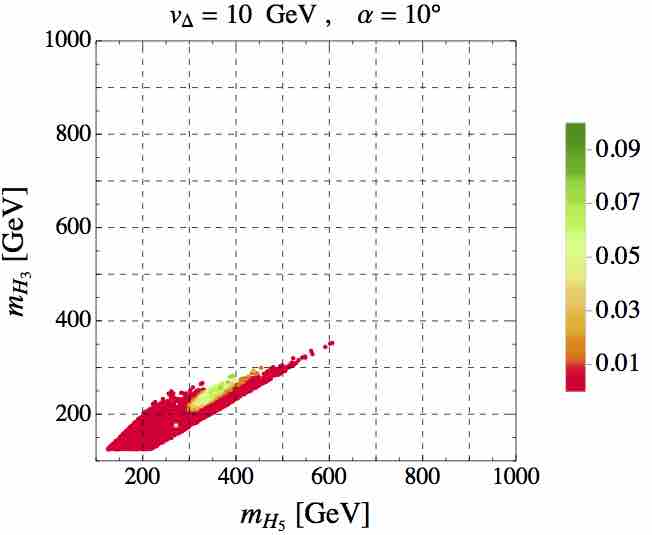}
\includegraphics[scale=0.16]{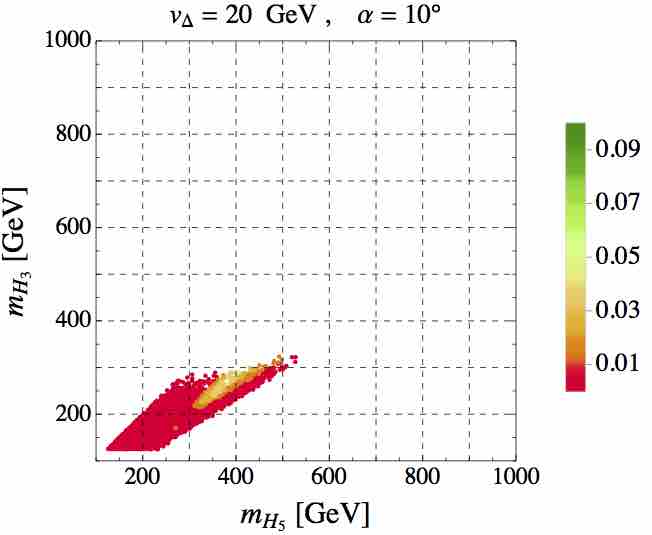}
\includegraphics[scale=0.16]{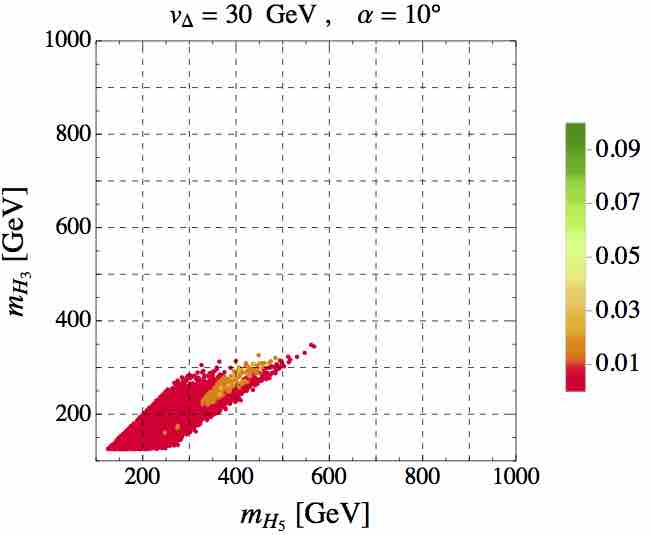}
\includegraphics[scale=0.16]{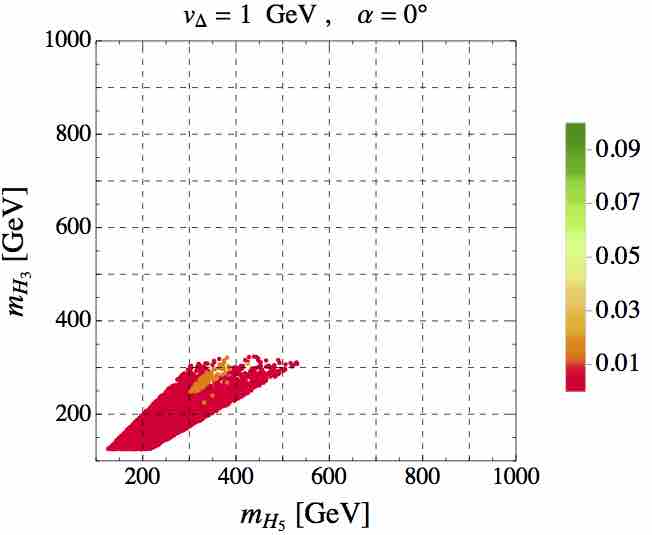}
\caption{$BR(H_5^{++} \rightarrow H_3^+ W^+(\rightarrow \ell^+ \nu_{\ell}) ) BR(H_3^+ \rightarrow h W^+(\rightarrow \ell^{\prime +} \nu_{\ell^{\prime}}))$ for various values of ($v_\Delta,\alpha$).}
\label{h5h3h}
\end{figure}

\begin{figure}[ht]
\centering
\includegraphics[scale=0.16]{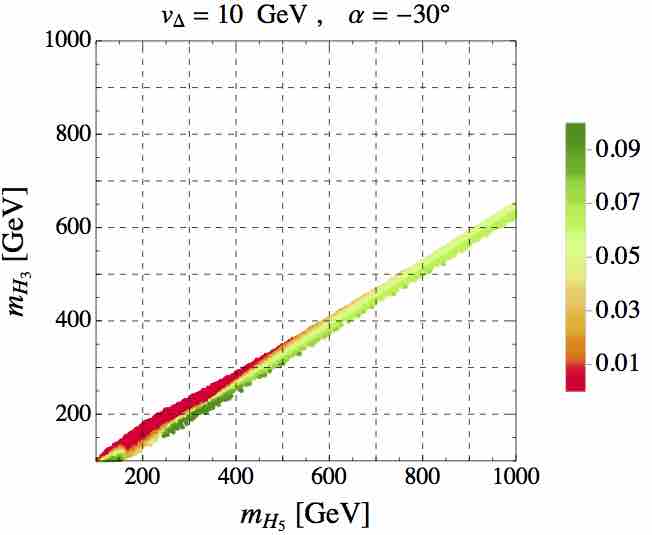}
\includegraphics[scale=0.16]{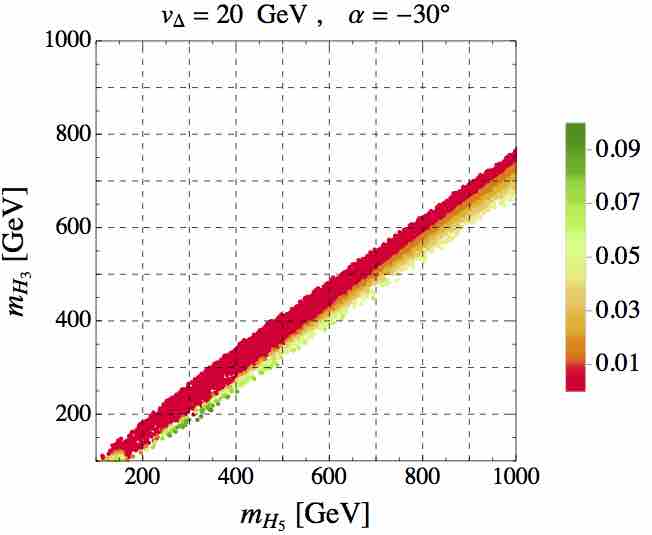}
\includegraphics[scale=0.16]{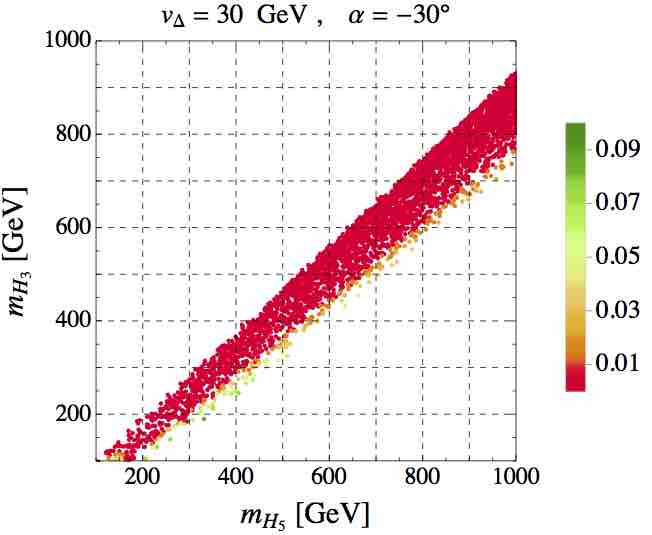}
\includegraphics[scale=0.16]{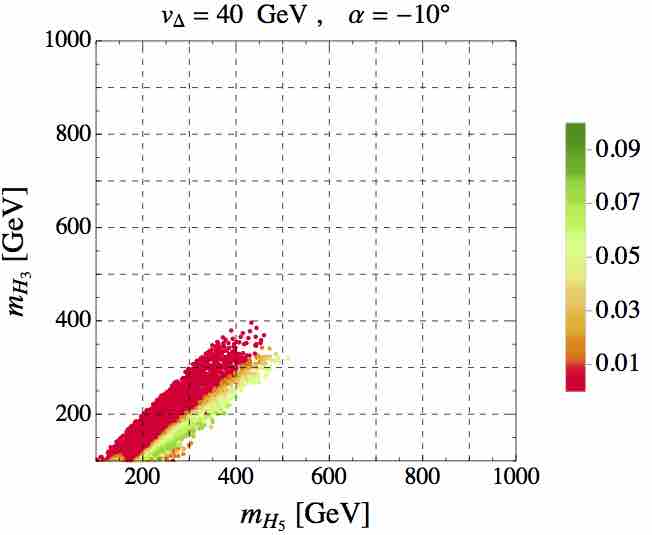}
\vspace{2mm}
\\
\includegraphics[scale=0.16]{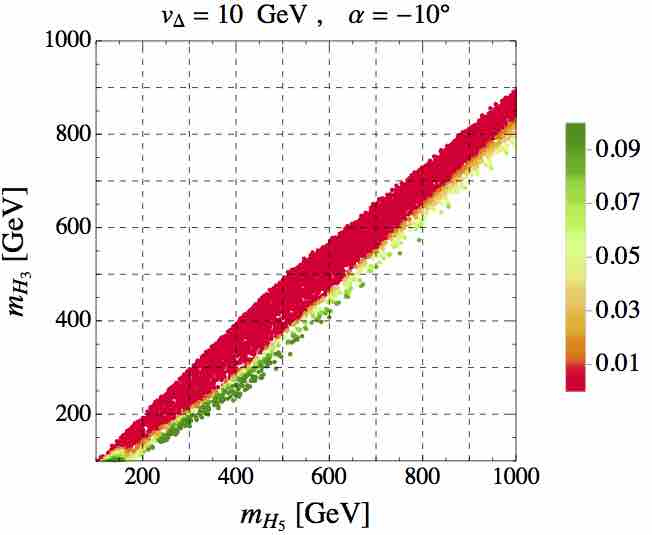}
\includegraphics[scale=0.16]{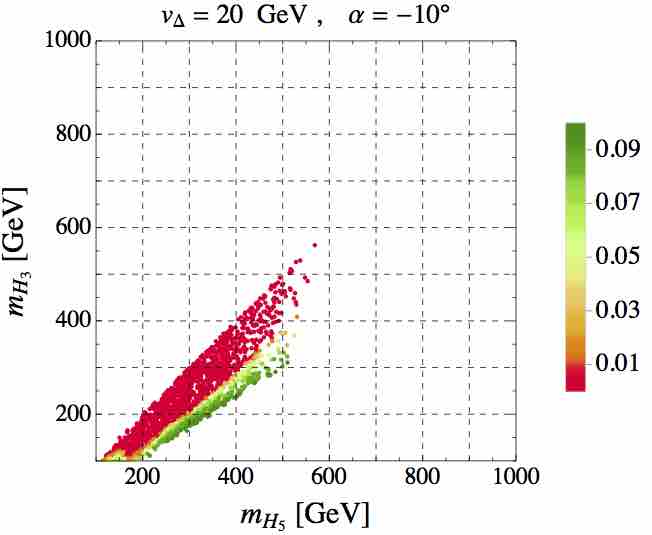}
\includegraphics[scale=0.16]{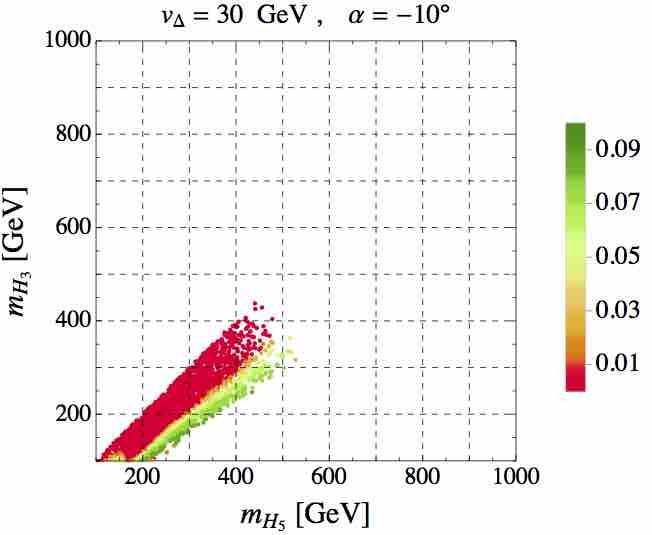}
\includegraphics[scale=0.16]{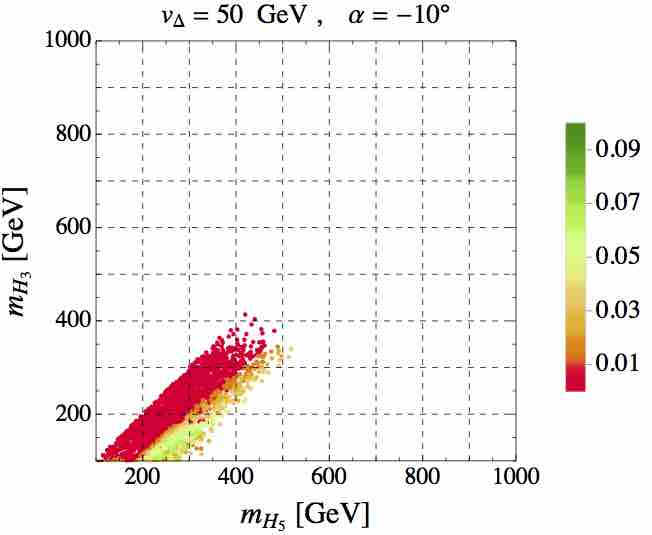}
\vspace{2mm}
\\
\includegraphics[scale=0.16]{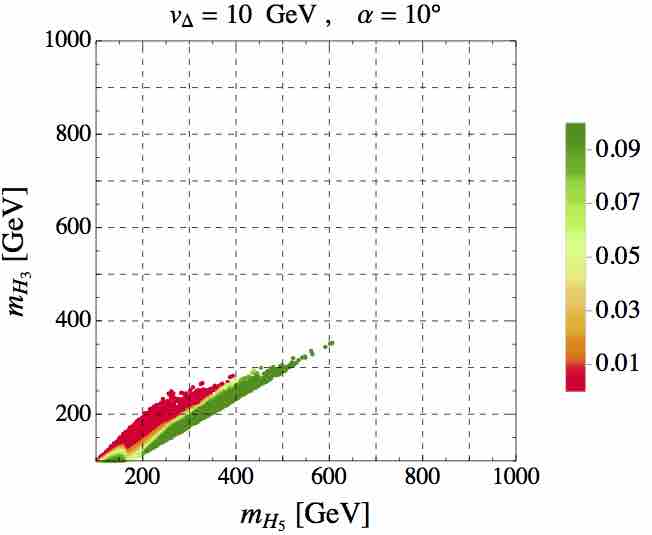}
\includegraphics[scale=0.16]{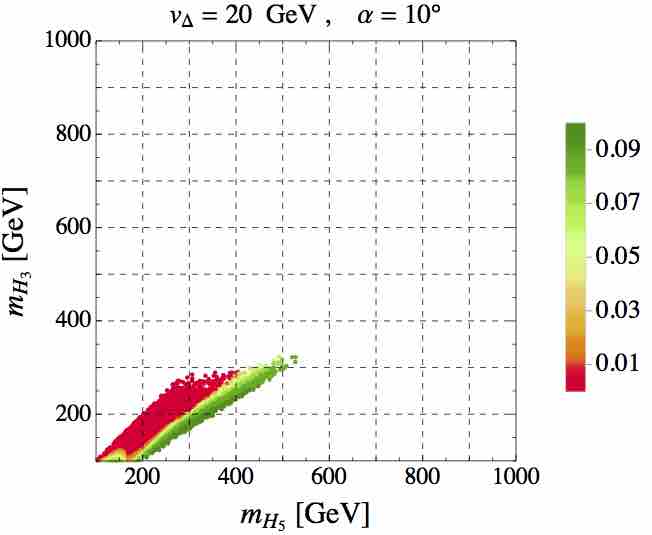}
\includegraphics[scale=0.16]{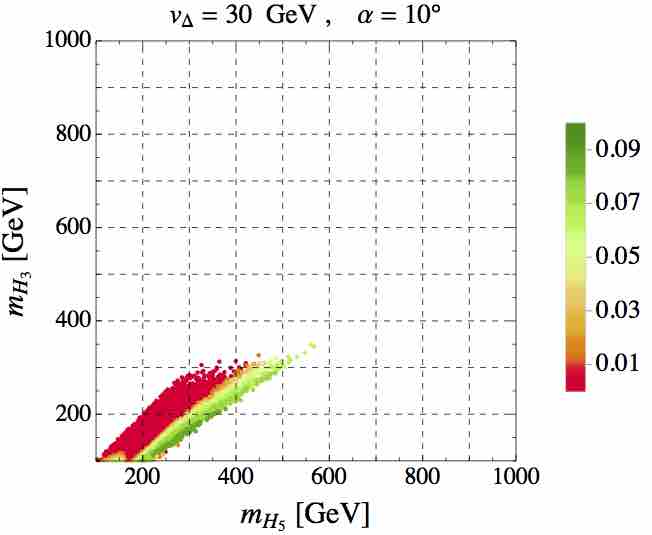}
\includegraphics[scale=0.16]{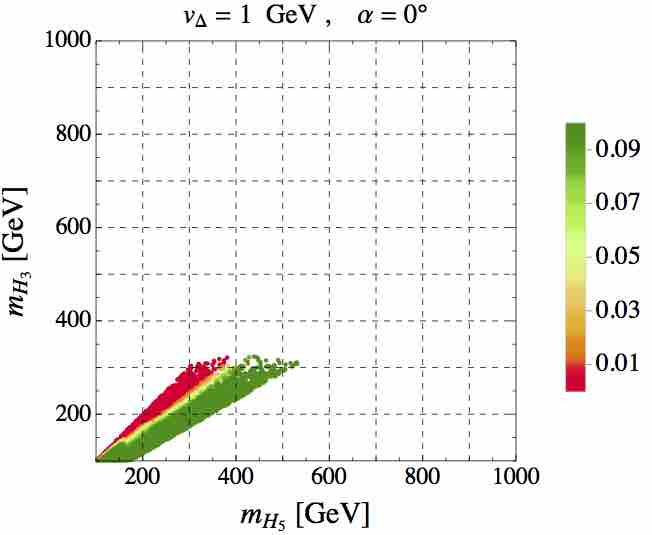}
\caption{$BR(H_5^{++} \rightarrow H_3^+ W^+(\rightarrow \ell^+ \nu_{\ell}) ) BR(H_3^+ \rightarrow H_1 W^+(\rightarrow \ell^{\prime +} \nu_{\ell^{\prime}}))$ for various values of ($v_\Delta,\alpha$).}
\label{h5h3h1}
\end{figure}

\begin{figure}[ht]
\centering
\includegraphics[scale=0.16]{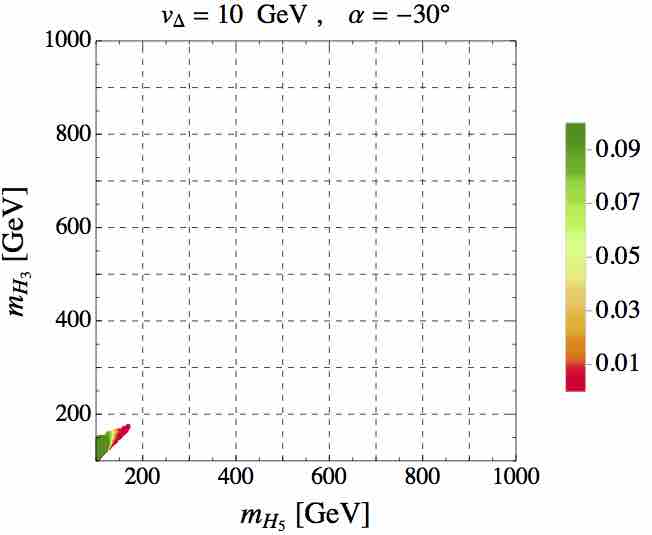}
\includegraphics[scale=0.16]{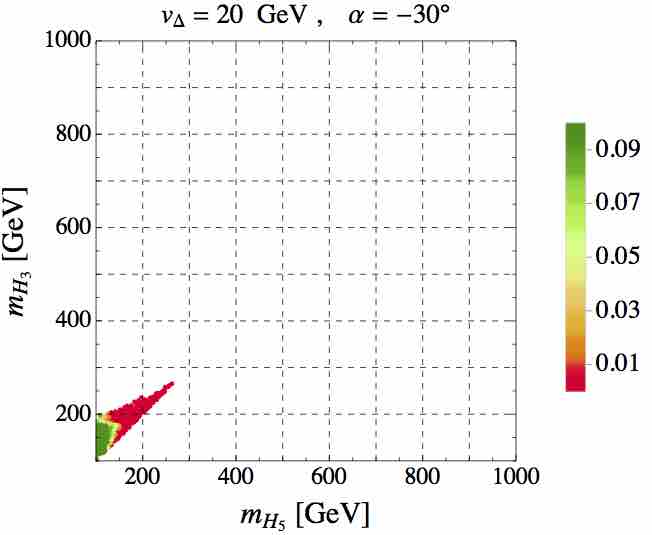}
\includegraphics[scale=0.16]{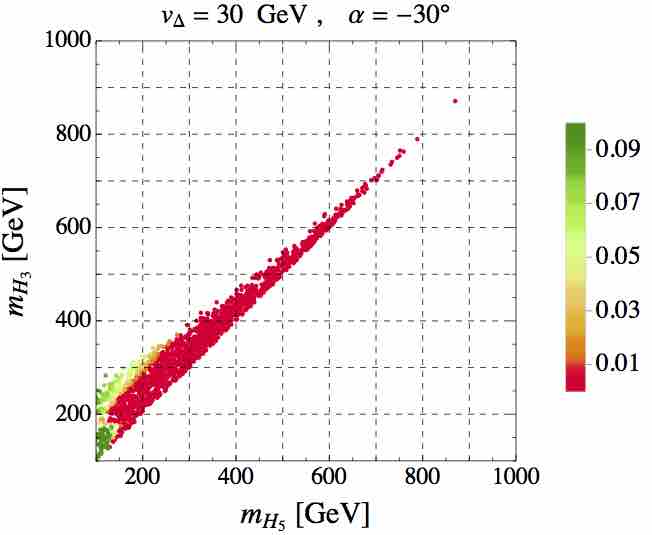}
\includegraphics[scale=0.16]{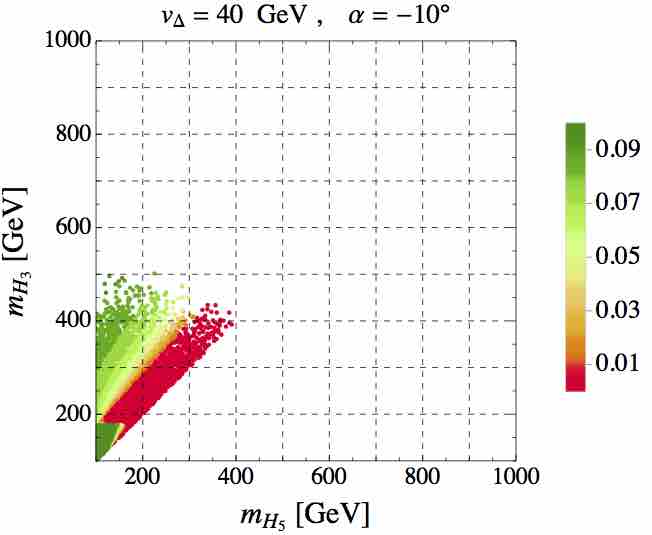}
\vspace{2mm}
\\
\includegraphics[scale=0.16]{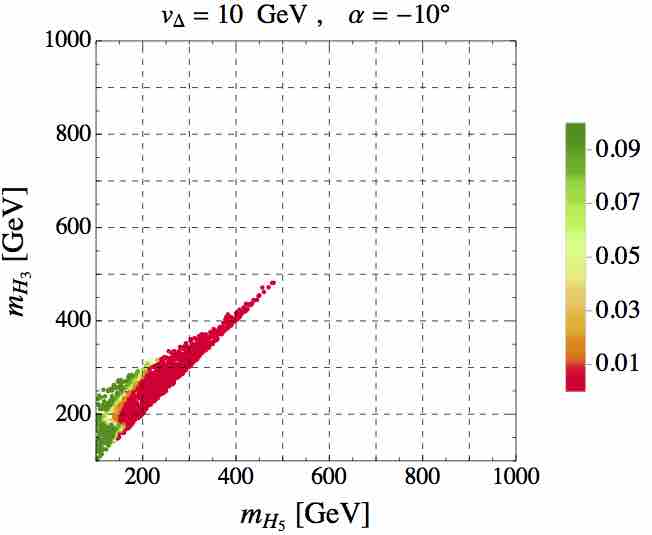}
\includegraphics[scale=0.16]{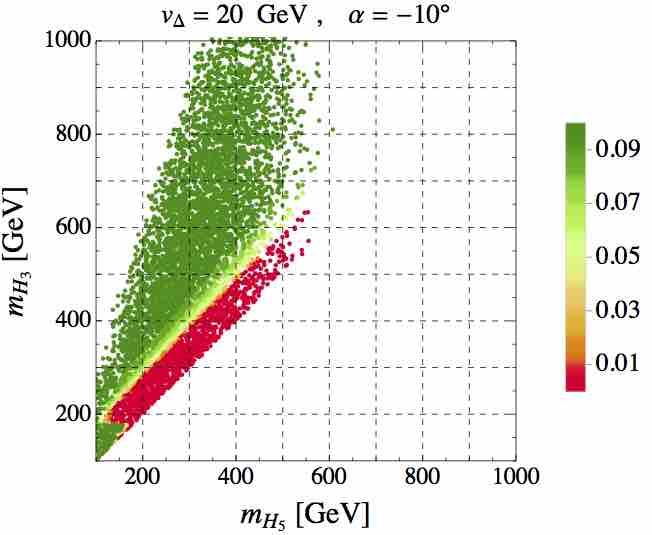}
\includegraphics[scale=0.16]{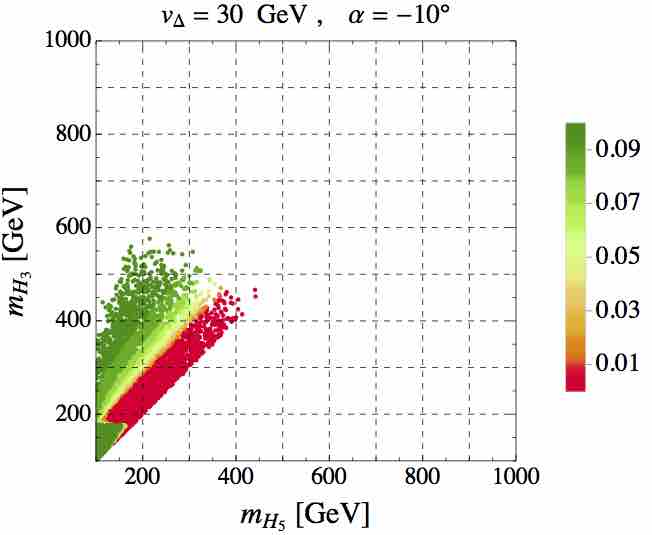}
\includegraphics[scale=0.16]{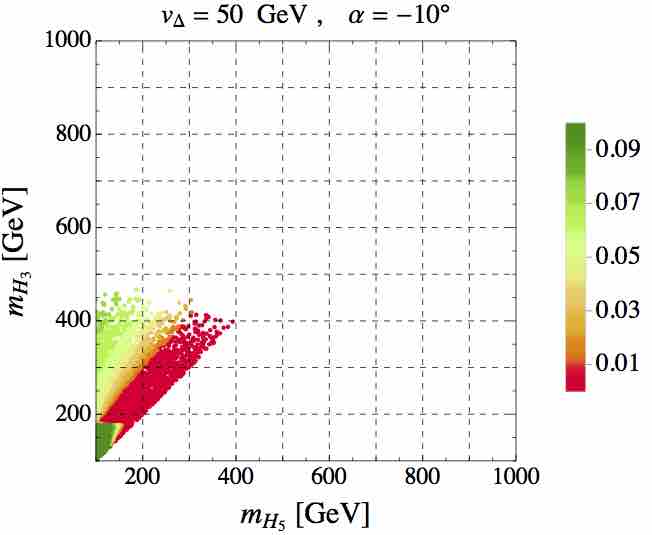}
\vspace{2mm}
\\
\includegraphics[scale=0.16]{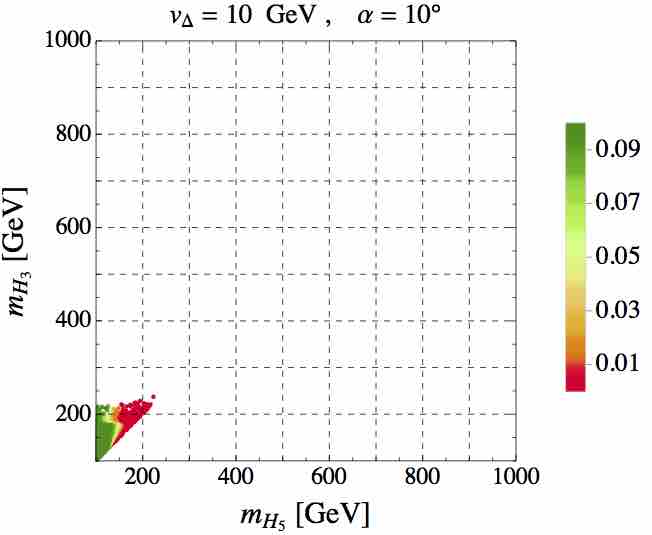}
\includegraphics[scale=0.16]{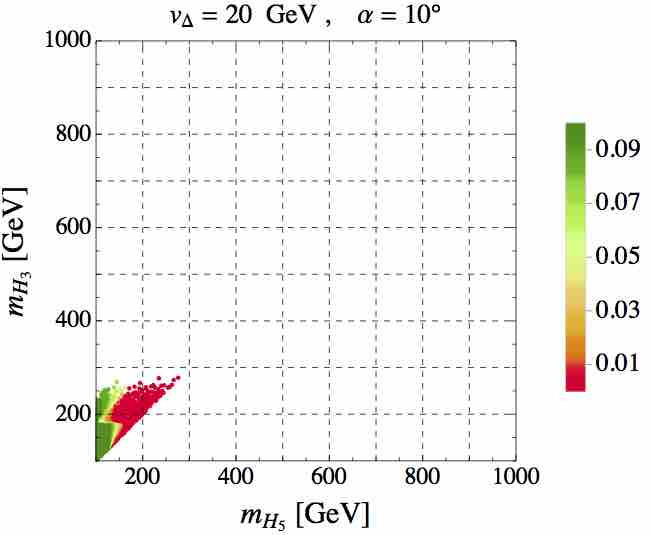}
\includegraphics[scale=0.16]{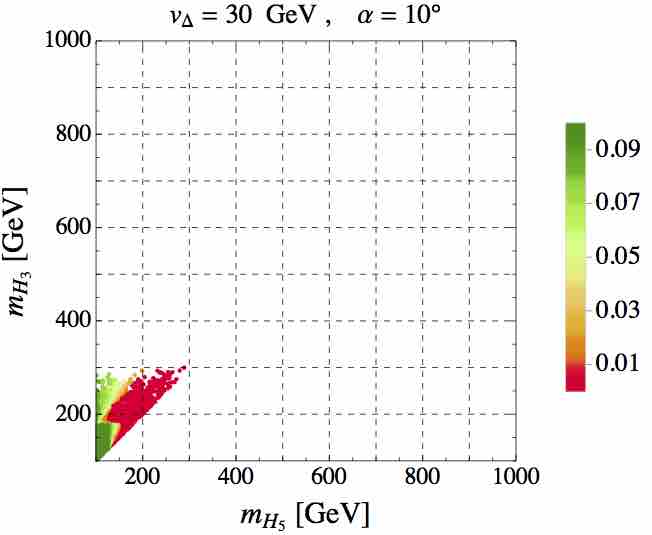}
\includegraphics[scale=0.16]{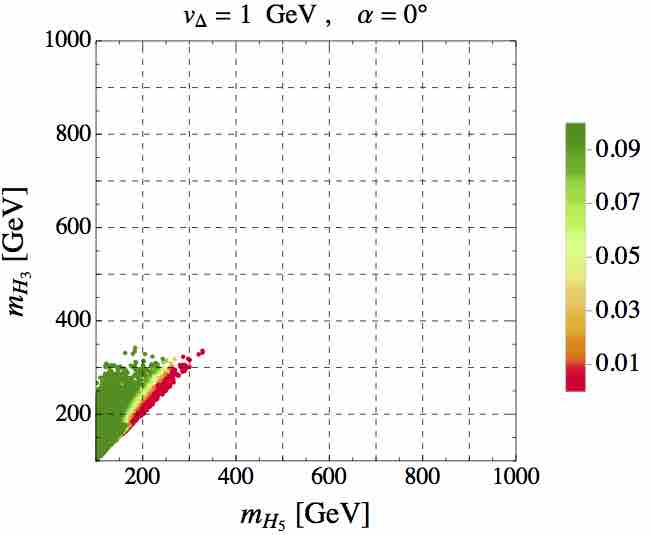}
\caption{$BR(H_3^+ \rightarrow H_5^{++} W^-) BR(H_5^{++} \rightarrow W^+ (\rightarrow \ell^+ \nu_{\ell})W^+(\rightarrow \ell^+ \nu_{\ell}))$ for various values of ($v_\Delta,\alpha$).}
\label{h3h5}
\end{figure}

\begin{figure}[ht]
\centering
\includegraphics[scale=0.16]{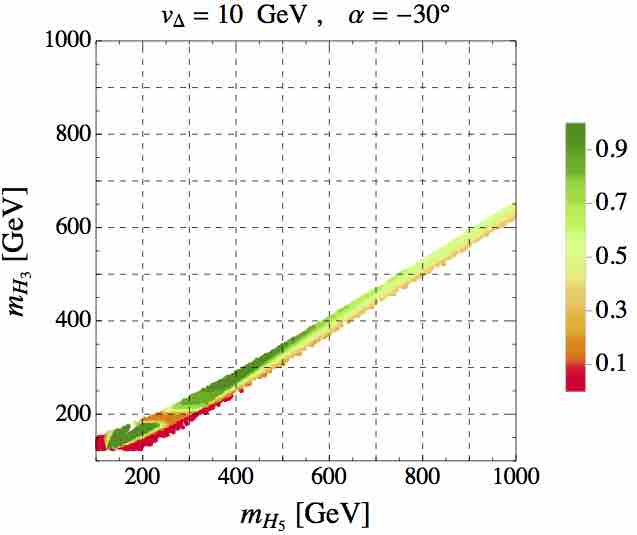}
\includegraphics[scale=0.16]{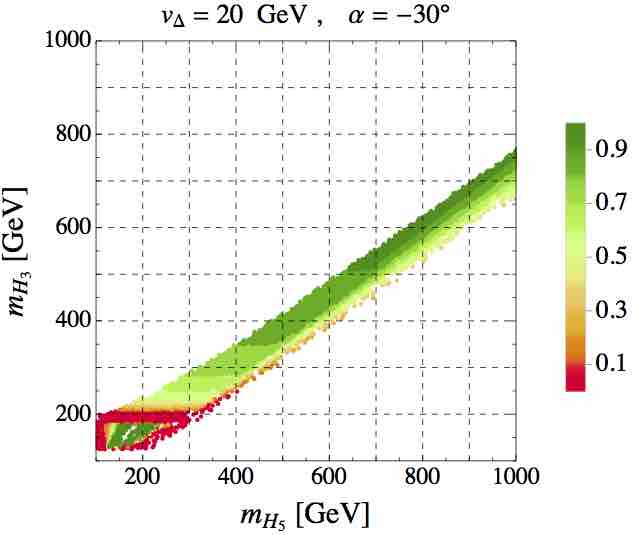}
\includegraphics[scale=0.16]{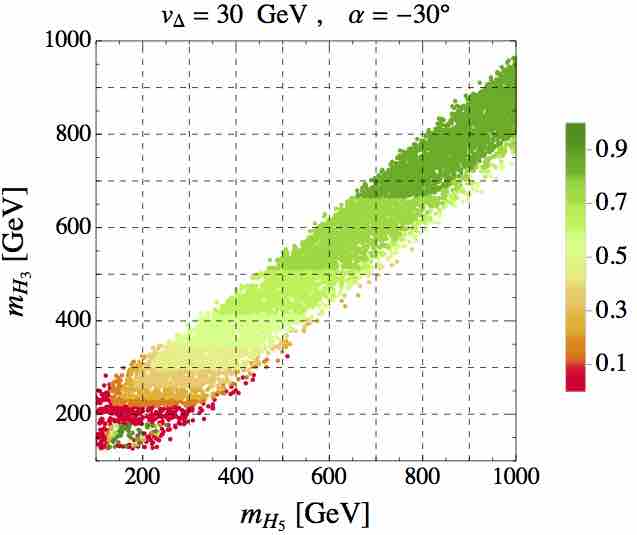}
\includegraphics[scale=0.16]{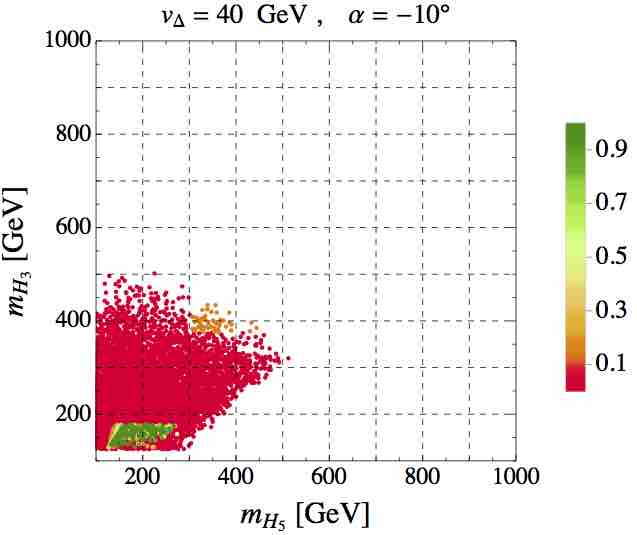}
\vspace{2mm}
\\
\includegraphics[scale=0.16]{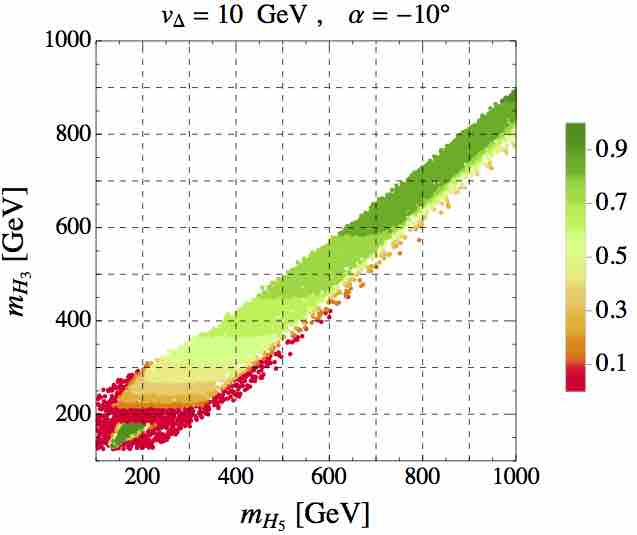}
\includegraphics[scale=0.16]{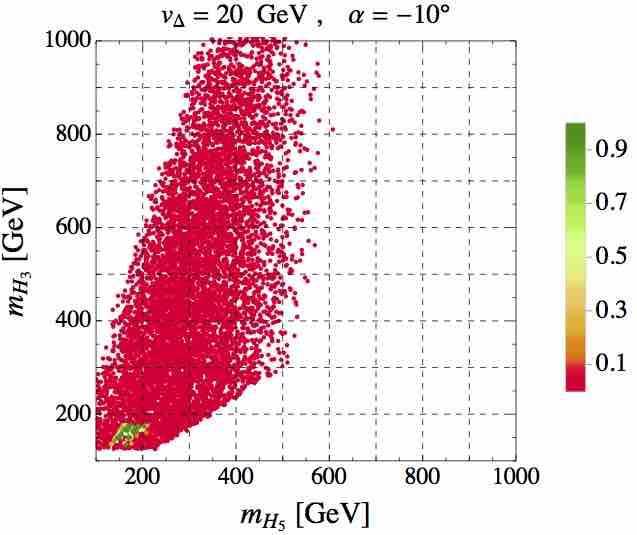}
\includegraphics[scale=0.16]{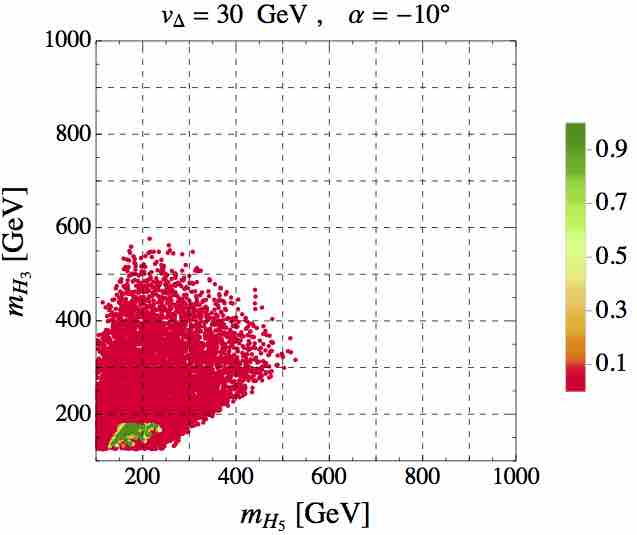}
\includegraphics[scale=0.16]{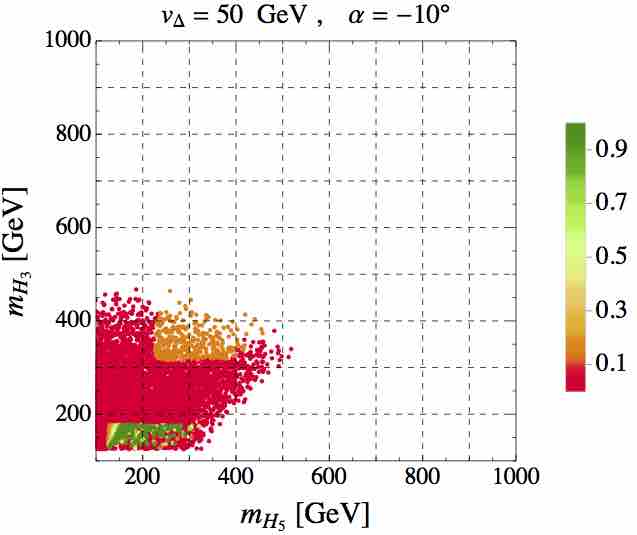}
\vspace{2mm}
\\
\includegraphics[scale=0.16]{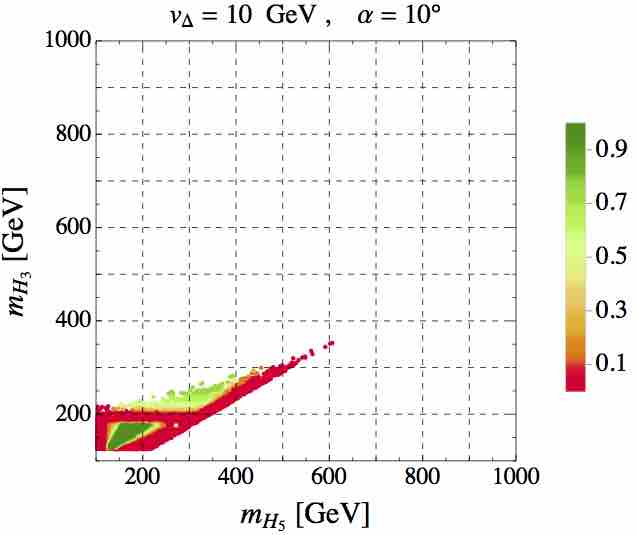}
\includegraphics[scale=0.16]{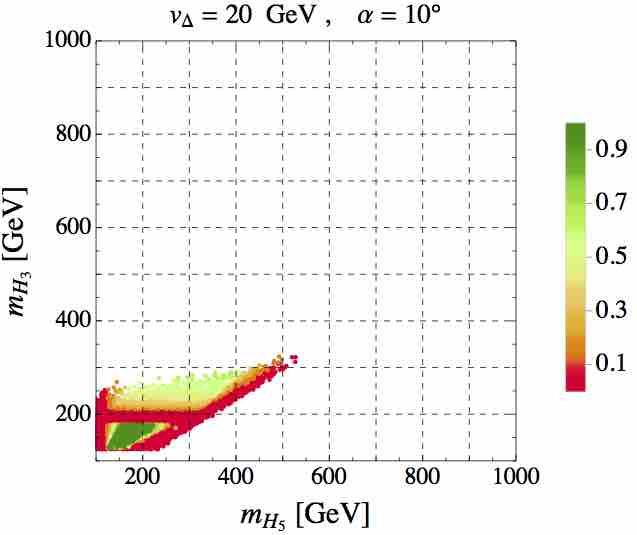}
\includegraphics[scale=0.16]{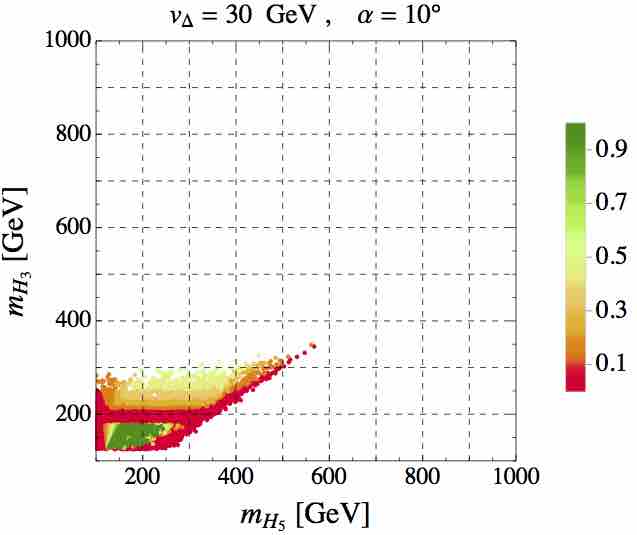}
\includegraphics[scale=0.16]{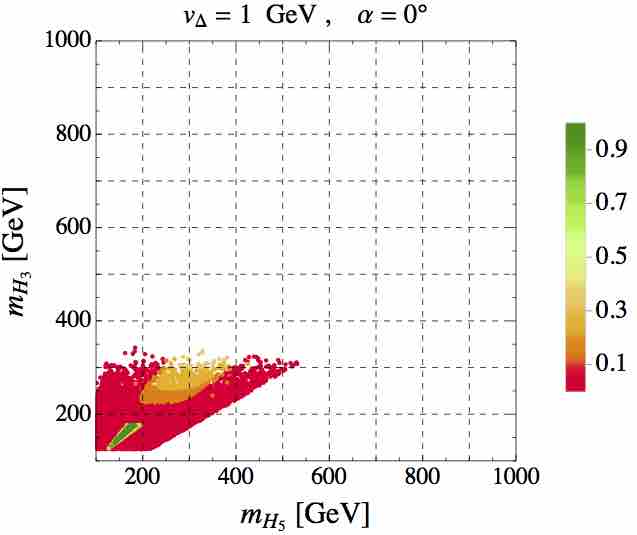}
\caption{$BR(H_3^+\to hW^+)$ for various values of ($v_\Delta,\alpha$).}
\label{H3phWm}
\end{figure}

\begin{figure}[ht]
\centering
\includegraphics[scale=0.16]{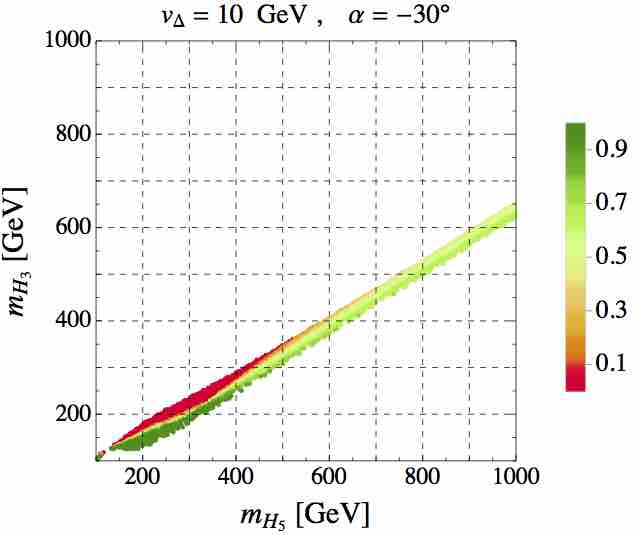}
\includegraphics[scale=0.16]{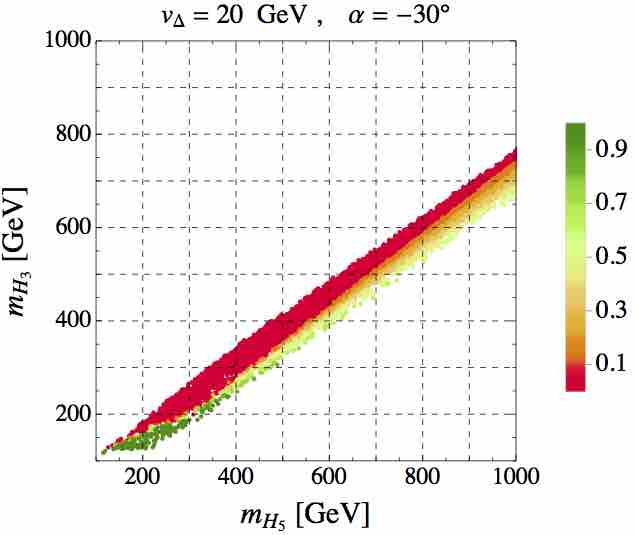}
\includegraphics[scale=0.16]{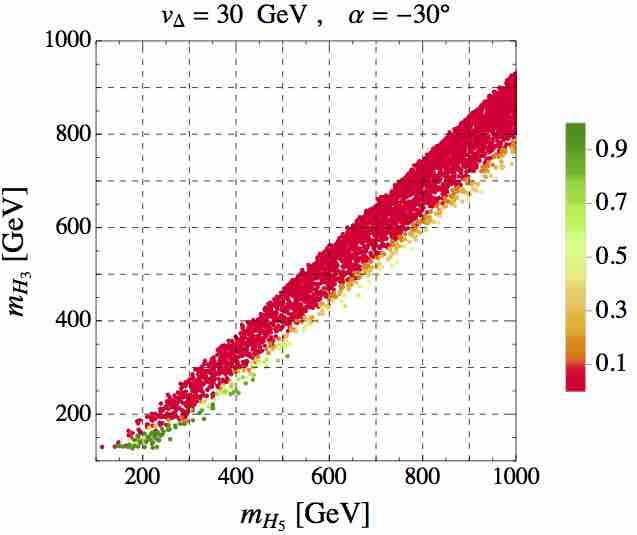}
\includegraphics[scale=0.16]{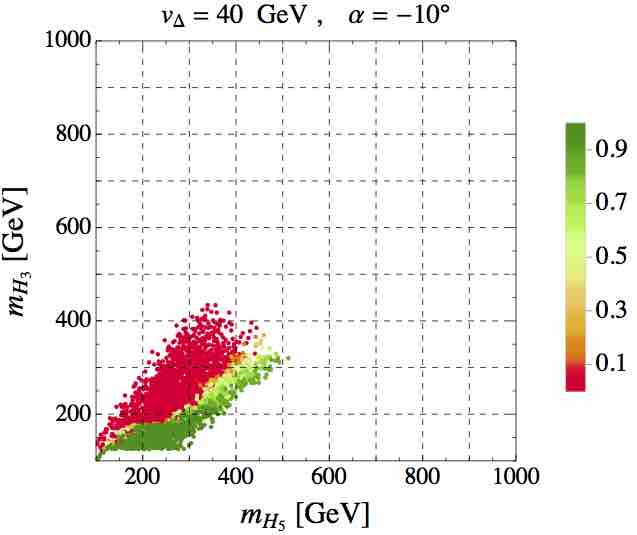}
\vspace{2mm}
\\
\includegraphics[scale=0.16]{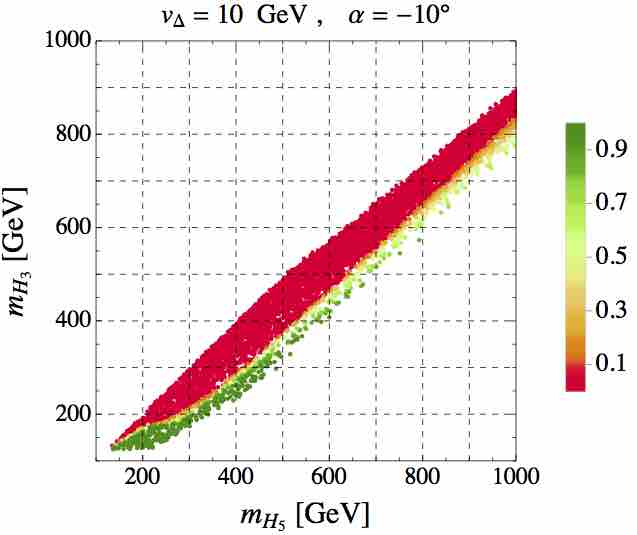}
\includegraphics[scale=0.16]{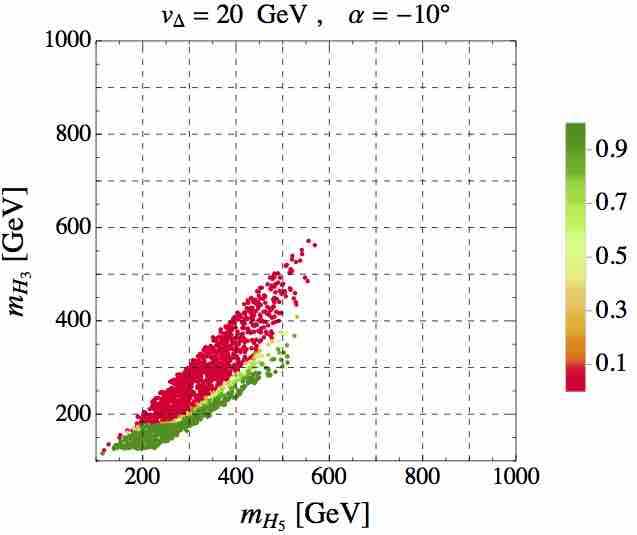}
\includegraphics[scale=0.16]{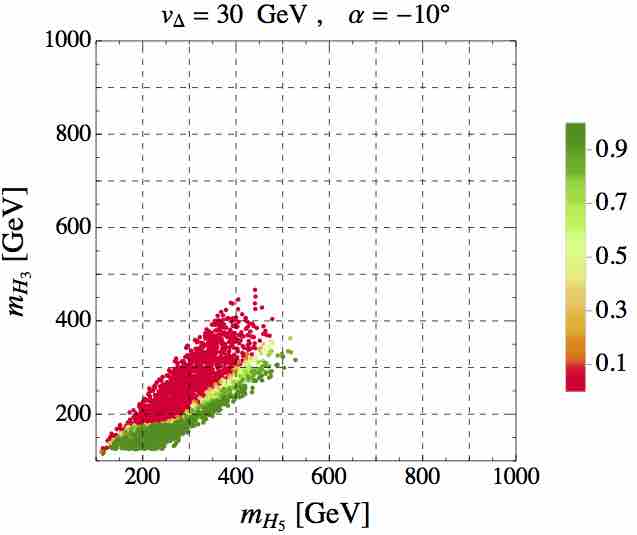}
\includegraphics[scale=0.16]{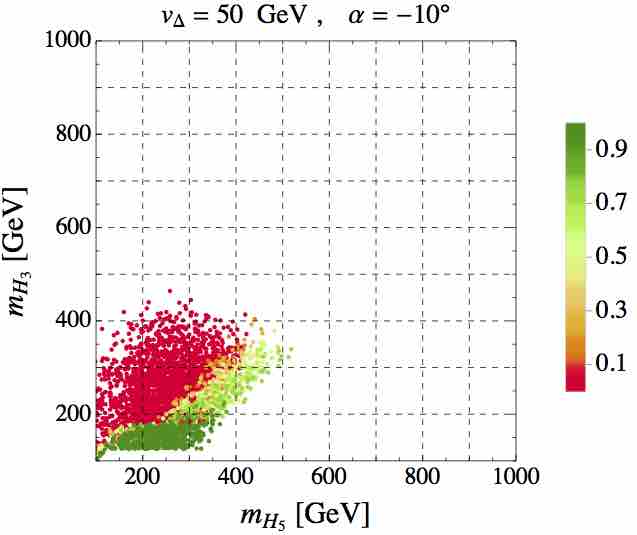}
\vspace{2mm}
\\
\includegraphics[scale=0.16]{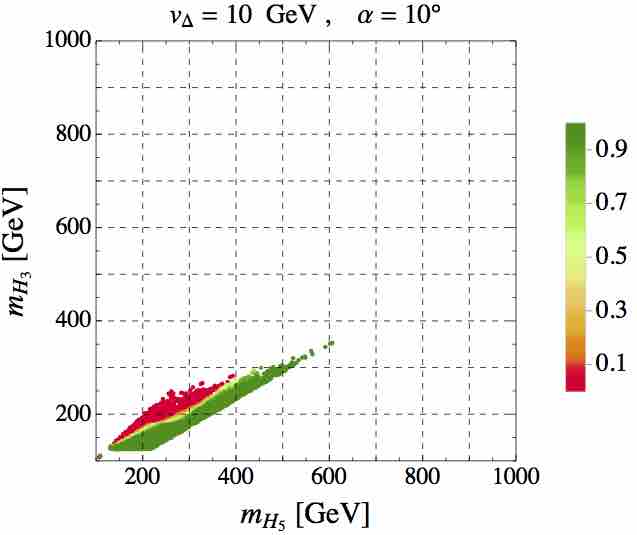}
\includegraphics[scale=0.16]{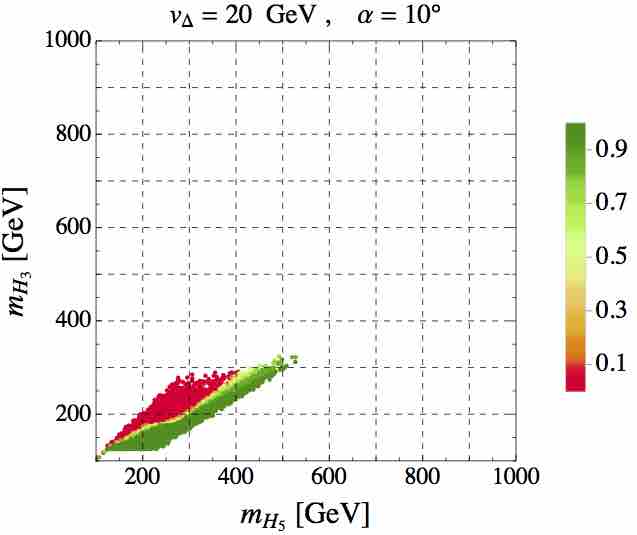}
\includegraphics[scale=0.16]{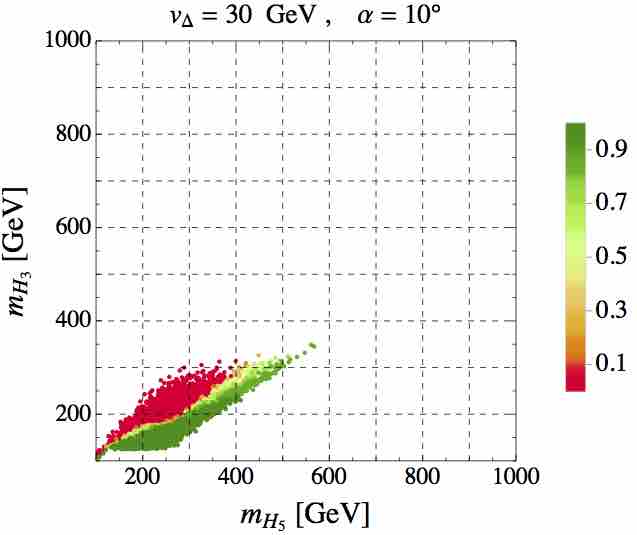}
\includegraphics[scale=0.16]{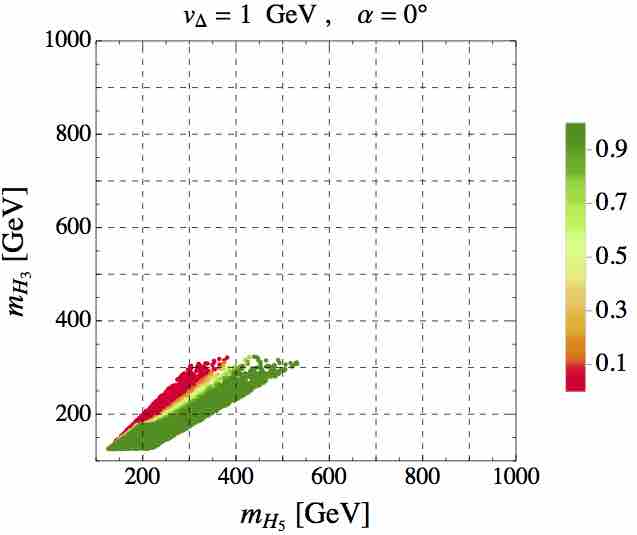}
\caption{$BR(H_3^+\to H_1W^+)$ for various values of ($v_\Delta,\alpha$).}
\label{H3pH1Wm}
\end{figure}

\begin{figure}[ht]
\centering
\includegraphics[scale=0.16]{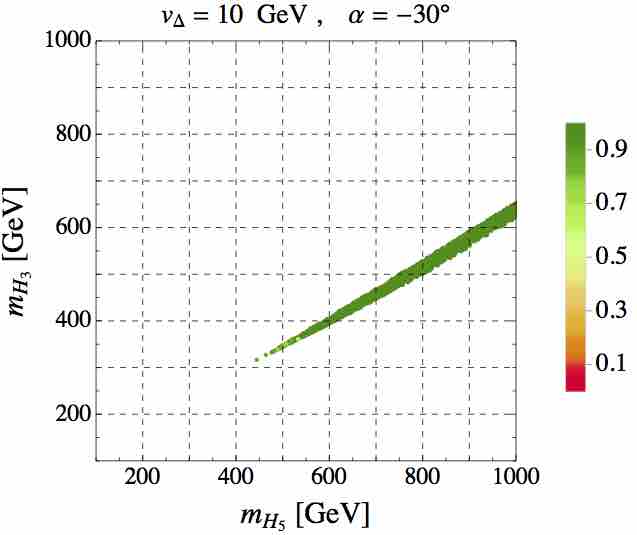}
\includegraphics[scale=0.16]{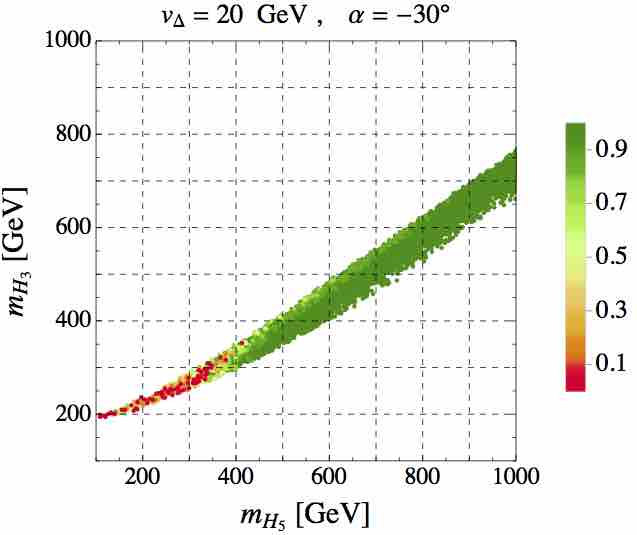}
\includegraphics[scale=0.16]{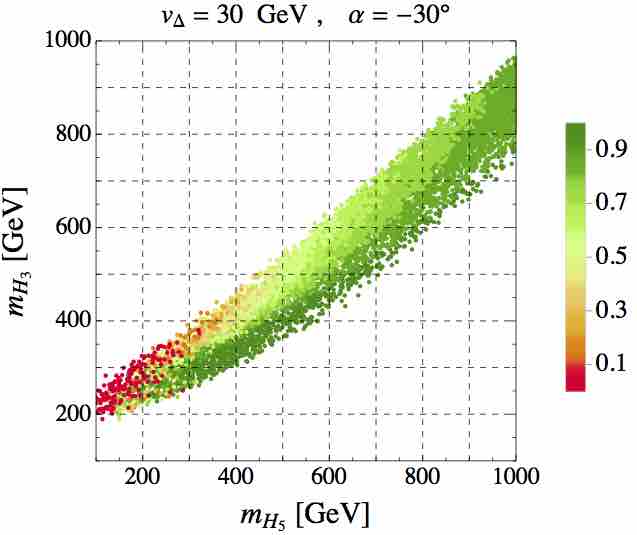}
\includegraphics[scale=0.16]{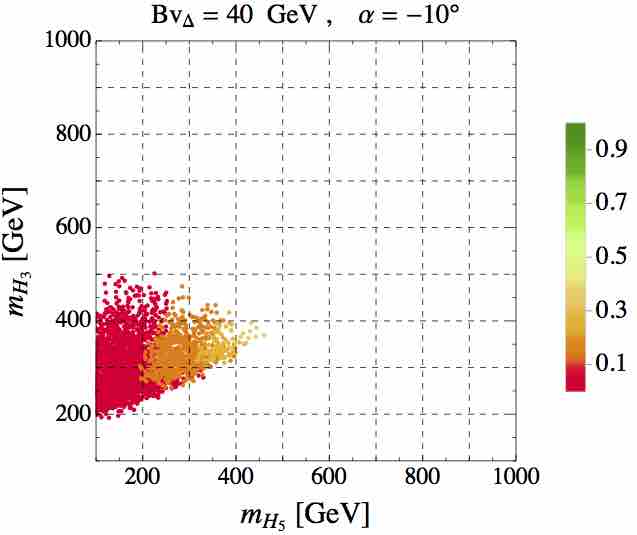}
\vspace{2mm}
\\
\includegraphics[scale=0.16]{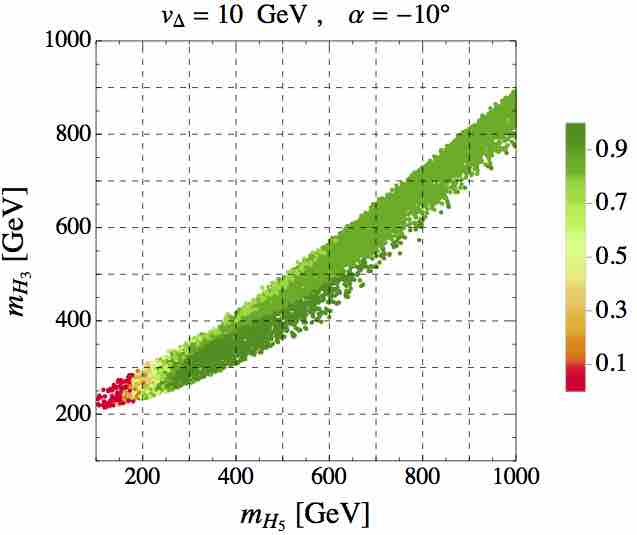}
\includegraphics[scale=0.16]{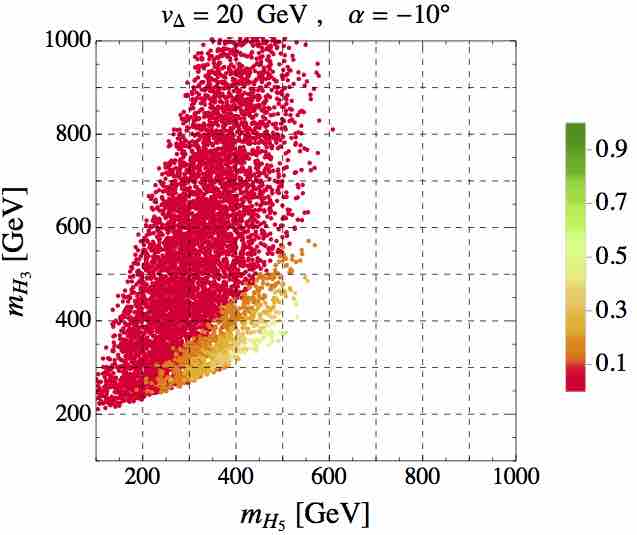}
\includegraphics[scale=0.16]{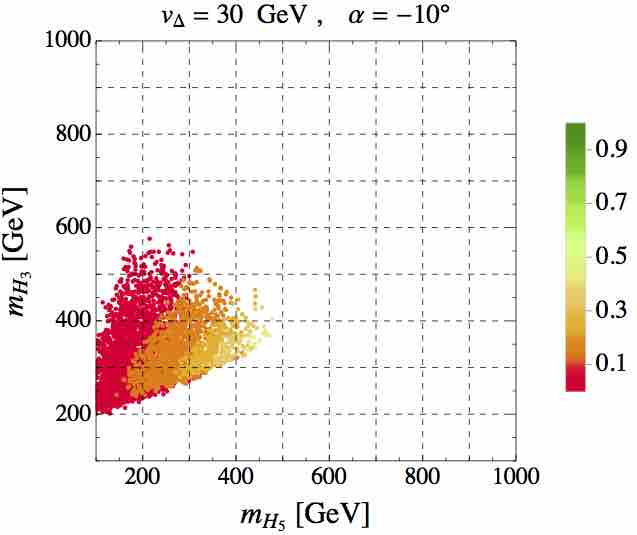}
\includegraphics[scale=0.16]{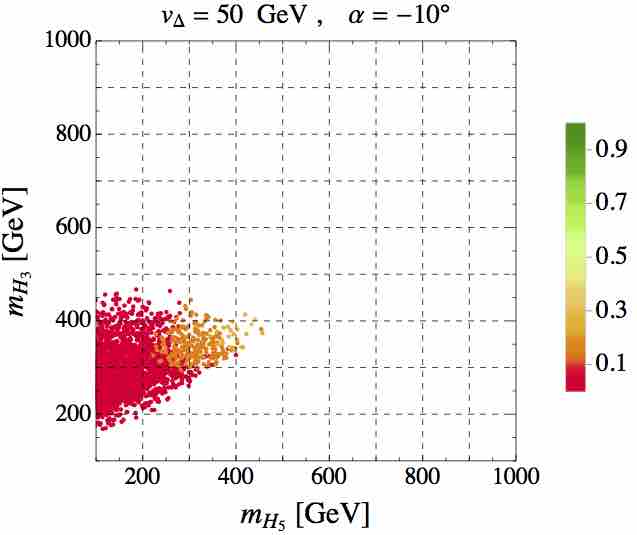}
\vspace{2mm}
\\
\includegraphics[scale=0.16]{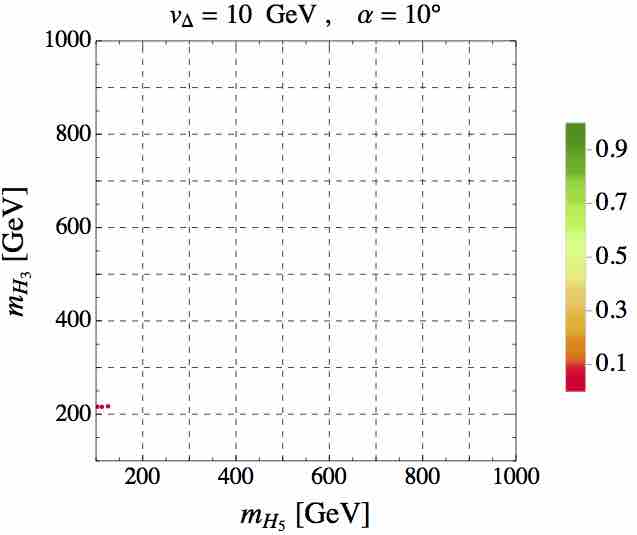}
\includegraphics[scale=0.16]{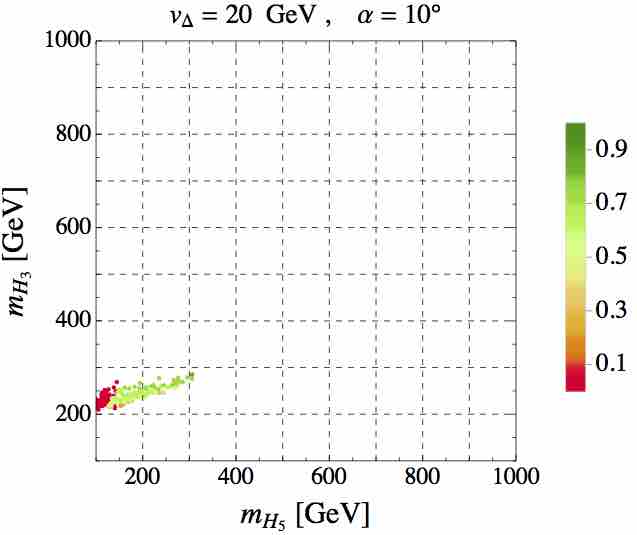}
\includegraphics[scale=0.16]{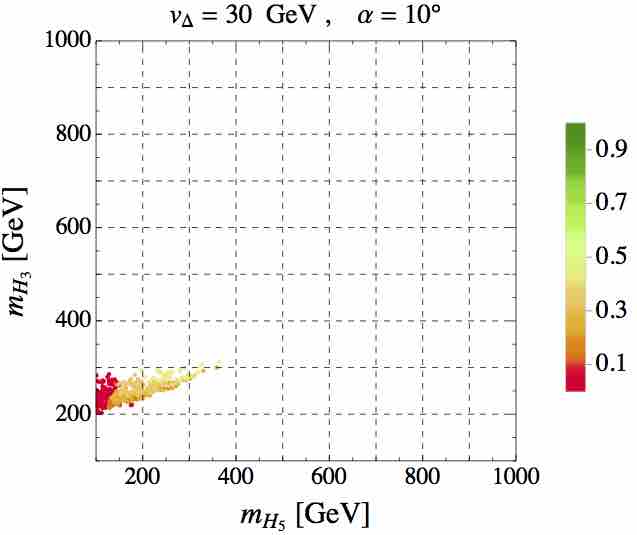}
\caption{$BR(H_1\to hh)$ for various values of ($v_\Delta,\alpha$).  We do not show the trivial plot for the case of $(v_\Delta,\alpha)=(1~\text{GeV},0^\circ)$.}
\label{H1hh}
\end{figure}

\begin{figure}[ht]
\centering
\includegraphics[scale=0.16]{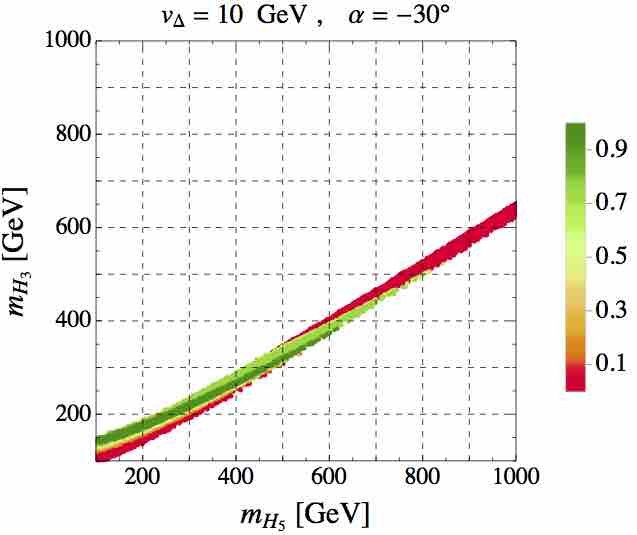}
\includegraphics[scale=0.16]{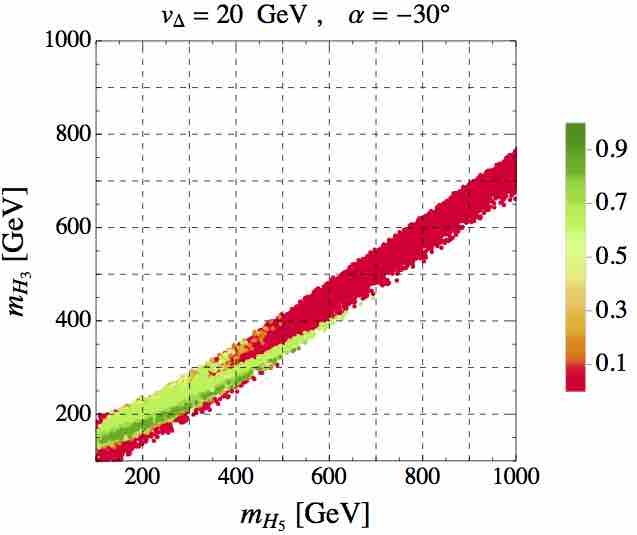}
\includegraphics[scale=0.16]{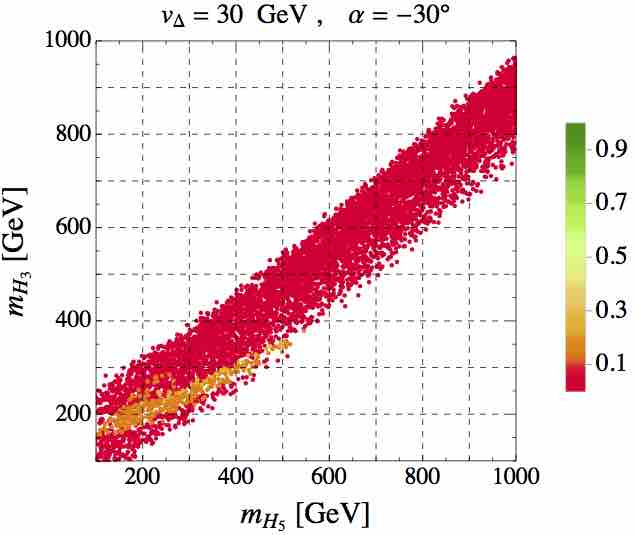}
\includegraphics[scale=0.16]{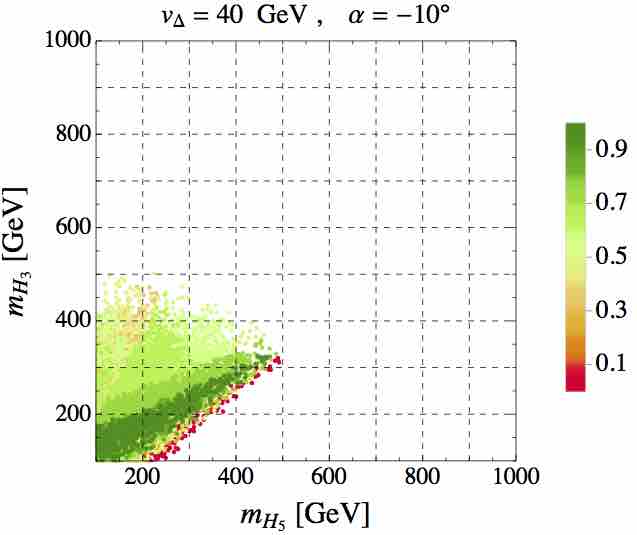}
\vspace{2mm}
\\
\includegraphics[scale=0.16]{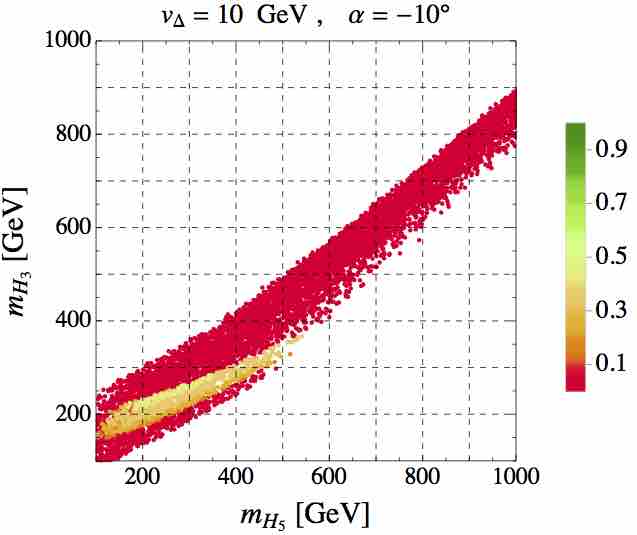}
\includegraphics[scale=0.16]{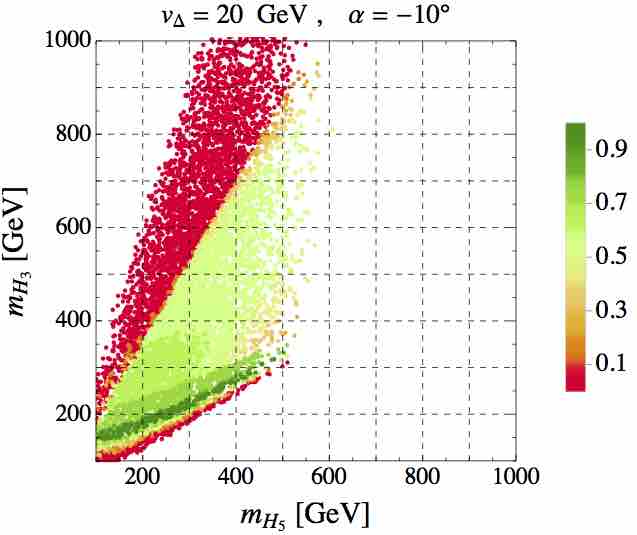}
\includegraphics[scale=0.16]{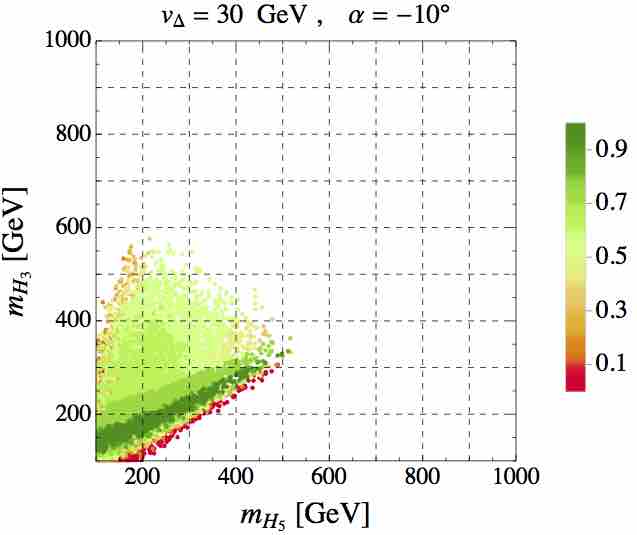}
\includegraphics[scale=0.16]{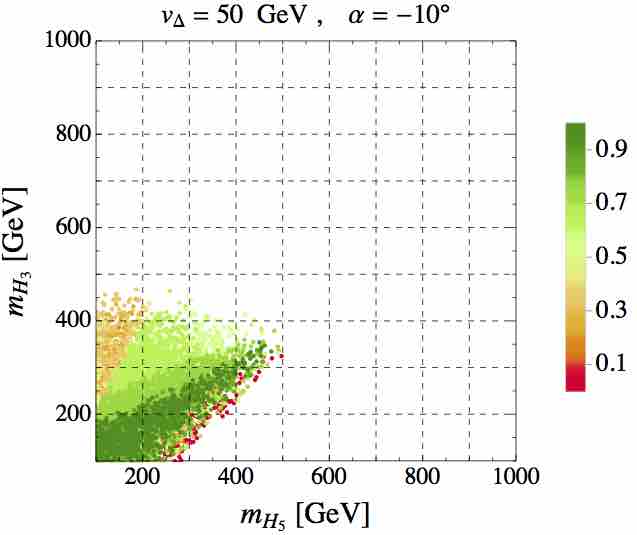}
\vspace{2mm}
\\
\includegraphics[scale=0.16]{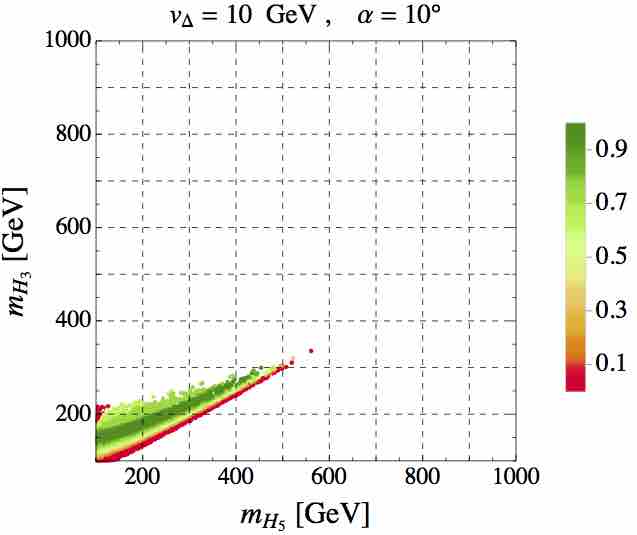}
\includegraphics[scale=0.16]{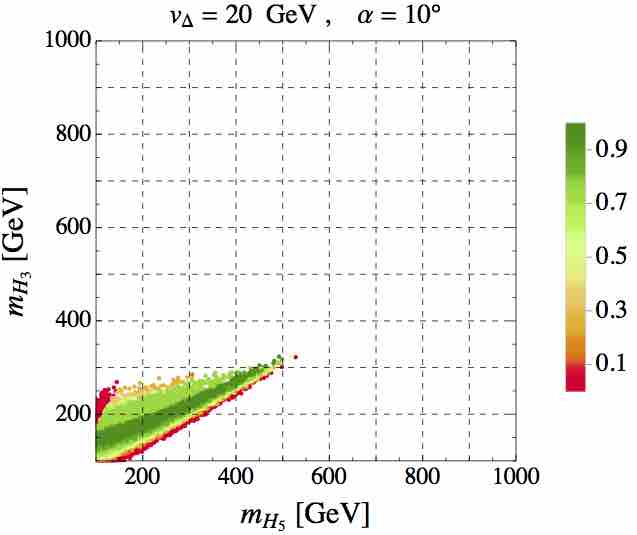}
\includegraphics[scale=0.16]{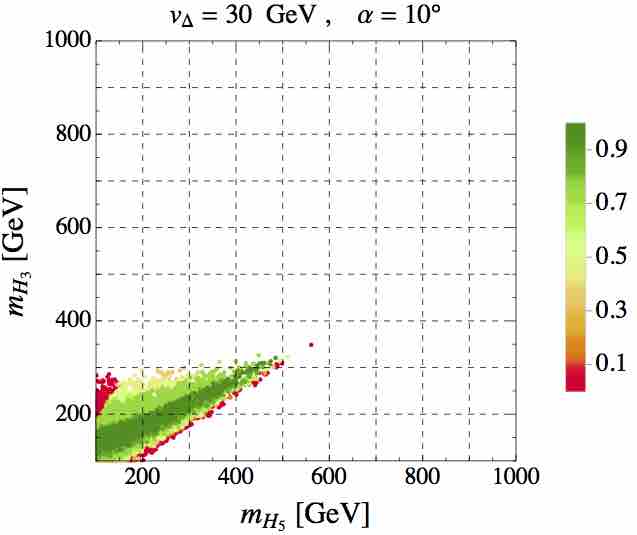}
\includegraphics[scale=0.16]{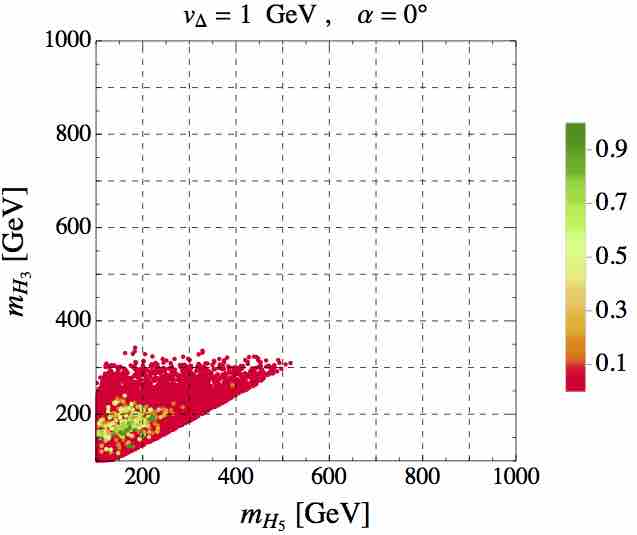}
\caption{$BR(H_1\to W^+W^-)$ for various values of ($v_\Delta,\alpha$).}
\label{H1WW}
\end{figure}

We now conduct a comprehensive parameter scan for viable exotic Higgs boson mass spectra of the GM model by using the most general set of parameters.  From randomly generated mass spectra, we select those that pass the theoretical constraints given in Eqs.~(\ref{stability}) and (\ref{unitarity}). 
Also, the mass spectra are required to satisfy at the $2\sigma$ level the experimental constraints derived from electroweak precision tests in Eq.~(\ref{Sparameter}) and the $Z b \bar{b}$ coupling measurement.  As discussed in the previous section, we adopt the twelve sets of $(v_\Delta,\alpha)$ selected based on Fig.~\ref{chisq} for further numerical analyses.
We calculate the following quantities for each viable mass spectrum and plot them on a two-dimensional plane spanned by the 5-plet mass $m_{H_5}$ and the 3-plet mass $m_{H_3}$.  In the case of decays, we only show the results for positively charged Higgs bosons while those for the conjugate particles should be obvious.
\begin{enumerate}
\item The mass of the heavier SU(2)$_V$ singlet $H_1$, $m_{H_1}$, shown in Fig.~\ref{h1mass}.
\item The signal strength of the GGF production of $h$ followed by a decay into $\gamma \gamma$,
\begin{align*}
\mu^{\text{GGF}}_{h\gamma\gamma} &= \frac{\sigma(g_{/p} g_{/p} \rightarrow h)_{{\rm GM}} BR(h \rightarrow \gamma \gamma)_{{\rm GM}}}{\sigma(g_{/p} g_{/p} \rightarrow h)_{{\rm SM}} BR(h \rightarrow \gamma \gamma)_{{\rm SM}}}~,
\end{align*}
shown in Fig.~\ref{muhdiphoton}.
\item The signal strength of the GGF production of $h$ followed by a decay into $\gamma Z$,
\begin{align*}
\mu^{\text{GGF}}_{h\gamma Z} &= \frac{\sigma(g_{/p} \, g_{/p} \rightarrow h)_{{\rm GM}} BR(h \rightarrow \gamma Z)_{{\rm GM}}}{\sigma(g_{/p} g_{/p} \rightarrow h)_{{\rm SM}} BR(h \rightarrow \gamma Z)_{{\rm SM}}}~,
\end{align*}
shown in Fig.~\ref{muhzphoton}.
\item The total decay widths of $H_5^{++}$ and $H_3^+$ divided by their corresponding masses, 
\begin{align*}
\frac{\Gamma_{H_5^{++}}}{m_{H_5}}~,  \ \ \frac{\Gamma_{H_3^+}}{m_{H_3}}~,
\end{align*}
shown respectively in Figs.~\ref{H5ppwidth} and \ref{H3pwidth}.
\item The branching ratio of the direct decay of $H_5^{++}$ into $W^+ W^+$ followed by the leptonic decay of each $W^+$ (summed over all flavors), including contributions from off-shell $W^+$,
\begin{align*}
BR(H_5^{++} \rightarrow W^+(\rightarrow \ell^+ \nu_{\ell}) \, W^+(\rightarrow \ell^{\prime +} \nu_{\ell^{\prime}}) )~,
\end{align*}
shown in Fig.~\ref{h5ww}.
\item The branching ratio of the cascade decay of $H_5^{++}$ into $H_3^+ W^+$ followed by the $H_3^+$ decay into $h W^+$ where each $W^+$ decays leptonically,
including contributions from off-shell $W^+$,
\begin{align*}
BR(H_5^{++} \rightarrow H_3^+ W^+(\rightarrow \ell^+ \nu_{\ell}) ) BR(H_3^+ \rightarrow h W^+(\rightarrow \ell^{\prime +} \nu_{\ell^{\prime}}))~,
\end{align*}
shown in Fig.~\ref{h5h3h}.  The plot is restricted to the region with $m_{H_5} > m_{H_3}$ where this process is possible.
\item The branching ratio of the cascade decay of $H_5^{++}$ into $H_3^+ W^+$ followed by the $H_3^+$ decay into $H_1 W^+$ where each $W^+$ decays leptonically, including contributions from off-shell $W^+$, 
\begin{align*}
BR(H_5^{++} \rightarrow H_3^+ W^+(\rightarrow \ell^+ \nu_{\ell}) ) BR(H_3^+ \rightarrow H_1 W^+(\rightarrow \ell^{\prime +} \nu_{\ell^{\prime}}))~,
\end{align*}
shown in Fig.~\ref{h5h3h1}.  The plot is restricted to the parameter points with $m_{H_5} > m_{H_3} > m_{H_1}$ where this process is possible.
\item The branching ratio of the cascade decay of $H_3^+$ into $H_5^{++} W^-$ followed by the $H_5^{++}$ decay into $W^+ W^+$ with each $W^+$ further decaying leptonically, including contributions from off-shell $W^{\pm}$,  
\begin{align*}
BR(H_3^+ \rightarrow H_5^{++} W^-) BR(H_5^{++} \rightarrow W^+ (\rightarrow \ell^+ \nu_{\ell})W^+(\rightarrow \ell^+ \nu_{\ell})) ~,
\end{align*}
shown in Fig.~\ref{h3h5}.  The plot is restricted to the region with $m_{H_3} > m_{H_5}$ where this process is possible.
\item The branching ratios of $H_3^+$ decaying into $hW^+$ and $H_1W^+$, including contributions from off-shell $W^+$,
\begin{align*}
BR(H_3^+\to hW^+)~,  \ \ BR(H_3^+\to H_1W^+)~,
\end{align*}
shown respectively in Figs.~\ref{H3phWm} and \ref{H3pH1Wm}.
\item The branching ratios of $H_1$ decaying into $hh$ and $W^+ W^-$, including contributions from off-shell $W^{\pm}$ in the latter case,
\begin{align*}
BR(H_1 \rightarrow hh)~,  \ \ BR(H_1 \rightarrow W^+ W^-)~,
\end{align*}
shown respectively in Figs.~\ref{H1hh} and \ref{H1WW}.
\end{enumerate}

One can use the plot of $\mu^{\text{GGF}}_{h\gamma\gamma}$ ({\it i.e.}, Fig.~\ref{muhdiphoton}) to compare the prediction of the GM model with the corresponding 7-TeV and 8-TeV LHC data, thereby studying which mass spectra are consistent with experiments.
The plot of $\mu^{GGF}_{h \gamma Z}$ ({\it i.e.}, Fig.~\ref{muhzphoton}) enables one to estimate the possibility of indirect search of the GM model through the $h \rightarrow Z \gamma$ process.
We will make use of the plots of $H_5^{++}$ branching ratios ({\it i.e.}, Figs.~\ref{h5ww}, \ref{h5h3h}, and \ref{h5h3h1}) to examine prospects for discovering the GM model at the 14-TeV LHC in the next section.

The method of our parameter scan is explicitly described as follows.
The electroweak VEV and the mass of the SM-like Higgs boson are fixed at $v=246$~GeV and $m_h=125$~GeV, respectively.
We choose the following seven independent parameters to scan: 
the VEV ratio $\tan\beta$, 
the mixing angle $\alpha$,
the three mass eigenvalues $m_{H_5}, m_{H_3}, m_{H_1}$, and 
the parameters $M_1^2$ and $M_2^2$. 
Note that the signs of $M_1^2$ and $M_2^2$ could be either positive or negative.
Using the twelve sets of ($v_{\Delta}$, $\alpha$) selected within the $2\sigma$ bound in Fig.~\ref{chisq}, we randomly generate the rest five parameters ($m_{H_5}$, $m_{H_3}$, $m_{H_1}$, $M_1^2$ and $M_2^2$), and check whether they satisfy the constraints in Eqs.~(\ref{stability}) and (\ref{unitarity}) and the $2\sigma$ bound of Eq.~(\ref{Sparameter}).  Note that at this stage, we do not assume any mass hierarchy among the Higgs bosons.  To be phenomenologically interesting at the LHC, the ranges of $m_{H_5}$ and $m_{H_3}$ are both fixed as 100~GeV $< m_{H_{3,5}} <$ 1~TeV.
The ranges of $m_{H_1}$, $M_1^2$ and $M_2^2$ are determined according to the generated values of $m_{H_5}, m_{H_3}$, by taking advantage of the following inequalities that are deduced from the mass formulas in Eq.~(\ref{massformulas}) and the theoretical constraints in Eqs.~(\ref{stability}) and (\ref{unitarity}):
\begin{align}
& \frac{2}{v^2} | M_1^2 - m_{H_3}^2 | <  \frac{8(\sqrt{3}+1)\pi}{3} ~, 
\label{M1} \\
& 0 < \frac{1}{3 \cos^2 \beta \, v^2} \left\{ m_{H_5}^2 - 3\sin^2 \beta \, m_{H_3}^2 + 2(m_{H_1}^2 \cos^2\alpha + m_h^2 \sin^2\alpha) \right\} < \pi ~, 
\label{mH1} \\
& 0 < \frac{1}{6 \cos^2 \beta \, v^2} \left\{ 4 m_{H_5}^2 - 12 \sin^2 \beta \, m_{H_3}^2 + 2(m_{H_1}^2 \cos^2\alpha + m_h^2 \sin^2\alpha) \right.
\nonumber \\
& \qquad\qquad\qquad\qquad\qquad \left. + 6 \sin^2 \beta \, M_1^2 -3 M_2^2 \right\} < \frac{4\pi}{3} ~. 
\label{M2}
\end{align}
We generate 8000 mass spectra for each set of $(v_\Delta,\alpha)$ and plot the results in Fig.~\ref{h1mass}, with different colors representing $m_{H_1}$ falling in different mass ranges.  The magenta colored points are those with $m_{H_1} \le m_h$.  It should be noted that for a given point $(m_{H_5},m_{H_3})$ in each scatter plot, the value of $m_{H_1}$ actually varies with $M_1^2$ and $M_2^2$ over a small range.  It is seen that the parameter spaces for $\alpha = 10^\circ$ or the close-to-decoupling limit $(v_\Delta,\alpha) = (1~\mbox{GeV},0^\circ)$ (plots on the right hand side of the figure) are relatively limited, with $m_{H_5} \lesssim 600$~GeV, $m_{H_3} \lesssim 350$~GeV and $m_{H_1} \lesssim 300$~GeV.  In a certain region of $(v_\Delta,\alpha)$ (upper left plots), some or all the exotic Higgs boson masses can be in the TeV regime.  These spectra serve as the basis of Figs.~\ref{muhdiphoton} to \ref{H1WW}.

There is an upper bound on $m_{H_3}$ when $\alpha > 0$, as one can readily observe in the plots with $\alpha=10^\circ$ in Figs.~\ref{h1mass} to \ref{H1WW}.  The origin of this bound is understood as follows.
The combination of the couplings $4\lambda_4 + \lambda_5$ can be expressed, with the help of Eq.~(\ref{massformulas}), as
\begin{align}
4\lambda_4 + \lambda_5 = \frac{2}{v^2} \left[ m_{H_3}^2 - \sqrt{\frac{2}{3}} \frac{\sin \alpha \cos \alpha}{\sin \beta \cos \beta} (m_h^2 - m_{H_1}^2) \right] ~.
\end{align}
Since $m_{H_1}^2 > 0$, we obtain the following inequality when $\alpha>0$:
\begin{align}
m_{H_3}^2 \leq \frac{1}{2}(4\lambda_4 + \lambda_5)v^2 + \sqrt{\frac{2}{3}} \frac{\sin \alpha \cos \alpha}{\sin \beta \cos \beta} m_h^2 ~.
\end{align}
Larger $\lambda_4$ or $\lambda_5$ would lead to violation of the vacuum stability conditions in Eq.~(\ref{stability}) or the perturbative unitarity conditions in Eq.~(\ref{unitarity}).  Therefore, $m_{H_3}$ is bounded from above, though the exact value of the upper bound cannot be expressed analytically.

Also explicitly shown in Fig.~\ref{h1mass} is that all the six mass hierarchies for the exotic Higgs bosons are possible according to the parameter scan.  Nevertheless, the most probable ones are either $m_{H_5}>m_{H_3}>m_{H_1}$, dubbed the normal hierarchy, or $m_{H_1}>m_{H_3}>m_{H_5}$, dubbed the inverted hierarchy.

The average signal strength of the SM-like Higgs boson production and decay into two photons from the ATLAS Collaboration~\cite{ATLASsignalstrength} and CMS Collaboration~\cite{CMSsignalstrength} is given by $\mu^{\text{GGF}}_{h\gamma\gamma}=1.12\pm0.22$.  In Fig.~\ref{muhdiphoton}, we see that the predicted signal strength ranges from $\sim 0.6 - 1.4$ for almost all the obtained mass spectra.  On the other hand, the constraint from $\mu^{GGF}_{h\gamma Z}=2.7^{+4.5}_{-4.3}$~\cite{ATLASsignalstrength} is not constraining at all.  Most of the predicted $\mu^{GGF}_{h\gamma Z}$ values tend to be bigger than or about 1.

The total decay widths of $H_5^{++}$ and $H_3^+$ normalized to their corresponding masses are shown in Figs.~\ref{H5ppwidth} and \ref{H3pwidth}.  Since $H_5^{++}$ is a quark-phobic scalar, its only decay channels are $H_3^+W^+$, $W^+W^+$ and $H_3^+H_3^+$ at tree level, provided allowed by kinematics.  It is seen that its decay width is $\lesssim 1\%$ of its mass in almost all cases.  On the other hand, $H_3^+$ is a gauge-phobic scalar boson.  The only decay channels of $H_3^+$ are $hW^+$, $H_1W^+$, $H_5^{++}W^-$ and $t\overline{b}$ at tree level as long as it is kinematically allowed.  Compared to $H_5^{++}$, the $H_3^+$ boson has a slightly larger value of the total width-to-mass ratio in most allowed $(v_\Delta,\alpha)$ space and can sometimes reach $\sim 10\%$.  In general, Figs.~\ref{H5ppwidth} and \ref{H3pwidth} verify that the narrow width approximation employed in our numerical analysis is valid in most spectra.

We show plots for the branching ratios of the decays of $H_5^{++}$ into several different final states in Fig.~\ref{h5ww} to \ref{h5h3h1}.  One distinct feature of the GM model is that thanks to the custodial symmetry, $v_\Delta$ can be larger compared to the model extended with only one complex Higgs triplet field.  Hence same-sign dilepton events coming from the process of $H_5^{++}\to W^+(\to\ell^+\nu_\ell)W^+(\to\ell'^+\nu_{\ell'})$ with the inclusive $BR(W^+\to\ell^+\nu)=10.86\%$~\cite{pdg} provide a distinguished way to test the model because the $H_5^{++}W^-W^-$ vertex is proportional to $v_\Delta$.  The resulting branching ratio for different sets of $v_\Delta$ and $\alpha$ are shown in Fig.~\ref{h5ww}.  It is observed that the $H_5^{++}\to W^+(\to\ell^+\nu_\ell)W^+(\to\ell'^+\nu_{\ell'})$ decay is often a major one in a significant portion of the allowed region in each plot. 
This is because most cases either do not have a hierarchy with $m_{H_5} > m_{H_3}$ in the mass spectrum or do not have sufficient mass splitting.
When cascade decays are allowed, $H_5^{++}$ can also decay into $H_3^+W^+$ with $H_3^+$ further decaying into $hW^+$ or $H_1W^+$.
Their corresponding branching ratios are respectively shown in Fig.~\ref{h5h3h} and Fig.~\ref{h5h3h1}, where only those mass spectra with $m_{H_5} > m_{H_3}$ are plotted.  A detailed collider phenomenology study of these scenarios is given in the next section.

If instead $m_{H_5}$ is sufficiently lighter than $m_{H_3}$, $H_3^+$ can decay into $H_5^{++}W^-$ with $H_5^{++}$ further decaying into $W^+W^+$.  Such a result is shown in Fig.~\ref{h3h5}, where only the mass spectra with $m_{H_5}<m_{H_3}$ are shown.  In addition, $H_3^+$ can also decay into $hW^+$ or $H_1W^+$ as shown in Fig.~\ref{H3phWm} and Fig.~\ref{H3pH1Wm}, respectively. In the latter case, $m_{H_1}$ must be smaller than $m_{H_3}$ while the mass relation between $m_{H_1}$ or $m_{H_3}$ and $m_{H_5}$ remains arbitrary in both cases.  As shown in Fig.~\ref{H3pH1Wm}, $BR(H_3^+\to H_1W^+)$ increases with $m_{H_5}$ for a fixed $m_{H_3}$.

Finally, we discuss the decay channels of $H_1$.  In addition to the same decay channels as the SM-like Higgs boson, it can also decay into $H_3^\pm W^{\mp}$ and/or a pair of other Higgs bosons, provided these are kinematically allowed.  As two promising channels in the search of an additional neutral Higgs boson, we discuss the $H_1 \to hh / W^+W^-$ decays.  The branching ratio of $H_1 \to hh$ is shown in Fig.~\ref{H1hh}, where we have omitted the plot for $(v_\Delta,\alpha)=(1~\text{GeV},0^\circ)$ because the value is diminishing in such a decoupling limit.  In the $\alpha = 10^\circ$ cases, there is only a very tiny portion of the allowed spectra that can have this decay channel, and the branching ratio is generally less than $\sim 40\%$.  In cases with negative $\alpha$, however, the branching ratio can sometimes reach above $\sim 90\%$.~
\footnote{See also Ref.~\cite{Godunov:2015lea} for a similar finding under the consideration of a simplified Higgs potential of the GM model.} 
The branching ratio of the $H_1$ decaying into $W^+W^-$ is plotted in Fig.~\ref{H1WW}.  The $H_1W^+W^-$ vertex has the coupling $(g^2/6)(3\sin\alpha\sin\beta+2\sqrt6\cos\alpha\cos\beta)$.  With appropriate $\alpha$ and $v_\Delta$, the $H_1 \to W^+W^-$ mode can be the dominant one.

\section{Prospects for Observing Signatures of Georgi-Machacek Model at 14-TeV LHC
\label{sec:numerical_results}}

\begin{figure}[t!]
\centering
\includegraphics[scale=0.55]{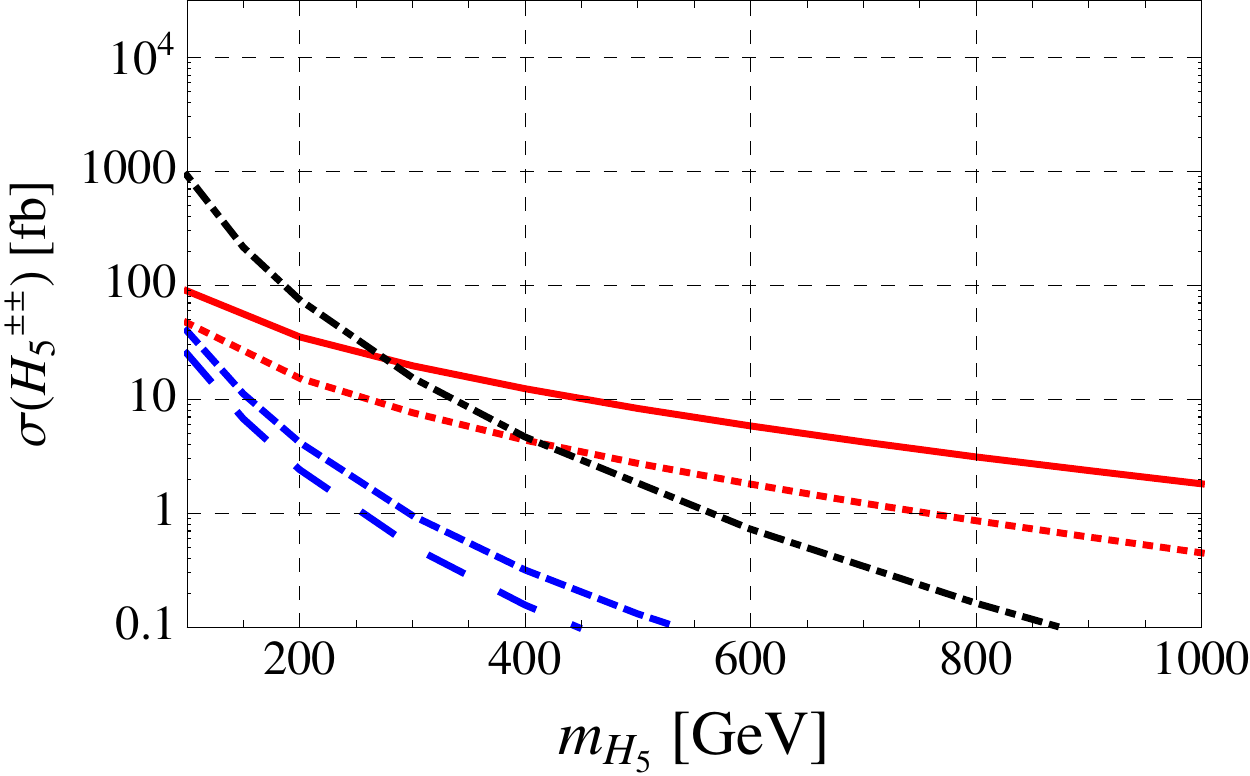}
\hspace{2mm}
\includegraphics[scale=0.55]{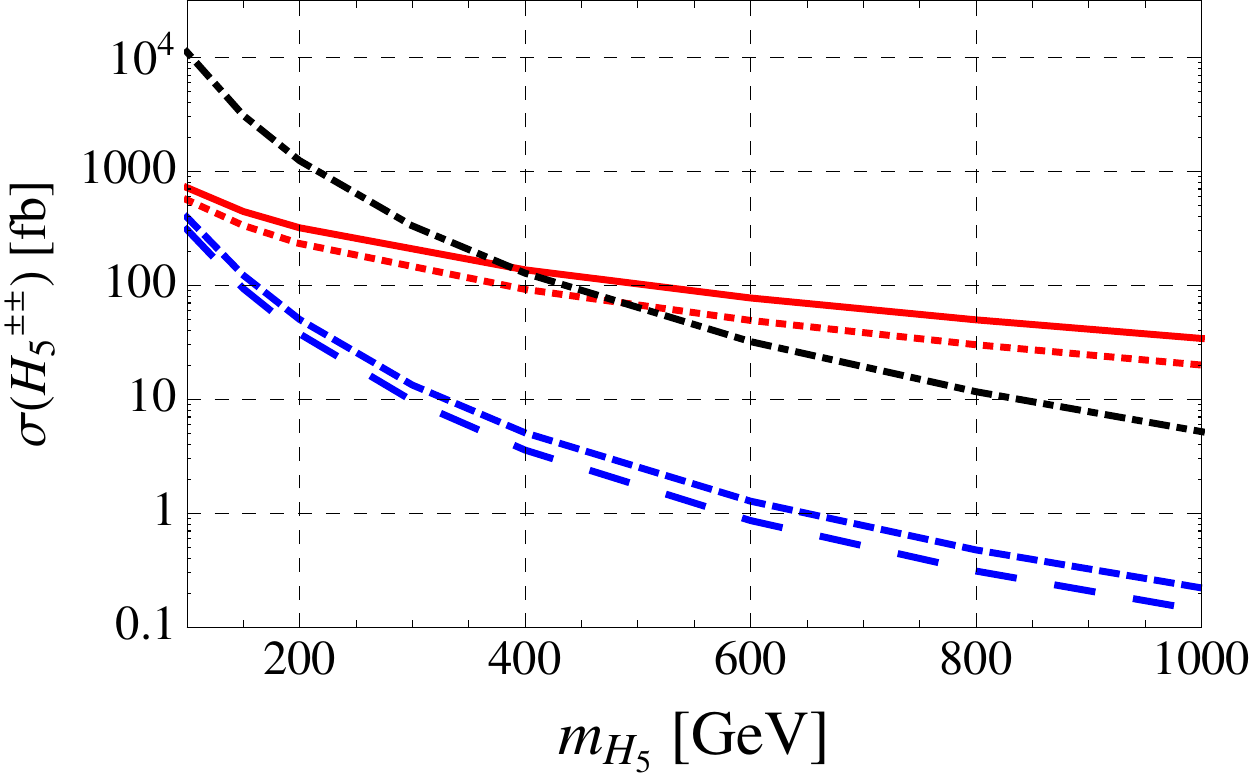}
\caption{Left: The production cross sections (fb) of $H_5^{\pm\pm}$ for various channels in $pp$ collisions with $\sqrt{s}=14$~TeV as a function of $m_{H_5}$.  The red curves correspond to those for the vector boson fusion, the blue curves to those for the associated production of $H_5^{\pm\pm} W^\mp$, and the black curve to that for the Drell-Yan production of $H_5^{++}H_5^{--}$.  The solid red and short-dashed blue curves are for $H_5^{++}$ while the dotted red and long-dashed curves are for $H_5^{--}$.  Here we take $v_{\Delta}=10$~GeV. 
Right: The same as left, but with $\sqrt{s}=100$~TeV.}
\label{fig:cross_section}
\end{figure}

We evaluate the prospects for observing a signature of the GM model in the 14-TeV run of the LHC.
One of the promising channels for discovering the GM model is the production of a doubly-charged Higgs boson $H_5^{\pm \pm}$ via the VBF process, followed by its decay into final states containing a pair of same-sign leptons.
This channel has three advantages:
First, the SM background for events with two same-sign leptons is suppressed compared to those with opposite-sign leptons or only one lepton.
Secondly, the VBF process is the dominant production mechanism of $H_5^{\pm \pm}$ if its mass is above $\sim 300$~GeV and $v_{\Delta} \gtrsim 10$~GeV (see the left plot of Fig.~\ref{fig:cross_section}).  Hence, it is a good strategy to concentrate on its production via the VBF process.
Thirdly, in the VBF production of $H_5^{\pm \pm}$, leptons possibly arise only from the decay of the singly-produced $H_5^{\pm \pm}$.
This fact allows a less parameter-dependent estimation of the acceptance and efficiency of events with two same-sign leptons, in comparison with the Drell-Yan (DY) production of $H_5^{\pm \pm} H_5^{\mp \mp}$ or $H_5^{\pm \pm} H_3^{\mp}$ and the associated production of $H_5^{\pm \pm} W^{\mp}$.
For these reasons, we hereafter focus on the process of the VBF production of $H_5^{\pm \pm}$ followed by its decay into a pair of same-sign leptons through same-sign $W$ bosons \footnote{We neglect the direct decays into like-sign leptons because the couplings of $H_5^{\pm \pm}$ to SM leptons are strongly suppressed when $v_{\Delta} \gtrsim 1$~GeV.}. The search for this process can be most easily done by selecting events containing a pair of same-sign light leptons, $\mu^\pm \mu^\pm$, $e^\pm e^\pm$ and $e^\pm \mu^\pm$.
It is not necessary to impose any selection cut on the jets associated with the VBF process, as our purpose is to observe the production of $H_5^{\pm \pm}$ rather than identifying the production process.

We can evaluate the significance of a signal of the GM model by comparing the number of events with two same-sign light leptons coming from the production and decay of $H_5^{\pm \pm}$, $N_{H_5^{++}; {\rm SS \, light \, leptons}}$, where $\ell,\ell'$ denote SM leptons, with the number of those coming from SM processes.  The former can be expressed as (the expression for the number of events with two negatively-charged leptons is similar):
\begin{align}
N_{H_5^{++}; {\rm SS \, light \, leptons}} \ &= \ L \, \sigma(p p \rightarrow H_5^{++} + X) \, BR(H_5^{++} \rightarrow \ell^{+} \ell^{\prime +} + X^{\prime})
\, (A \times \epsilon)~,
\label{numberofsignals}
\end{align}
 where $L$ denotes the integrated luminosity, $\sigma(p p \rightarrow H_5^{++} + X)$ the inclusive production cross section of $H_5^{++}$, $BR(H_5^{++} \rightarrow \ell^{+} \ell^{\prime +} + X^{\prime})$ the branching ratio of $H_5^{++}$ decaying into final states containing two same-sign leptons, and $A \times \epsilon$ the acceptance times efficiency for events with two same-sign leptons arising from processes involving $H_5^{++}$ with certain event selection criteria.  When multiple processes contribute to such events, one should take an average over these processes for the calculation of $A \times \epsilon$.  We include the decays into tau leptons in the definition of $BR(H_5^{++} \rightarrow \ell^{+} \ell^{\prime +} + X^{\prime})$, where the tau leptons further decay leptonically.

At this stage, we do not specify the production process or the decay process of $H_5^{\pm \pm}$.
Later on, however, we will find numerically that the dominant production process is the VBF mechanism,
 and the dominant decay channel whose final state involves two same-sign leptons is either
 $H_5^{\pm \pm} \rightarrow W^{\pm}(\rightarrow \ell^{\pm} \nu) W^{\pm}(\rightarrow \ell^{\prime \pm} \nu)$ or
 $H_5^{\pm \pm} \rightarrow W^{\pm}(\rightarrow \ell^{\pm} \nu) H_3^{\pm}, \, H_3^{\pm} \rightarrow W^{\pm}(\rightarrow \ell^{\prime \pm} \nu) h/H_1$.
By focusing on these specific production and decay channels, it becomes simple and straightforward to estimate the acceptance times efficiency, $A \times \epsilon$, that eventually yields $N_{H_5^{++}; {\rm SS \, light \, leptons}}$.

In Fig.~\ref{fig:cross_section}, we display the cross sections for the VBF production of $H_5^{\pm \pm}$ in $pp$ collisions with $\sqrt{s}=14$~GeV (left plot) and $100$~TeV (right plot), as well as the cross sections for the DY production of $H_5^{\pm \pm} H_5^{\mp \mp}$ and the associated production of $H_5^{\pm \pm} W^{\mp}$.  The DY production of $H_5^{\pm\pm}H_3^\mp$ is not taken into account in our analysis.  All these production cross sections are independent of $\alpha$, as $\alpha$ is the mixing angle between the two singlets.  Here we do not impose any selection cut on the jets associated with the VBF process.  The figure tells us that the VBF mechanism is the dominant production process for $H_5^{++}$ and $H_5^{--}$ when $v_{\Delta}$ is above 10~GeV and $m_{H_5}$ is above $\sim 300$ (400)~GeV and $\sim400$ (500)~GeV with $\sqrt s=14$ $(100)$~GeV, respectively.  
The cross sections for the VBF production and the associated production with different values of $v_\Delta$ can be readily obtained by rescaling, since both of them are proportional to $v_\Delta^2$.  On the other hand, the cross sections for the Drell-Yan production of $H_5^{++}H_5^{--}$ are independent of $v_\Delta$.

Regarding the calculation of $BR(H_5^{\pm \pm} \rightarrow \ell^{\pm} \ell^{\pm} + X^{\prime})$, we note that $H_5^{\pm \pm}$ has only two decay channels for sufficiently large $v_\Delta$. 
It decays into either $W^{\pm} W^{\pm}$ or $H_3^{\pm} W^{\pm}$, where $W^{\pm}$ can be off-shell, and $H_3^{\pm}$ further decays into SM particles, possibly involving $H_1$ at an intermediate stage.
The $W^{\pm}$ boson and the decay products of $H_3^{\pm}$ can decay into SM leptons, thereby giving rise to two-same-sign-lepton events.
In Fig.~\ref{h5ww}, we present scatter plots of the branching ratio of $H_5^{++}$ decaying into $W^{+} W^{+}$ and the $W^+$'s further decaying leptonically, $BR(H_5^{++} \rightarrow W^+(\rightarrow \ell^+ \nu_{\ell}) \, W^+(\rightarrow \ell^{\prime +} \nu_{\ell^{\prime}}) )$, with one of the $W^{+}$'s possibly off-shell, on the plane spanned by $m_{H_5}$ and $m_{H_3}$ for various values of $\alpha$ and $v_{\Delta}$.
The $H_3^{\pm}$ boson has a variety of decay channels, but the one with the dominant branching fraction is either $H_3^{+} \rightarrow h W^+$ or $H_3^{+} \rightarrow H_1 W^+$, depending on the mass spectrum and other parameters.
We thus present in Fig.~\ref{h5h3h} and \ref{h5h3h1} scatter plots of the products of branching ratios, $BR(H_5^{++} \rightarrow H_3^+ W^+(\rightarrow \ell^+ \nu_{\ell}) ) BR(H_3^+ \rightarrow h W^+(\rightarrow \ell^{\prime +} \nu_{\ell^{\prime}}))$ and $BR(H_5^{++} \rightarrow H_3^+ W^+(\rightarrow \ell^+ \nu_{\ell}) ) BR(H_3^+ \rightarrow H_1 W^+(\rightarrow \ell^{\prime +} \nu_{\ell^{\prime}}))$, 
 where the $W^{+}$'s can be off-shell.

We estimate the acceptance times efficiency, $A \times \epsilon$, for the processes of $H_5^{\pm \pm}$ production followed by its decay into final states containing two same-sign leptons.
We define the acceptance times efficiency as
\begin{align}
A \times \epsilon \equiv \frac{N_{\text{pass}}}{N_{\text{all}}} ~,
\end{align}
where $N_{\text{pass}}$ is the number of events that pass the selection criteria (a) through (e) defined below, and $N_{\text{all}}$ is the number of events for the processes of $p p \rightarrow H_5^{\pm \pm} + X, \, H_5^{\pm \pm} \rightarrow \ell^{\pm} \ell^{\prime \pm} + X^{\prime}$.
Note that $A \times \epsilon$ is almost the same for both $H_5^{++}$ and $H_5^{--}$.
In our simulation study, we consider the following criteria for selecting events with two same-sign leptons, based on which we estimate $A \times \epsilon$.  These criteria mimic part of the selection criteria used in the analysis of Ref.~\cite{sslepton}: \\

\noindent
(a) An electron is identified when its transverse momentum satisfies $p_{eT} > 10$~GeV and its pseudo-rapidity satisfies $\vert \eta_e \vert < 1.37$ or $1.52 < \vert \eta_e \vert < 2.47$.
A muon is identified when its transverse momentum satisfies $p_{\mu T} > 10$~GeV and its pseudo-rapidity satisfies $\vert \eta_{\mu} \vert < 2.5$.
\\
(b) The event should contain $e^{\pm} e^{\pm}$, $e^{\pm} \mu^{\pm}$, or $\mu^{\pm} \mu^{\pm}$.  The harder lepton $\ell_1$ should have a transverse momentum above 25~GeV, $p_{\ell_1 T} > 25$~GeV, and the other lepton $\ell^{\prime}_2$ should have a transverse momentum above 20~GeV, $p_{\ell^{\prime}_2 T} > 20$~GeV.
\\
(c) If the event further contains a lepton with a sign opposite to the same-sign lepton pair found above, then the event is vetoed.
\\
(d) The invariant mass of the two same-sign leptons should be larger than 15~GeV, $m(\ell, \ell^{\prime}) > 15$~GeV.
\\
(e) If the two same-sign leptons are electrons, their invariant mass should be below 70~GeV or above 110~GeV, $m(\ell_1, \ell_2) \, <$~70~GeV or $m(\ell_1, \ell_2) \, >$~110~GeV.
\\

\noindent
Based on the selection criteria (a) through (e), we estimate $A \times \epsilon$ for the following processes:
\begin{align}
p \, p \ &\rightarrow \ H_5^{++} \, j \, j~, \ \ \ H_5^{++} \ \rightarrow \ W^+(\rightarrow \ell^+ \nu) \, W^+(\rightarrow \ell^{\prime +} \nu)~; \label{wwdecay} \\
p \, p \ &\rightarrow \ H_5^{++} \, j \, j~, \ \ \ H_5^{++} \ \rightarrow \ H_3^+ \, W^+(\rightarrow \ell^+ \nu)~, \ \ \ H_3^+ \, \rightarrow \, h \, W^+(\rightarrow \ell^{\prime +} \nu)~; \label{wwhdecay} \\
p \, p \ &\rightarrow \ H_5^{++} \, j \, j~, \ \ \ H_5^{++} \ \rightarrow \ H_3^+ \, W^+(\rightarrow \ell^+ \nu)~, \ \ \ H_3^+ \, \rightarrow \, H_1 \, W^+(\rightarrow \ell^{\prime +} \nu)~. \label{wwh1decay}
\end{align}
Here $j$ denotes a jet originating from a quark involved in the VBF process, and $\ell, \ell^{\prime}$ represent an electron, a muon or a tau lepton, where the tau lepton will further decay leptonically.
Contributions from decays involving an off-shell $W$ boson are also taken into account.
We neglect the decay products of $h$ and $H_1$.  Although electrons and muons that come from the decay of $h$ or $H_1$ may affect the chance for an event to pass the selection criteria, we expect that their impact is negligibly small because more than 90\% of $h$'s decay into final states without electron or muon.
As for $H_1$, it mainly decays into $hh$, $t \bar{t}$, $W^+ W^-$ and $ZZ$ when $m_{H_1}$ is below $m_{H_3}$ and $m_{H_5}$.  Thus, more than 60\% of $H_1$'s decay into final states without electron or muon.
To evaluate $A \times \epsilon$, we perform a realistic detector-level simulation by using the Monte Carlo event generator {\tt MadGraph5}~\cite{mg5}, interfaced with {\tt PYTHIA6}~\cite{pythia6} for simulating parton showering and hadronization and with {\tt DELPHES3}~\cite{delphes3} for simulating detector responses and object reconstruction.
Our simulation of $H_5^{\pm \pm}$ production events is based on the leading-order (LO) QCD calculation of matrix elements.  
In the following, we present the acceptance times efficiency for the processes of Eq.~(\ref{wwdecay}) and Eq.~(\ref{wwhdecay}).

\begin{figure}[!ht]
\centering
\includegraphics[scale=0.6]{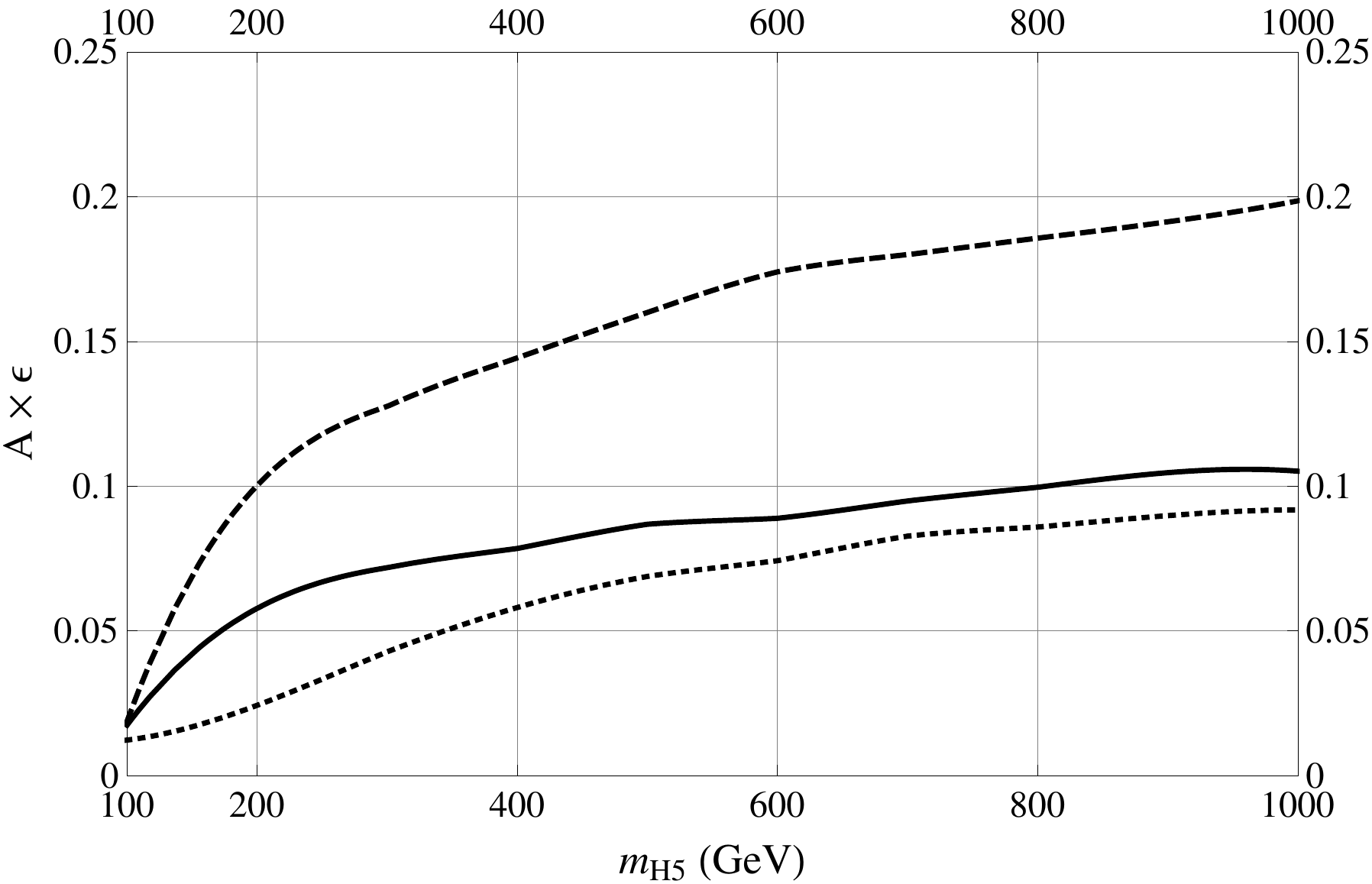}
\caption{The acceptance times efficiency, $A \times \epsilon$, for the process of Eq.~(\ref{wwdecay}) in 14-TeV $pp$ collisions with the selection criteria (a) through (e) as a function of $m_{H_5}$.
The solid curve corresponds to the final state with $\mu^\pm \mu^\pm$, the dashed curve to that with $\mu^\pm e^\pm$, and the dotted curve to that with $e^\pm e^\pm$.
\label{Ae1}}
\end{figure}

In Fig.~\ref{Ae1}, we plot the acceptance times efficiency for the process of Eq.~(\ref{wwdecay}) in 14-TeV $pp$ collisions with the selection criteria (a) to (e) as a function of the $H_5^{\pm \pm}$ mass.
In Table~\ref{Ae2}, we present $A \times \epsilon$ for the process of Eq.~(\ref{wwhdecay}) in 14-TeV $pp$ collisions with the selection criteria (a) to (e), for various benchmark values of $(m_{H_5}, \, m_{H_3})$, all of which can be realistic mass spectra consistent with all theoretical and experimental constraints, as can be read from Fig.~\ref{h1mass} and other figures.
The acceptance times efficiency is calculated for the three channels with $e^{\pm} e^{\pm}$, $e^{\pm} \mu^{\pm}$ and $\mu^{\pm} \mu^{\pm}$.
We do not show $A \times \epsilon$ for the process of Eq.~(\ref{wwh1decay}) because it depends on $m_{H_5}$, $m_{H_3}$, and $m_{H_1}$ and hence is highly mass spectrum-dependent.  This fact renders the process of Eq.~(\ref{wwh1decay}) an ineffective channel for the discovery of the GM model.

\begin{table}
\begin{center}
\begin{tabular}{|c|c|c|c|c|c|} \hline
$(m_{H_5}, \, m_{H_3})$~GeV                           & (300, 200) & (400, 250) & (400, 300) & (500, 300) & (500, 350) \\ \hline
$A \times \epsilon$ for $e^{\pm} e^{\pm}$ channel     & 0.021      & 0.032      & 0.029      & 0.046      & 0.041      \\ \hline
$A \times \epsilon$ for $e^{\pm} \mu^{\pm}$ channel   & 0.067      & 0.11       & 0.10       & 0.13       & 0.12       \\ \hline
$A \times \epsilon$ for $\mu^{\pm} \mu^{\pm}$ channel & 0.038      & 0.063      & 0.059      & 0.075      & 0.074      \\ 
\end{tabular}
\begin{tabular}{|c|c|c|c|c|c|} \hline \hline
$(m_{H_5}, \, m_{H_3})$~GeV                           & (500, 400) & (600, 400) & (600, 450) & (600, 500) & (700, 500) \\ \hline
$A \times \epsilon$ for $e^{\pm} e^{\pm}$ channel     & 0.036      & 0.051      & 0.050      & 0.040      & 0.058      \\ \hline
$A \times \epsilon$ for $e^{\pm} \mu^{\pm}$ channel   & 0.11       & 0.14       & 0.13       & 0.12       & 0.15       \\ \hline
$A \times \epsilon$ for $\mu^{\pm} \mu^{\pm}$ channel & 0.069      & 0.081      & 0.076      & 0.069      & 0.083      \\ 
\end{tabular}
\begin{tabular}{|c|c|c|c|c|c|} \hline \hline
$(m_{H_5}, \, m_{H_3})$~GeV                           & (700, 550) & (700, 600) & (800, 600) & (800, 650) & (800, 700) \\ \hline
$A \times \epsilon$ for $e^{\pm} e^{\pm}$ channel     & 0.053      & 0.045      & 0.062      & 0.056      & 0.046      \\ \hline
$A \times \epsilon$ for $e^{\pm} \mu^{\pm}$ channel   & 0.14       & 0.13       & 0.15       & 0.14       & 0.13       \\ \hline
$A \times \epsilon$ for $\mu^{\pm} \mu^{\pm}$ channel & 0.083      & 0.074      & 0.087      & 0.081      & 0.077      \\ \hline
\end{tabular}
\caption{$A \times \epsilon$ for the process of Eq.~(\ref{wwhdecay}) in 14-TeV $pp$ collisions with the selection criteria (a) to (e) for various benchmark values of $(m_{H_5}, \, m_{H_3})$~GeV.}
\label{Ae2}
\end{center}
\end{table}

Finally, we estimate the number of background events that arise from SM processes and pass the selection criteria (a) through (e).
The dominant sources of background events are the production of $W^{\pm} Z$ and $ZZ$ followed by their decays into leptons including tau leptons:
\begin{align}
p \, p \ &\rightarrow \ W^+(\rightarrow \ell^+ \nu) \, Z(\rightarrow \ell^{\prime +} \ell^{\prime -})~, \label{wzbkg} \\
p \, p \ &\rightarrow \ Z(\rightarrow \ell^+ \ell^-) \, Z(\rightarrow \ell^{\prime +} \ell^{\prime -})~, \label{zzbkg}
\end{align}
 where $\ell, \ell^{\prime}$ represent an electron, a muon or a tau lepton.
We further take into account the background coming from the $W^{\pm} W^{\pm}$ production process:
\begin{align}
p \, p \ &\rightarrow \ W^+(\rightarrow \ell^+ \nu) \, W^+(\rightarrow \ell^{\prime +} \nu) \, j \, j~, \label{wwbkg} 
\end{align}
where $j$ denotes a jet originating from a quark in the subprocesses of $u u \rightarrow W^+ W^+ d d$ or $d d \rightarrow W^- W^- u u$.  Although its contribution is subdominant, this process gives an irreducible background to the same-sign lepton signal defined in terms of the selection criteria (a) to (e).
We note that charge misidentification of electrons can be another dominant source of backgrounds for the channels involving an electron. 
Nevertheless, we do not estimate its contribution as it is beyond the scope of our theoretical study.
To assess the number of background events, we also perform a realistic detector-level simulation by using MadGraph5 \cite{mg5} interfaced with PYTHIA6 \cite{pythia6} and DELPHES3 \cite{delphes3}.
The generation of background events is based on the LO QCD calculation of matrix elements, but we take into account the next-to-leading order (NLO) QCD effects by multiplying the number of background events with a $K$-factor derived as the ratio of the $W^{\pm} W^{\pm}$, $W^{\pm} Z$ or $ZZ$ production cross section calculated at the NLO divided by the corresponding one at the LO.
In Table~\ref{bkg}, we present the LO and NLO production cross sections, $\sigma_{{\rm LO}}$ and $\sigma_{{\rm NLO}}$, the $K$-factor estimated as above, and the number of background events that pass the selection criteria (a) through (e) divided by the integrated luminosity, $N_{bkg}/L$ (the $K$-factor is multiplied already), for each of the $W^{\pm} W^{\pm}$, $W^{\pm} Z$ and $ZZ$ production processes, for each final state containing $e^{\pm}e^{\pm}$, $e^{\pm}\mu^{\pm}$ or $\mu^{\pm}\mu^{\pm}$.

\begin{table}
\begin{center}
\begin{tabular}{|c|c|c|c|c|c|} \hline
Process              & $W^{+} Z$            & $W^{-} Z$            & $ZZ$                 & $W^{+} W^{+}$ & $W^{-} W^{-}$ \\ \hline
$\sigma_{{\rm LO}}$  & $1.72\times 10^4$~fb & $1.05\times 10^4$~fb & $1.06\times 10^4$~fb & 257~fb        & 113~fb        \\ \hline
$\sigma_{{\rm NLO}}$ & $3.01\times 10^4$~fb & $1.94\times 10^4$~fb & $1.52\times 10^4$~fb & 340~fb        & 161~fb        \\ \hline
$K$-factor           & 1.75                 & 1.84                 & 1.43                 & 1.33          & 1.42          \\ \hline
$N_{bkg}(e^{\pm}e^{\pm})/L$     & 245~fb    & 158~fb               & 110~fb               & 6.70~fb       & 3.17~fb       \\ \hline
$N_{bkg}(e^{\pm}\mu^{\pm})/L$   & 723~fb    & 466~fb               & 379~fb               & 19.3~fb       & 9.12~fb       \\ \hline
$N_{bkg}(\mu^{\pm}\mu^{\pm})/L$ & 370~fb    & 239~fb               & 178~fb               & 11.5~fb       & 5.45~fb       \\ \hline
 \end{tabular}
\caption{The production cross sections calculated at the LO, $\sigma_{{\rm LO}}$, and at the NLO, $\sigma_{{\rm NLO}}$, the estimated $K$-factor, and the number of background events that pass the selection criteria (a) to (e) divided by the integrated luminosity, $N_{bkg}/L$ (with the $K$-factor multiplied) for each of the $W^{\pm} W^{\pm}$, $W^{\pm} Z$ and $ZZ$ production processes and for each final state containing $e^{\pm}e^{\pm}$, $e^{\pm}\mu^{\pm}$ or $\mu^{\pm}\mu^{\pm}$ at the 14-TeV LHC.} 
 \label{bkg}
\end{center}
\end{table}

With Figs.~\ref{muhdiphoton},~\ref{muhzphoton},~\ref{h5ww},~\ref{h5h3h},~\ref{fig:cross_section},~\ref{Ae1} and Tables~\ref{Ae2},~\ref{bkg}, we can evaluate the significance of the same-sign lepton signal for the most general mass spectra of the GM model.
This is done in the following manner.
First, we take a set of parameters $(v_{\Delta}, \alpha, m_{H_5}, m_{H_3})$ for which we want to study the discovery potential of the GM model at the LHC.
From Figs.~\ref{muhdiphoton} and \ref{muhzphoton}, we check if there exists a mass spectrum that satisfies all the theoretical and experimental constraints.
We then look up the corresponding values of $BR(H_5^{++} \rightarrow W^+(\rightarrow \ell^+ \nu_{\ell}) \, W^+(\rightarrow \ell^{\prime +} \nu_{\ell^{\prime}}) )$, $BR(H_5^{++} \rightarrow H_3^+ W^+(\rightarrow \ell^+ \nu_{\ell}) ) BR(H_3^+ \rightarrow h W^+(\rightarrow \ell^{\prime +} \nu_{\ell^{\prime}}))$ in Figs.~\ref{h5ww} and \ref{h5h3h} and the VBF production cross section of $H_5^{\pm \pm}$ in $pp$ collisions in Fig.~\ref{fig:cross_section}.
The acceptance times efficiency $A\times \epsilon$ for these values of $m_{H_5}, m_{H_3}$ can be estimated using Fig.~\ref{Ae1} and Table~\ref{Ae2} for the two processes of $p p \rightarrow H_5^{++} \, j \, j, \, H_5^{++} \rightarrow W^+(\rightarrow \ell^+ \nu) \, W^+(\rightarrow \ell^{\prime +} \nu)$ and $p p \rightarrow H_5^{++} \, j \, j, \, H_5^{++} \rightarrow H_3^+ \, W^+(\rightarrow \ell^+ \nu), \, H_3^+ \rightarrow h \, W^+(\rightarrow \ell^{\prime +} \nu)$, respectively.
Finally, we evaluate the number of events with a same-sign light lepton pair arising from the production and decay of $H_5^{\pm \pm}$ by Eq.~(\ref{numberofsignals}), and compare it with the number of SM background events that can be extracted from Table~\ref{bkg} to derive the significance of the signal for some value of the integrated luminosity.
We note in passing that the true value of the significance of the same-sign light lepton signal can be larger than evaluated above, because the vector boson associated and Drell-Yan productions of $H_5^{\pm \pm}$ as well as the process of Eq.~(\ref{wwh1decay}), which are neglected in the above evaluation, also contribute to the signal.
Hence the significance evaluated following the above-described procedure actually corresponds to the lower bound.

The results given in the previous section can be extended to a 100 TeV hadron collider.  Since the cross section for the VBF production of $H_5^{++}$ is much larger than that for the associated production (see Fig.~\ref{fig:cross_section}), the process of $pp\to H_5^{++}jj$ followed by $H_5^{++}$ decays can be used to estimate the observability of $H_5^{++}$.  The transverse momenta of $H_5^{++}$'s in the VBF process tend to zero.  Therefore, the acceptance times efficiency for the decay products of $H_5^{++}$ depends only on the mass of $H_5^{++}$.  Note that no selection cuts are put on the two forward jets in the VBF process.  Otherwise, it will depend on the collision energy.  Thus, we can safely assume that the acceptance times efficiency does not vary significantly from 14-TeV to 100-TeV colliders.  In other words, the values of the acceptance times efficiency given in Fig.~\ref{Ae1} can simply be applied to the case with $\sqrt{s}=100$~TeV. 
For a future $pp$ collider with 100-TeV collision energy, it is sufficient to observe the production of $H_5^{++}$ to test the GM model.

\section{Conclusions \label{sec:summary}}

In this work, we have studied the most general mass spectrum of the exotic Higgs bosons of the Georgi-Machacek model that is allowed by theoretical and experimental constraints.  
As theoretical constraints, we have taken into account the unitarity of the perturbation theory and the stabiliy of the electroweak symmetry breaking vacuum.
On the other hand, the experimental constraints we have considered are the electroweak precision tests, the $Z b \bar{b}$ vertex measurement and the Higgs boson signal strength data.
Here we used the latest Higgs boson signal strength data to find the allowed region at $1\sigma$ and $2\sigma$ confidence level on the plane of the triplet VEV $v_\Delta$ and the singlet mixing angle $\alpha$. 
The diphoton channel was not included in the above analysis, because its strength depends on the charged Higgs boson mass spectrum, in addition to $v_\Delta$ and $\alpha$.
From the $2\sigma$ region, we identified twelve example sets of $(v_\Delta,\alpha)$ for subsequent analyses.
We have performed a comprehensive parameter scan for the exotic Higgs boson mass spectrum allowed by these constraints.  
We found that the most probable spectra are either the normal hierarchy ($m_{H_5} > m_{H_3} > m_{H_1}$) or the inverted hierarchy ($m_{H_5} < m_{H_3} < m_{H_1}$), though other scenarios are generally possible as well. 
We worked out the signal strengths of diphoton and $Z\gamma$ channels of the SM-like Higgs boson via the gluon-gluon fusion process, the decay widths of $H_5^{\pm\pm}$ and $H_3^\pm$, branching ratios of various cascade decays of $H_5^{++}$ and $H_3^{+}$, and branching ratios of the $H_1 \to hh / W^+W^-$ decay.

As to collider signatures of the model, we focused on the production and decays of the $H_5^{\pm\pm}$ boson.
We computed the cross sections of the vector boson fusion, vector boson associated, and Drell-Yan productions of the $H_5^{\pm\pm}$ boson for the 14-TeV LHC and a future 100-TeV $pp$ collider.  
In accord with our selection criteria set for signal events, acceptance times efficiency for signal events was evaluated for the three processes: $p p \rightarrow H_5^{++} j j$ with $H_5^{++} \rightarrow W^+(\rightarrow \ell^+ \nu) W^+(\rightarrow \ell^{\prime +} \nu)$; 
$p p \rightarrow H_5^{++} j j$ with $H_5^{++} \rightarrow H_3^+ W^+(\rightarrow \ell^+ \nu)$ and $H_3^+ \rightarrow h W^+(\rightarrow \ell^{\prime +} \nu)$; and 
$p p \rightarrow H_5^{++} j j$ with $H_5^{++} \rightarrow H_3^+ W^+(\rightarrow \ell^+ \nu)$ and $H_3^+ \rightarrow H_1 W^+(\rightarrow \ell^{\prime +} \nu)$.  This part was done at the leading order in QCD and with a realistic detector-level simulation.  For background events, a similar simulation was carried out at the leading order of QCD and then scaled by the $K$-factor to the next-to-leading order.
We have argued that, by combining our estimates on the production cross section of $H_5^{\pm\pm}$, acceptance times efficiency for the signal and the branching ratios of cascade decays of $H_5^{\pm\pm}$, and by comparing them with SM background estimates,
 we can evaluate prospects for observing a signal of the GM model for its most general mass spectrum.

Finally, we have also made a remark that the same acceptance times efficiency that we computed for the 14-TeV LHC can be applied to the case of a future 100-TeV $pp$ collider to a good approximation because no forward jet tagging is required in our proposed selection cuts.

{\it Note Added:} 
Recently, the ATLAS and CMS Collaborations have observed through the diphoton decay mode a hint of a resonance at about 750~GeV and a width of about 45~GeV~\cite{ATLAS,CMS:2015dxe}.  In the GM model, $H_1$ is a good candidate for the 750-GeV resonance because it can be produced in the s-channel in $pp$ collisions and decay into diphotons.  The mass of $H_1$ can be read from Fig.~\ref{h1mass}.  For example, there is some parameter space in the plot for $v_\Delta = 10$~GeV and $\alpha = -10^\circ$ that gives $m_{H_1} \simeq 750$~GeV.  A quick estimate shows that its decay width is about 10~GeV and the cross section of the diphoton channel at the 13-TeV LHC is about 1~fb.  We leave the detailed analysis to a future work.

\acknowledgments

This work was supported in part by the Ministry of Science and Technology of Taiwan under Grant Nos. MOST-100-2628-M-008-003-MY4, 104-2628-M-008-004-MY4, and 103-2811-M-008-058,
and by the National Research Foundation of Korea (NRF) Research Grant NRF-2015R1A2A1A05001869 (TY).

\appendix

\end{document}